\documentclass[pra, notitlepage, showpacs, floatfix, reprint, nobalancelastpage,citeautoscript]{revtex4-1}

\usepackage[T1]{fontenc}
\usepackage{rotating} 
\usepackage{caption}
\usepackage{subcaption}
\usepackage[latin9]{inputenc}
\usepackage[strict]{changepage}
\usepackage{amsmath}
\usepackage{amssymb}
\usepackage{bbm}
\usepackage{braket}
\usepackage{xcolor}
\usepackage[shortlabels]{enumitem}
\usepackage{array, makecell}
\usepackage{pifont}
\newcommand{\cmark}{\ding{51}}
\newcommand{\xmark}{\ding{55}}
\usepackage{float}
\usepackage{ulem}
\allowdisplaybreaks

\usepackage{graphicx}
\usepackage[colorlinks=true]{hyperref}  
\hypersetup{
    bookmarks=true,         
    unicode=false,          
    pdftoolbar=true,        
    pdfmenubar=true,        
    pdffitwindow=false,     
    pdfstartview={FitH},    
    pdftitle={Wess-Zumino-Witten terms in twisted bilayer graphene},    
    pdfauthor={Christos, Sachdev, Scheurer},     
    pdfsubject={},   
    pdfcreator={},   
    pdfproducer={}, 
    pdfkeywords={} {} {}, 
    pdfnewwindow=true,      
    colorlinks=true,       
    linkcolor=magenta, 
    citecolor=blue,        
    filecolor=magenta,      
    urlcolor=blue           
}

\newcommand{\equref}[1]{Eq.~(\ref{#1})}

\newcommand{\secref}[1]{Sec.~\ref{#1}}
\newcommand{\figref}[1]{Fig.~\ref{#1}}
\newcommand{\refcite}[1]{Ref.~\onlinecite{#1}}
\newcommand{\refscite}[1]{Refs.~\onlinecite{#1}}

\newcommand{\tableref}[1]{Table~\ref{#1}}
\newcommand{\appref}[1]{Appendix~\ref{#1}}

\newcommand{\pdagger}{{\phantom{\dagger}}}
\newcommand{\diff}{\mathrm{d}}

\renewcommand{\approx}{\simeq}

\renewcommand{\vec}[1]{\boldsymbol{#1}}

\definecolor{wrongultramarine}{rgb}{1,0.5,0}

\begin{document}

\begin{flushright}
\href{https://arxiv.org/abs/2007.00007}{arXiv:2007.00007}
\end{flushright}

\title{Superconductivity, correlated insulators, and\\ Wess-Zumino-Witten terms in twisted bilayer graphene}

\author{Maine Christos}
\affiliation{Department of Physics, Harvard University, Cambridge MA 02138, USA}

\author{Subir Sachdev}
\affiliation{Department of Physics, Harvard University, Cambridge MA 02138, USA}

\author{Mathias S.~Scheurer}
\affiliation{Department of Physics, Harvard University, Cambridge MA 02138, USA}

\begin{abstract}
Recent experiments on twisted bilayer graphene have shown a high-temperature parent state with massless Dirac fermions and broken electronic flavor symmetry; superconductivity and correlated insulators emerge from this parent state at lower temperatures. We propose that the superconducting and correlated insulating orders are connected by Wess-Zumino-Witten terms, so that defects of one order contain quanta of another order and skyrmion fluctuations of the correlated insulator are a `mechanism' for superconductivity. We present a comprehensive listing of plausible low-temperature orders, and the parent flavor symmetry breaking orders. The previously characterized topological nature of the band structure of twisted bilayer graphene plays an important role in this analysis.
\end{abstract}

\maketitle



\section{Introduction}

A number of recent experimental studies of twisted bilayer graphene (TBG) \cite{macdonald2019bilayer,SenthilJournalClub,Efetov19,Young19,Li20,PabloNematicity,WSe2TBG} have explored its phase diagram as a function of electron density and temperature, and found correlated insulating states at integer filling fractions separating the superconducting domes at low temperatures. Complementary information has emerged from scanning probe measurements \cite{Yazdani19,Ilani19}, showing a cascade of phase transitions with `Dirac revivals' at the integer filling fractions $\nu$: while the bare flat bands of TBG only exhibit Dirac cones around charge neutrality ($\nu=0$), additional flavor symmetry breaking is argued by the authors to lead to the re-emergence of Dirac cones at non-zero integer $\nu$; this defines the high-temperature ``parent state out of which the more fragile superconducting and correlated insulating ground states emerge'' \cite{Ilani19}.

Here we propose a common origin for the superconducting and correlated insulating states. We will connect these orders by Wess-Zumino-Witten (WZW) terms \cite{WESS197195,WITTEN1983422} with quantized co-efficients. The WZW term associates a Berry phase with spatiotemporal textures of the different order parameters. Textures or defects in one order parameter contain quanta of the other order, leading to proximate phases in which different order parameters condense and break associated symmetries.
In condensed matter systems in two spatial dimensions, WZW terms first appeared \cite{TanakaHu05,SenthilFisher06,ABANOV2000685,Ma:2019ysf,Nahum:2019fjw,Mong20,He:2020azu} in studies of the interplay between the antiferromagnetic N\'eel and valence bond solid order parameters on the square lattice \cite{PhysRevLett.62.1694}. They also appeared earlier in the interplay between these orders in one dimension \cite{PhysRevB.36.5291}. Indeed, studies of field theories with WZW and related terms have been crucial to our global understanding of the phase diagrams of quantum spin systems in both one and two spatial dimensions \cite{PhysRevB.36.5291,Wang2017}. 

Grover and Senthil \cite{PhysRevLett.100.156804} extended these ideas to include the superconducting order for fermions with Dirac dispersion on the honeycomb lattice, and this will be relevant for our analysis here. They showed that skyrmion textures in the order parameter complementary to superconductivity are then electrically charged. The mechanism of associating electric charge with a topological texture allows for electrical transport coming from skyrmion defects; this has been discussed, in particular, in quantum Hall ferromagnets \cite{PhysRevLett.64.1313,PhysRevB.47.16419,PhysRevB.51.5138}, which could also be relevant for the description of TBG \cite{2020PhRvB.101p5141C}. The skyrmion fluctuations of the complementary order are then a `mechanism' for superconductivity, analogous to skyrmion fluctuations ({\it i.e.\/} hedgehogs) in the N\'eel order being a mechanism for valence bond solid order in square lattice antiferromagnets \cite{PhysRevLett.62.1694}.

TBG has massless Dirac fermions at charge neutrality \cite{2011PNAS..10812233B,CastroNeto2011,PhysRevX.8.031089}, and these extend all the way to a `chiral limit' \cite{2019PhRvL.122j6405T} when the bands are exactly flat and Landau-level like.
Interestingly, WZW terms can also be obtained from exactly flat Landau levels \cite{2015PhRvL.114v6801L}. The arguments of Yao and Lee \cite{PhysRevB.82.245117} show that the same quantized WZW term is obtained from a theory which focuses on the vicinity of dispersing Dirac nodes, as would be obtained from a theory which considers the flat (or nearly flat) band across the entire moir\'e Brillouin zone. We will choose to use the first method here, and employ the theory of linearly-dispersing Dirac fermions at all momenta, while imposing the symmetry constraints arising from their embedding in the moir\'e Brillouin zone. This approach will allow us to account for the `Dirac revivals' observed in recent experiments \cite{Yazdani19,Ilani19} in a relatively straightforward manner.

We will begin in \secref{ModelsSymmetries} by introducing the Dirac fermion model of TBG, and discuss its symmetry and topological properties. \secref{SuperconductingStates} will list possible spin-singlet superconducting states. \secref{WZWNoOtherOrders} will introduce the partner order parameters $m_j$, which combine in WZW terms with the superconducting orders. Without additional flavor symmetry breaking by a parent (or `high temperature') order $M$, these $m_j$ characterize the correlated insulators near $\nu=0$. A discussion of the parent orders $M$, which are responsible for the Dirac revivals \cite{Yazdani19,Ilani19}, appears in \secref{SymmetryBreakingAtHighT}. These $M$ can combine with suitable $m_j$ to form correlated insulators near $\nu = \pm 2$. The extension to superconductors with triplet pairing appears in \secref{TripletPairing}.

While our work was in progress, we learnt of the work of Khalaf {\it et al.\/} \cite{2020arXiv200400638K}, which contains some related ideas; we discuss the connections to their work further in Appendix~\ref{ConnectionToAshvin}.

\section{Model and symmetries}
\label{ModelsSymmetries}

To construct the superconducting order parameters, we could, in principle, start with a tight-binding model \cite{PhysRevX.8.031088,VafekModel,FuModel,PhysRevX.8.031089} of the quasi-flat (and necessary auxiliary) bands, write down pairing terms on the lattice, and then project onto the Dirac cones. Since we, however, do not have a clear understanding (other than symmetry) how the pairing states should look like in real space, we here proceed differently by working entirely in momentum space. Denoting the projection (implemented by operator $P$) of the non-interacting Hamiltonian, $H_0$, onto the quasi-flat bands by $H_{\text{FB}} = P H_0 P$, neglecting any coupling between the different valleys of the original graphene layers (associated with index $v=\pm$), and neglecting spin-orbit coupling, the Hamiltonian must be of the form
\begin{equation}
	H_{\text{FB}} = \sum_{\vec{k}} c^\dagger_{\vec{k},\sigma,v,s} \left[ \delta_{ss'}\epsilon_{\vec{k}v}+ \vec{g}_v(\vec{k}) \cdot \vec{\rho}_{ss'} \right] c^\pdagger_{\vec{k},\sigma,v,s'}. \label{GeneralFormOfModel}
\end{equation}
Here $\sigma$ denotes the spin of the electrons, $s$ and $\rho_{x,y,z}$ are indices and Pauli matrices in ``generalized sublattice space'' that gives rise to the Dirac cones to be discussed below.

\begin{figure}[tb]
     \centering
     \begin{subfigure}[b]{0.28\textwidth}
         \centering
         \includegraphics[width=\textwidth]{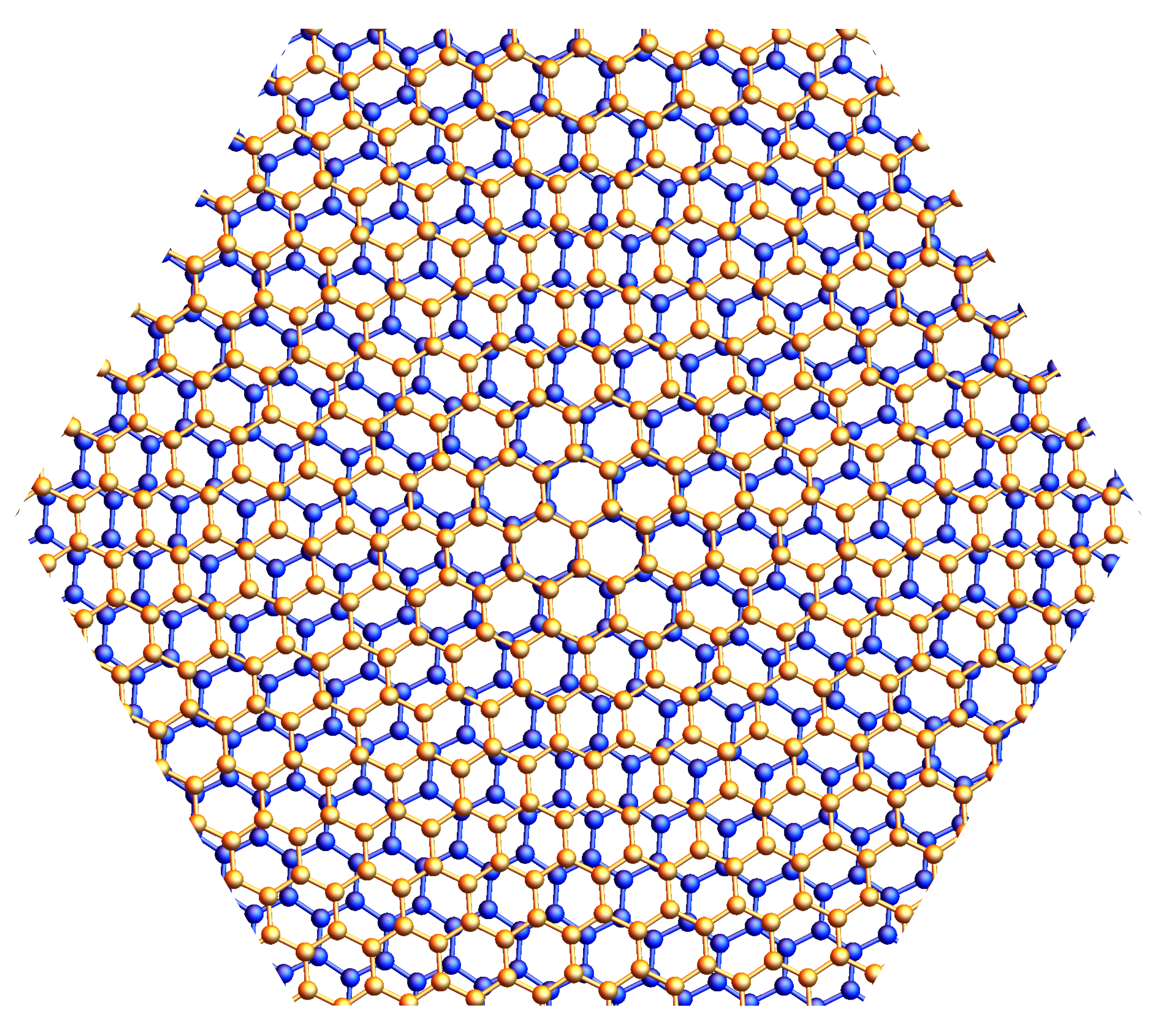}
         \caption{TBG lattice at small twist angle $\theta$.}
         \label{bilayer}
     \end{subfigure}
     \hfill
    \begin{subfigure}[b]{0.15\textwidth}
         \centering
         \includegraphics[width=\textwidth]{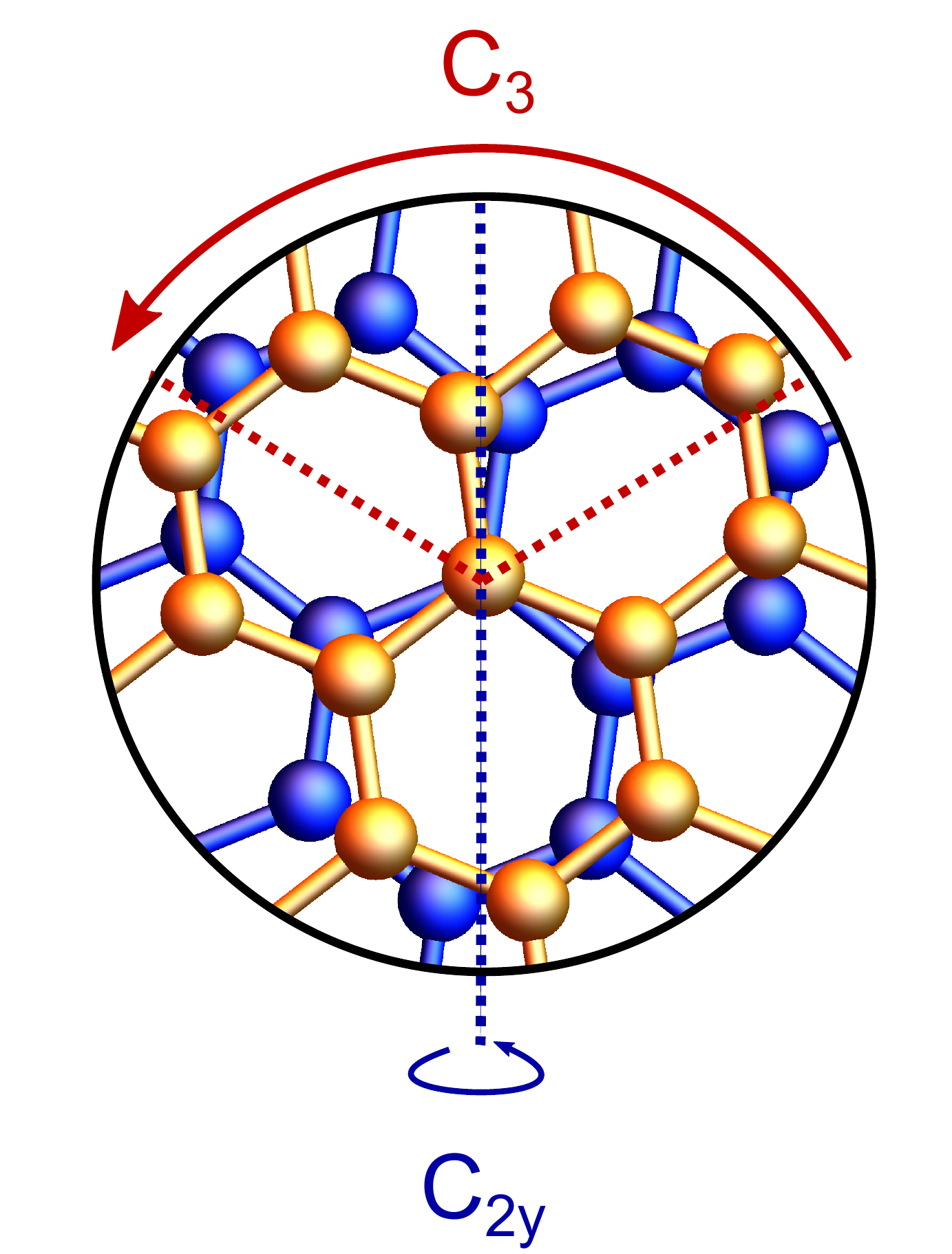}
         \caption{Action of exact spatial symmetries $C_3$ and $C_{2y}$.}
         \label{spacesymm}
     \end{subfigure}
     \begin{subfigure}[b]{0.25\textwidth}
         \centering
         \includegraphics[width=\textwidth]{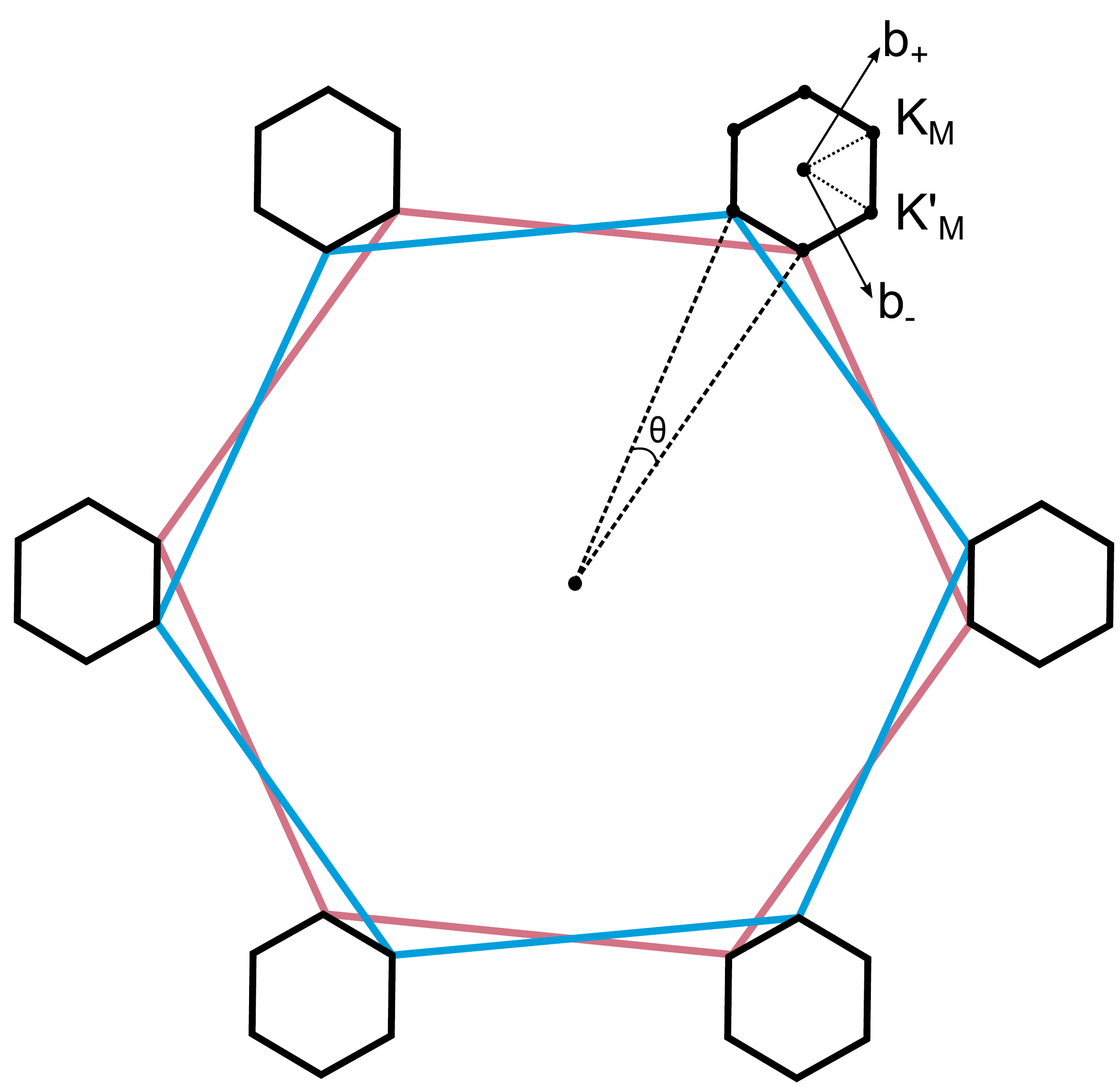}
         \caption{Brillouin zones for layer 1 (blue) and layer 2 (red) and mini Brillouin zone.}
         \label{MBZ}
     \end{subfigure}
     \begin{subfigure}[b]{0.2\textwidth}
         \centering
         \includegraphics[width=\textwidth]{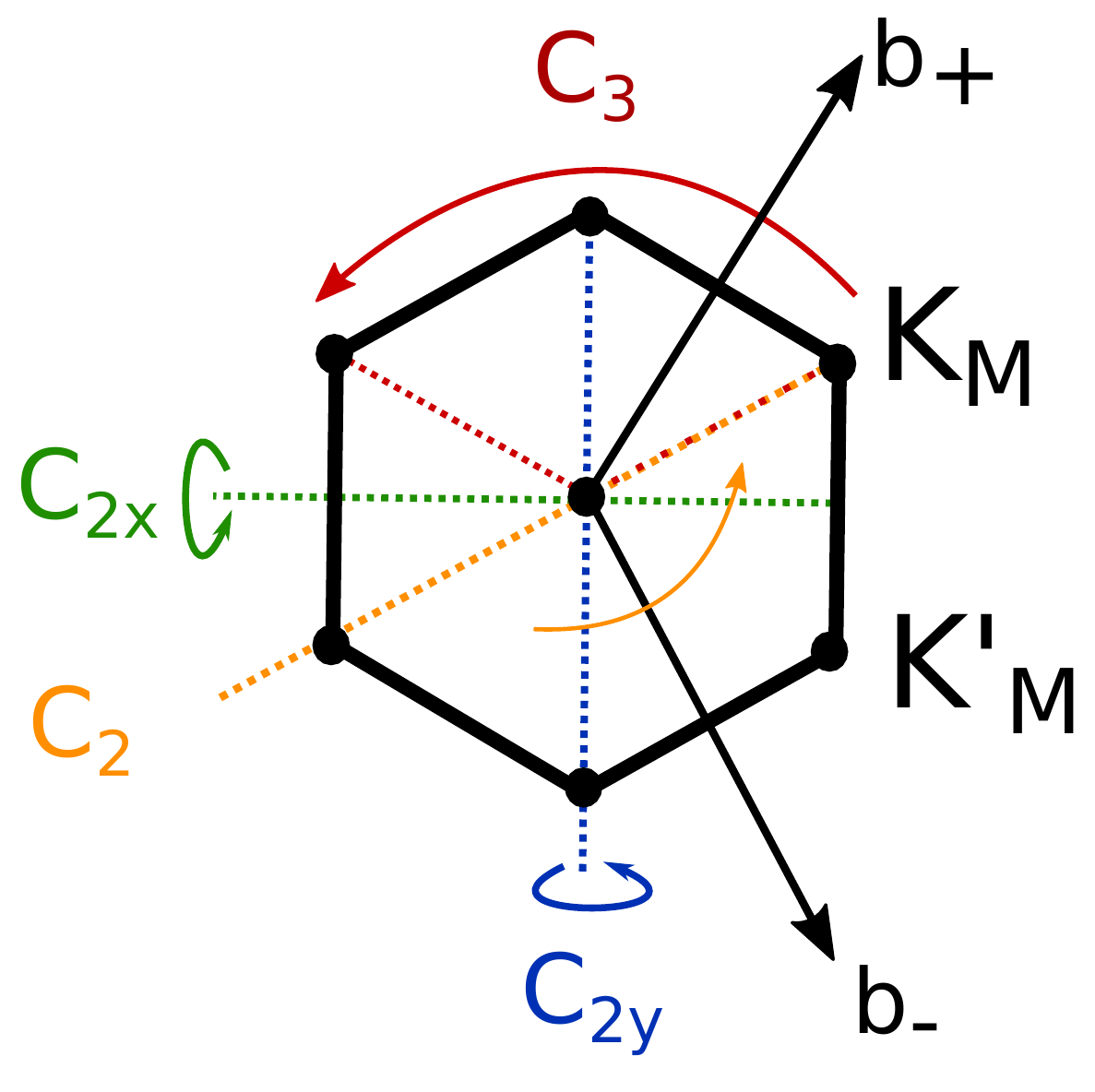}
         \caption{Action of exact and emergent symmetries in the moir\'e reciprocal lattice with reciprocal lattice vectors $b_{\pm}=\frac{2\pi}{a\sqrt{3}}(1,\pm\sqrt{3})$.}
         \label{BZsymm}
     \end{subfigure}
        \caption{Lattice geometry and symmetries. As discussed in the main text, we impose $C_2$ as an emergent symmetry. The primitive vectors of the moir\'e Bravais lattice will be denoted by $\vec{a}_r = a(\cos \pi/6, r\,\sin \pi/6)$, $r=\pm$, with moir\'e lattice constant $a$, which we will set to $a=1$.}
        \label{fig:Lattice}
\end{figure}

As can be seen in \figref{fig:Lattice}, the lattice does not have an exact $C_2$ symmetry, but we will impose it as it emerges approximately at small twist angles; this follows naturally from the fact that, at small twist angles, the difference between the twist axis going through an AA-site or through the center of a hexagon, which leads to an exact $C_2$ symmetry, vanishes asymptotically. To specify the basis for the $\rho_{x,y,z}$ matrices in \equref{GeneralFormOfModel}, we choose the representation $R_{C_2} = \rho_x \tau_x \sigma_0$ with $\tau_j$ and $\sigma_j$ acting in valley and spin space, respectively; as required, $C_2$ flips the valley and we choose it to flip generalized sublattice as well (to resemble the representation of $C_2$ in single-layer graphene). This fixes the representation, $\Theta$, of time-reversal: it has to act between different valleys ($\propto \tau_{x,y}$) and we want it to be ``on-site'' in generalized sublattice space ($\propto \rho_{0,z}$); out of these four options, only $\Theta = \sigma_y \rho_0 \tau_x \mathcal{K}$ (where $\mathcal{K}$ is the complex conjugation operator) is consistent with $\Theta^2 = -\mathbbm{1}$ and $[R_{C_2},\Theta]=0$.

Combining these two symmetries, we get $R_{C_2}\Theta = \sigma_y \rho_x\tau_0 \mathcal{K}$, which acts locally in $\vec{k}$-space and forces \equref{GeneralFormOfModel} to have the form
\begin{equation}
	H_{\text{FB}} = \sum_{\vec{k}} c^\dagger_{\vec{k},\sigma,v,s} \left[\rho_0\epsilon_{\vec{k}v}+g^x_v(\vec{k}) \rho_x+g^y_v(\vec{k}) \rho_y\right]_{ss'} c^\pdagger_{\vec{k},\sigma,v,s'},
\end{equation}
where 
\begin{equation}
	g^x_v(\vec{k}) = g^x_{-v}(-\vec{k}), \quad g^y_v(\vec{k}) = -g^y_{-v}(-\vec{k}), \quad \epsilon_{\vec{k}v} = \epsilon_{-\vec{k}-v}, \label{ConstraintFromC2}
\end{equation}
due to the $C_2$ symmetry. A topological aspect of twisted bilayer graphene, which is crucial for the structure of the WZW terms, is that it has two Dirac cones per spin and valley at the $\text{K}_\text{M}$ and $\text{K'}_\text{M}$ points (in the following referred to as ``mini valley'') of the moir\'e Brillouin zone with \textit{the same} chirality \cite{CastroNeto2011,PhysRevX.8.031089,Bernevig19}, {\it i.e.\/}, $\vec{g}^{xy}_v := (g^x_v(\vec{k}),g^y_v(\vec{k}))$ vanishes at these momentum points and winds around once for any contour surrounding $\text{K}_\text{M}$ or $\text{K'}_\text{M}$. Due to \equref{ConstraintFromC2}, the chirality must be opposite in the other (non-mini) valley. We, thus, consider the following low-energy theory where we only keep the Dirac cones at $\text{K}_\text{M}$ ($p=+$) and at $\text{K'}_\text{M}$ ($p=-$):
\begin{align}\begin{split}
	H_{\text{LE}} =  \,\sum_{\vec{q}}^{\Lambda} &f^\dagger_{\vec{q},\sigma,v,s,p} \Bigl[ v\, \nu^{pv}_x  q_x \rho_x  \\ &\qquad + \nu^{pv}_y q_y  \rho_y + \epsilon_{pv} \Bigr]_{ss'} f^\pdagger_{\vec{q},\sigma,v,s',p}, \label{FinalLowEnergyTheory}
\end{split}\end{align}
where the velocities $\nu_{x,y}^{pv}$ and $\epsilon_{pv}$ only depend on the product $p\cdot v = \pm$ and momenta $\vec{q}$ (cut-off as $|\vec{q}| < \Lambda$) are measured relative to the respective Dirac point. Using $\mu_{x,y,z}$ to denote Pauli matrices in the mini-valley space, the representations of all physical symmetries of the system are summarized in \tableref{RepresentationsOfSymmetries}. Note that these symmetries further imply $\nu^{pv}_x = \nu^{pv}_y = \nu$ and $\epsilon_{pv} = \epsilon$, independent of $pv$. Suppressing indices and setting $\epsilon=0$ without loss of generality, \equref{FinalLowEnergyTheory} can thus be written as
\begin{equation}
	H_{\text{LE}} = \nu \sum_{\vec{q}}^{\Lambda} f^\dagger_{\vec{q}} \left( q_x \gamma_x + q_y  \gamma_y \right) f^\pdagger_{\vec{q}}, \label{FinalLowEnergyTheory2}
\end{equation}
where $\nu$ is the velocity of the moir\'e Dirac cones and $\gamma_x=\tau_z\rho_x$ and $\gamma_y=\rho_y$ are 16$\times$16 matrices with $\tau_i$ acting on valley, $\mu_i$ on mini-valley, $\sigma_i$ on spin, and $\rho_i$ on generalized sublattice space. 

The choices of $\gamma_{x,y}$ in \equref{FinalLowEnergyTheory2}, and the symmetry transformations in \tableref{RepresentationsOfSymmetries}, are sufficient to account for the topological character of the TBG band structure for our purposes.
Specifically, the Dirac chiralities of the 2 mini-valleys in \equref{FinalLowEnergyTheory2} are the same, and this will play central role in the structure of the WZW term.

\begin{table}[tb]
\begin{center}
\caption{Here we show the representations of the relevant symmetries in the basis used in \equref{FinalLowEnergyTheory}, with $\tau_i$, $\mu_i$, $\sigma_i$, and $\rho_i$, $i=0,x,y,z$, acting in valley, mini-valley, spin, and sublattice space, respectively. For convenience of the reader, we show more than a minimal set of generators. Here, $T_{\vec{a}_r}$ denotes moir\' e-lattice translation by $\vec{a}_r$ defined in \figref{fig:Lattice}, SU(2)$_s$ and U(1)$_v$ are conventional spin rotation and valley-U(1). The 2D space group of the full system is $p6mm$ and that of a single valley is the magnetic space group 183.188 \cite{litvin2013magnetic}.}
\label{RepresentationsOfSymmetries}
 \begin{ruledtabular}
 \begin{tabular} {c|ccc}
    $g$   & $\vec{q}=(q_x,q_y)$ & $R_g$ & consequences in \equref{FinalLowEnergyTheory} \\ \hline
$C_2$    & $-\vec{q}$   & $\rho_x \tau_x \mu_x$  & ---\\ 
$C_3$    & $C_3\vec{q}$   & $e^{-i \frac{2\pi}{3}\rho_z\tau_z}$  & $\nu^{vp}_x=\nu^{vp}_y$ \\ 
$C_{2x}$    & $(q_x,-q_y)$   & $\rho_x\mu_x$  & $\nu^{+}_{x,y}=\nu^{-}_{x,y}$, $\epsilon_{+}=\epsilon_{-}$ \\  \hline
$\Theta$    & $-\vec{q}$   & $\sigma_y\tau_x \mu_x\mathcal{K}$  & --- \\
$C_2\Theta$    & $\vec{q}$   & $\sigma_y \rho_x\mathcal{K}$  & --- \\  \hline
$T_{\vec{a}_r}$  & $\vec{q}$ & $e^{i\vec{q}\vec{a}_r} e^{\frac{2\pi i}{3}\mu_zr}$ & --- \\
SU(2)$_s$ & $\vec{q}$ & $e^{i\vec{\varphi}\vec{\sigma}}$ & --- \\
U(1)$_v$ & $\vec{q}$ & $e^{i\varphi\tau_z}$ & --- 
 \end{tabular}
 \end{ruledtabular}
\end{center}
\end{table}

\section{Possible spin-singlet pairing states}\label{SuperconductingStates}
We will begin our analysis by listing the possible superconducting order parameters within the low-energy theory introduced above, which we organize according to the irreducible representations (IRs) of the symmetry group. For a detailed classification of the pairing states in the full Brillouin zone and the consequences associated with emergent symmetries and the behavior once these are weakly broken, we refer to \refcite{2019arXiv190603258S}.
To describe superconductivity, we here use the notation
\begin{equation}
	H_{\text{SC}} = \sum_{\vec{q}} f^\dagger_{\vec{q}} \, \Delta_{\vec{q}} T f^\dagger_{-\vec{q}} + \text{H.c.} 
	\label{GeneralFormOfPairing}
\end{equation}
and refer to $\Delta_{\vec{q}}$ as the superconducting order parameter. Here, $T=i \sigma_y \tau_x \mu_x$ is the unitary part of the anti-unitary time-reversal operator, $\Theta=T\mathcal{K}$, and the superconducting order parameter $\Delta_{\vec{q}}$ is a matrix in spin, valley, mini-valley, and (generalized) sublattice space; it must satisfy
\begin{equation}
	T^\dagger\Delta_{-\vec{q}}^T T = \Delta_{\vec{q}},
\end{equation}
due to Fermi-Dirac statistics,
with the superscript in $\Delta^T$ representing matrix transpose.

To organize the discussion and narrow down the multitude of possible superconducting order parameters, we will first concentrate on singlet pairing, \textit{i.e.}, we have $\Delta_{\vec{q}} \propto \Delta_{\vec{q}}^s \sigma_0$, where $\Delta_{\vec{q}}^s$ is a matrix only in valley, mini-valley, and (generalized) sublattice space; we will come back to triplet pairing in \secref{TripletPairing} below.

Let us focus on pairing of electrons at opposite momenta $\vec{k}$ and $-\vec{k}$ in the moir\'e-Brillouin zone, thus, preserving moir\'e translational symmetry, and allowing us to classify the pairing states accoring to the IRs of the point group only. 
At least in the presence of time-reversal symmetry, this is expected to be energetically most favorable; it corresponds to pairing between $p$ and $-p$ only (inter-mini-valley pairing). By the same token, it seems natural to focus on intervalley pairing. Note that the U(1)$_v$ symmetry, associated with valley-charge conservation, forbids mixing between intra- and intervalley pairing \cite{2019arXiv190603258S}. Taken together, we choose $\Delta_{\vec{q}}^s$ such that only matrix elements with $p'=-p$ and $v'=-v$ are non-zero. 
Furthermore, let us restrict the discussion to the leading-order expansion of $\Delta_{\vec{q}}^s$ in $\vec{q}$, \textit{i.e.}, just the constant term, $\Delta_{\vec{q}}^s \rightarrow \Delta^s$, since we are interested in the vicinity of the Dirac points.

While this seems like a lot of constraints, there are, in fact, still eight different independent pairing terms which can realize almost all irreducible representations (only $E_1$ is missing) of the point group $D_6$ of the system, see \tableref{DifferentIndependentPairingTerms}. 

\begin{table}[tb]
\begin{center}
\caption{Summary of different singlet pairing states in the low-energy Dirac theory (\ref{FinalLowEnergyTheory2}). The fact that there are two different sets of order parameters transforming under $E_2$ means that the corresponding basis functions can mix. We also indicate, in the last column, whether the pairing states will gap the Dirac cones, and refer the reader to \refcite{2019arXiv190603258S} for a discussion of the gap structure in the full Brillouin zone at non-integer $\nu$.}
\label{DifferentIndependentPairingTerms}\begin{ruledtabular}
 \begin{tabular} {c|ccc}
   Order parameter  $\Delta^s$   & transform as & IR of $D_6$ & gap \\ \hline
$\mathbbm{1}$    & const., $z^2$   & $A_1$ & \cmark  \\ 
$\tau_z\mu_z$    & $z$   & $A_2$ & \cmark  \\ 
$\tau_z\mu_z\rho_z$    & $x(x^2-3y^2)$   & $B_1$ & \xmark  \\ 
$\rho_z$    & $y(3x^2-y^2)$   & $B_2$ & \xmark \\ 
$(\rho_x,-\tau_z\rho_y)$    & $(x^2-y^2,2xy)$   & $E_2$ & \xmark \\ 
$(\mu_z\rho_y,\tau_z\mu_z\rho_x)$    & $(x^2-y^2,2xy)$   & $E_2$ & \xmark
 \end{tabular}
 \end{ruledtabular}
\end{center}
\end{table}

The fact that we have two different pairs of $\Delta_{\vec{q}}$ that transform under $E_2$ means that they will, in general, mix. In other words, the superconducting partner functions for IR $E_2$ have the form $\chi^{E_2}_{\vec{k},1} = a \rho_x+ b \mu_z\rho_y$ and $\chi^{E_2}_{\vec{k},2} = -a \tau_z\rho_y+ b \tau_z\mu_z\rho_x$, where $a$ and $b$ are some undetermined, real parameters that depend on microscopic details. Since $E_2$ is a two-dimensional IR, the associated superconducting order parameter has the form $\Delta = \sum_{\mu=1,2} \eta_\mu \chi^{E_2}_{\vec{k},\mu}$, where the $\eta_\mu$ are constrained by symmetry and can only assume discrete values.

We also point out that it was previously shown \cite{2019arXiv190603258S} that the singlet states odd under $C_2$ cannot give rise to a finite gap at generic momentum points, where the (potentially spin and valley degenerate) bands are separated---this is related to the fact that $C_2$ $C_2$ simply flips the sign of momenta (and valley here) in 2D, exactly as time-reversal does  \cite{scheurer2017selection}. We here see that this also holds around the Dirac cones, since the states transforming under $B_1$ and $B_2$ in \tableref{DifferentIndependentPairingTerms} will not induce a gap.

\section{Wess-Zumino-Witten terms without additional orders}\label{WZWNoOtherOrders}
Having established the notation, the non-interacting model, and the different superconducting states, we are now in a position to look for natural WZW terms of those superconducting states with other order parameters, {\it e.g.\/}, associated with the correlated insulator. There will be no `high temperature' or `parent' orders $M$ in this section.

\subsection{Procedure for finding WZW terms}\label{ProcedureOfFindingTerms}
WZW terms have previously been studied in the context of Dirac theories \cite{ABANOV2000685,PhysRevB.82.245117} and we will make use of these results here.
To this end, let us first define the Nambu spinor
\begin{equation}
\Psi_{\vec{q}} = \begin{pmatrix} f_{\vec{q}\uparrow} \\ f^\dagger_{-\vec{q}\downarrow}   \end{pmatrix}, \label{FirstNambuSpinor}
\end{equation}
which is non-redundant, {\it i.e.\/}, a complex rather than a Majorana fermion and the results of \refscite{ABANOV2000685,PhysRevB.82.245117} apply.
With the new field in \equref{FirstNambuSpinor}, we can write the action associated with the above superconducting Dirac theory as
\begin{eqnarray}
 \mathcal{S} &=& \int \diff t \int \diff^2 \vec{q} \Bigl[ \Psi_{\vec{q}}^\dagger \left( \partial_t + q_1 \Gamma_1 + q_2 \Gamma_2 \right) \Psi_{\vec{q}}  \nonumber \\
&~&~~~~~~~~~~~~~~~~~~+  \sum_{a=3}^7 n_{a-2} \,  \Psi_{\vec{q}}^\dagger \mathcal{M}_a  \Psi^\pdagger_{\vec{q}}   \Bigr]
\label{NambuDiracTheory}
\end{eqnarray}
where $\Gamma_1=\tau_z\rho_x$, and $\Gamma_2=\eta_z\rho_y$, and $\eta_i$ are Pauli matrices in Nambu space. In \equref{NambuDiracTheory}, $\mathcal{M}_a$, $a=3,\dots N_s+2$, capture the superconducting states (with $N_s$ real components). Our goal is to systematically find the remaining $\mathcal{M}_a$, $a=N_s+3,\dots,7$, in the particle-hole channel, \textit{i.e.}, of the form
\begin{eqnarray}
    \mathcal{O}_j = \sum_{\vec{q}}\Psi_{\vec{q}}^\dagger \mathcal{M}_{2+j+N_s} \Psi_{\vec{q}}^\pdagger = \sum_{\vec{q}} f^\dagger_{\vec{q}} m_j f^\pdagger_{\vec{q}}, \label{PartnerOrderParamsDef}
\end{eqnarray}
that will give rise to a joint WZW term for the unit length field $n_a$ conjugate to the order parameters; $n_a$ is assumed constant in (\ref{NambuDiracTheory}).  We will refer to the associated $m_j$, $j=1,\dots 5-N_s$, as the \textit{partner order parameters} of the superconducting state. These will be our candidates for the correlated insulators found in experiment. 

We know from \refscite{ABANOV2000685,PhysRevB.82.245117} that a WZW will be generated if
\begin{align}
    \text{tr}\left[ \Gamma_{i_1} \Gamma_{i_2} \mathcal{M}_{a_1}\mathcal{M}_{a_2}\mathcal{M}_{a_3}\mathcal{M}_{a_4}\mathcal{M}_{a_5} \right] = 8 \mathcal{N} \, \epsilon_{i_1i_2a_1a_2a_3a_4a_5},\label{model3_coef}
\end{align}
with non-zero $\mathcal{N}$. The integer $\mathcal{N}$ determines the co-efficient of the WZW term. The WZW term can be written in an explicit form preserving all symmetries only by extending the field $n_a$ to 4 dimensional spacetime $(u,\tau,x,y)$ with an additional dimension $u$:
\begin{eqnarray}
\mathcal{S}_{\text{WZW}} &=& i \frac{2 \pi \mathcal{N}}{\Omega_4} \int_0^1 du \int d \tau dx dy 
\sum_{abcde=1}^5 \epsilon_{abcde} \nonumber \\
&~&~~~~~~~~~~\times n_a \partial_u n_b \partial_\tau n_c \partial_x n_d \partial_y n_e\,, \label{SWZW}
\end{eqnarray}
where $\Omega_4 = 8 \pi^2/3$ is the surface area of a unit sphere in 5 dimensions.
We are assuming here that the combined order parameters have 5 components. In models with larger symmetry, there could be additional order parameter components which would combine to yield a sum of terms like those in (\ref{SWZW}) but with a larger overall symmetry \cite{XuThomson}.

While the Nambu basis in \equref{FirstNambuSpinor} allows to bring all pairing states in \tableref{DifferentIndependentPairingTerms} in the form of the mass terms in \equref{NambuDiracTheory}, it constrains the possible partner orders we can study: we will only be able to write down $m_j$ that are diagonal in spin ($\propto \sigma_0,\sigma_z$). One straightforward way to generalize the analysis proceeds by considering several alternative choices of non-redundant Nambus spinors, such as 
\begin{equation}
\label{AlternativeNambuSpinors}
    \Psi_{\vec{q}} = \begin{pmatrix} f_{\vec{q},p=+} \\ f^\dagger_{-\vec{q},p=-}   \end{pmatrix}, \quad \Psi_{\vec{q}} = \begin{pmatrix} f_{\vec{q},v=+} \\ f^\dagger_{-\vec{q},v=-}   \end{pmatrix}.
\end{equation}
The first option allows to write down any inter-mini-valley pairing (singlet and triplet), which again includes all of the pairing states we are interested in. Moreover, partner order parameters in the particle-hole channel with arbitrary spin polarization (only restricted to intra-mini-valley, which means moir\' e-translation-invariant states) can be captured. The second choice will still allow to write down all of our pairing terms, the inter-valley pairing order parameters; as for the partner order parameters, we can now write down density-wave terms, that break the moir\' e translational symmetry, but cannot write down any inter-valley-coherent states.  
Clearly, many more choices are possible, such as $\Psi_{\vec{q}} = ( f_{\vec{q},\rho=1}, f^\dagger_{-\vec{q},\rho=2})^T$; since the kinetic terms in our Hamiltonian are off diagonal in sublattice space, for the sublattice Nambu spinor a unitary transformation must first be applied to the Hamiltonian to bring it to a form where the pairing term is off diagonal in sublattice space and the kinetic terms are diagonal ($e^{i\frac{\pi}{4}\rho_x\tau_x}$ and $e^{i\frac{\pi}{4}\rho_x\tau_y\mu_y}$ for $A_1$ and $A_2$ pairings respectively).

However, a more efficient criterion that is equivalent to \equref{model3_coef} for any such choice of Nambu spinor can be derived, see \appref{DerivationWZWTerm}: the partner orders $m_j$, Dirac matrices $\gamma_i$, and the superconducting order parameter must obey 
\begin{subequations}\begin{align}
&\gamma_i {\Delta}T = -{\Delta}T \gamma_i^T \neq 0, \qquad i=1,2, \label{FirstPartOfCond} \\
&m_j {\Delta}T = {\Delta}T m_j^T \neq 0, \qquad j=1,2,3, \\
&\text{tr}[\gamma_{i_1}\gamma_{i_2}m_{j_1}m_{j_2}m_{j_3}]\propto \epsilon_{i_1i_2j_1j_2j_3}.\label{LastPartOfGenCrti}
\end{align}\label{CriterionAlt}\end{subequations}

Anticipating that this will be the only relevant case below, we have here already assumed that $N_s=2$, \textit{i.e.}, only one-component complex superconducting order parameters (two fluctuating real components) play a role. Note that the third condition requires the partner orders to anti-commute with the kinetic terms in our Dirac Hamiltonian, implying they will gap out the Dirac cones.

We finally note that, although we began by considering a non-redundant basis, the conditions which account for every possible non-redundant Nambu basis are equivalent to \equref{model3_coef} in a redundant extended Nambu basis. This and the criterion \ref{CriterionAlt}, are derived in \appref{DerivationWZWTerm}.

\begin{table*}[tb]
\begin{center}
\caption{Possible partner order parameters, $m_j$, $j=1,2,3$, see \equref{PartnerOrderParamsDef}, for singlet pairing, assuming that all symmetries in \tableref{RepresentationsOfSymmetries} of the bare model (\ref{FinalLowEnergyTheory2}) are preserved. We further defined $\mu_{\pm} = (\mu_x \pm \mu_y)/2$, $\tau_{\pm}=(\tau_x \pm \tau_y)/2$, $\omega_{\pm} = \exp({\pm 2\pi i/3})$, and provide the associated IRs of the point group $D_6$. In the second to last column, we indicate how we denote these states in this work, including quantum spin Hall (QSH), a moir\'e density wave, which breaks moir\'e translation invariance and is even (MDW$_{+}$) or odd (MDW$_{-}$) under time-reversal, time-reversal even/odd intervalley-coherent phases (IVC$_{\pm}$), a sublattice polarized state (SP), and a valley-polarized state (VP). In the last column, we denote the high-temperature orders $M$ of \secref{HighTemperatureOrders} (see also \tableref{HighTemperatureOrders}) for which the order will survive projection to one of the energy eigenspaces. $M$'s which will lead to additional Fermi surfaces as described in \tableref{HighTemperatureOrders} and are thus less likely are denoted with brackets. The partner order parameters listed here are possible at $\nu=0$ (without high-temperature orders $M$) and at $\nu=\pm 2$ (given one of the listed $M$ is present 
).}
\label{SummaryOrderParamters}
\begin{ruledtabular}
 \begin{tabular} {c|cccccccc} 
Pairing  & $m_j$ & IR & $\Theta$  & $T_{\vec{a}_r}$ & U(1)$_v$ & SU(2)$_s$ & Type & $M$ \\ \hline
$A_1$ &  $(\tau_+,\tau_-)\rho_x$; $\rho_z$ & $A_1/B_1$; $B_2$ & $+$ & $1$ & $m=1$; $m=0$ & $\vec{1}$ & IVC$_+$; SP & $\mu_x;[\mu_z\sigma_z]$\\
$A_1$ &  $(\mu_+,\mu_-)\rho_z$; $\rho_z\tau_z\mu_z$ & $B_2$/$A_1$; $B_1$ & $+$ & $(\omega_r,\omega_r^*)$; $1$ & $m=0$ & $\vec{1}$ & MDW$_+$; VP & $\tau_x\rho_y\mu_z;\tau_z\sigma_z$\\
$A_1$ &  $(\sigma_x,\sigma_y,\sigma_z)\tau_z\rho_z$ & $A_2$ & $+$ & $1$ & $m=0$ & $\vec{3}$ & QSH & $\mu_x;\tau_x\rho_y\mu_z;[\tau_z\mu_z]$ \\ \hline 
$A_2$ &  $(\tau_+,\tau_-)\mu_z\rho_x$; $\rho_z$ & $B_2/A_2$; $B_2$ & $-$; $+$ & $1$ & $m=1$; $m=0$ & $\vec{1}$ & IVC$_-$; SP & $\tau_z\mu_x;[\mu_z\sigma_z]$\\
$A_2$ &  $(\mu_+,\mu_-)\tau_z\rho_z$; $\rho_z\tau_z\mu_z$ & $A_2$/$B_1$; $B_1$ & $-$; $+$ & $(\omega_r,\omega_r^*)$; $1$ & $m=0$ & $\vec{1}$ & MDW$_-$; VP & $\tau_z\sigma_z;\tau_x\rho_y$ \\
$A_2$ &  $(\sigma_x,\sigma_y,\sigma_z)\tau_z\rho_z$ & $A_2$ & $+$ & $1$ & $m=0$ & $\vec{3}$ & QSH  & $\tau_z\mu_x;\rho_y\tau_x;[\tau_z\mu_z]$ 
 \end{tabular}
 \end{ruledtabular}
\end{center}
\end{table*}

\subsection{Possible partner orders}
Using the procedure outlined above, we can systematically study all possible partner order parameters for the different superconducting states. All of these orders will have $\mathcal{N}=2$ in Eqs.~(\ref{model3_coef}) and (\ref{SWZW}). The value $\mathcal{N}=2$ implies that a skyrmion in the partner order has charge $\pm 4e$ \cite{PhysRevLett.100.156804}. The anisotropies in the free energy of the partner orders can allow stable half skyrmions ({\it i.e.\/} merons) of charge $\pm 2e$ \cite{2020arXiv200400638K}, and condensation of merons or skyrmions leads to superconductivity.

Out of the pairing states in \tableref{DifferentIndependentPairingTerms}, only those transforming under $A_1$ or $A_2$ allow for partner order parameters with WZW terms given that our alternative criterion (\ref{CriterionAlt}) requires that the pairing multiplied with $T$ must commute with the anti-symmetric $\gamma_y=\rho_y$ and anti-commute with the symmetric  $\gamma_x=\rho_x\tau_z$. Of the possible pairings, only $A_1$ and $A_2$ satisfy this condition. In fact, it is no coincidence that this is correlated with whether these superconducting order parameters will lead to a gap around the Dirac cones. Equation (\ref{FirstPartOfCond}) implies that the superconducting order parameters anti-commute with the kinetic terms in the Nambu Hamiltonian and, as such, gap out the spectrum. 
Consequently, only states transforming under one-dimensional IRs remain, leading to $N_s=2$, as mentioned above. 
We note that for non-integer fillings, such that the chemical potential does not go through the Dirac nodes of the normal-state bandstructure, the $A_1$ pairing state can remain gapless, while the $A_2$ state will necessarily have $6$ nodal points on any Fermi surface enclosing the $\Gamma$ point \cite{2019arXiv190603258S}.

For each of $A_1$ and $A_2$, we have derived the complete list of mathematically possible sets of partner order parameters satisfying \equref{CriterionAlt}; these are listed in \appref{FullSetOfOrderParameters}. However, only a small fraction of them are physically meaningful options if we assume that none of the symmetries in \tableref{RepresentationsOfSymmetries} are broken in the parent Hamiltonian for superconductivity and the partner order parameters. 

To understand the reduction of possibilities resulting from symmetries, consider the mathematically possible choice of
\begin{equation}
m_1 = \tau_x\rho_x, \quad m_2 = \tau_y\mu_x\rho_x,\quad m_3 = \mu_x\rho_z \label{ExampleOfThreems}
\end{equation}
for the partner order parameters in \equref{PartnerOrderParamsDef}. As long as we have U(1)$_v$, $\tau_x\rho_x$ must ``fluctuate with'' $\tau_y\rho_x$; more precisely, any low-energy field theory containing a field coupling to $m_1$ must also contain another field that describes fluctuations of $\tau_y\rho_x$. However, we have already exhausted the number of three particle-hole order parameters forming a WZW term with superconductivity.
This would already be enough to discard this choice of partner order parameters as it is incomplete from a symmetry perspective. We note that it is also incomplete due to translational symmetry which requires (at least to quadratic order) that, {\it e.g.\/}, $\mu_x\rho_z$ fluctuates with $\mu_y\rho_z$. This means that $m_j$ in \equref{ExampleOfThreems} constitute a valid set of partner orders only if both moir\'e translation and U(1)$_v$ are broken.

Applying such an analysis to all of the mathematically possible sets of partner order parameters, we find the remaining, physically relevant options summarized in \tableref{SummaryOrderParamters}. We note that the symmetries leading to the reduction of possibilities for $m_j$, such as translation and U(1)$_v$ for \equref{ExampleOfThreems}, do not involve the spatial rotation symmetry $C_6$. Consequently, the presence of lattice strain and/or nematic order \cite{PasupathySTM,NadjPergeSTM,AndreiSTM,YazdaniSTM,PabloNematicity,STMReview} will not lead to additional options.

\section{High-energy symmetry breaking at half-filling}\label{SymmetryBreakingAtHighT}
Next, let us take into account the additional symmetry breaking, associated with an order $M$, that is believed to set in at much higher temperatures than superconductivity and the correlated insulator, as found in a recent experiments \cite{Yazdani19,Ilani19}. At present, the microscopic form of the underlying order parameters is not known and so we will systematically analyze different possibilities. For concreteness, we focus here on the vicinity of half filling of the conduction or valence band, \textit{i.e.}, $\nu=\pm 2$.
\begin{figure}[tb]
     \centering
     \begin{subfigure}[b]{0.4\textwidth}
         \centering
         \includegraphics[width=\textwidth]{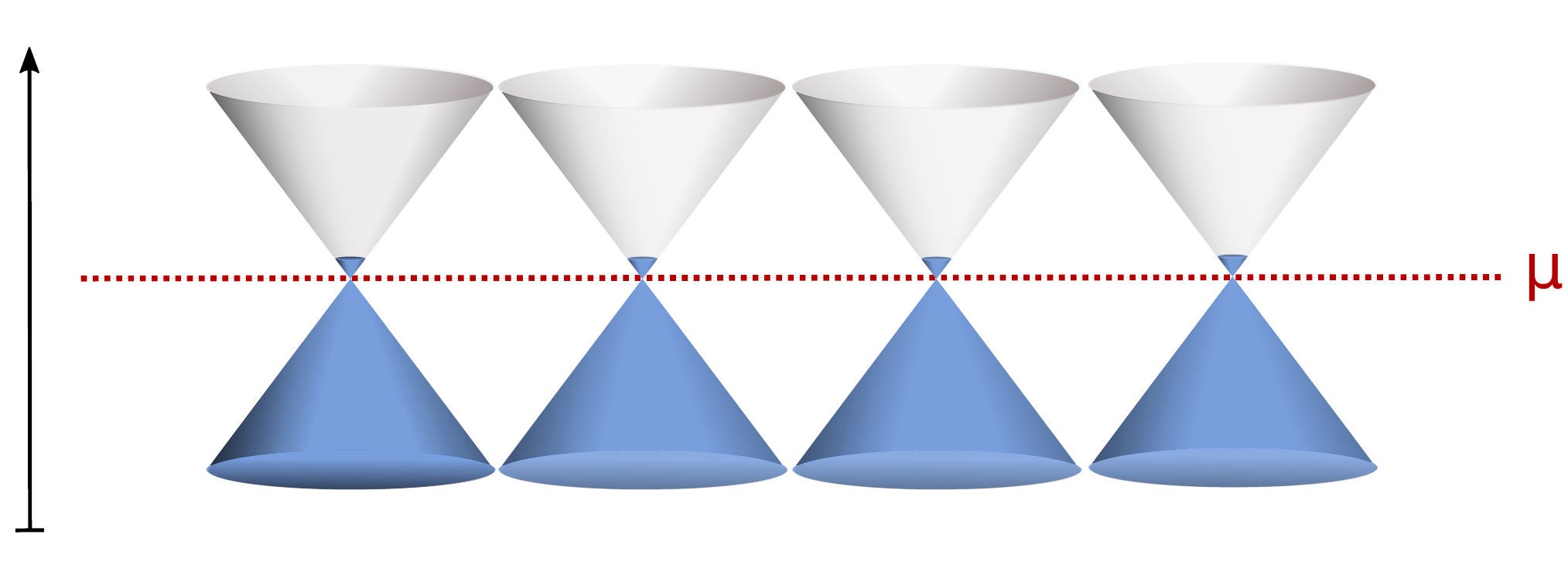}
         \caption{Degeneracy for $\nu=0$.}
         \label{nueq0}
     \end{subfigure}
     \hfill
    \begin{subfigure}[b]{0.4\textwidth}
         \centering
         \includegraphics[width=\textwidth]{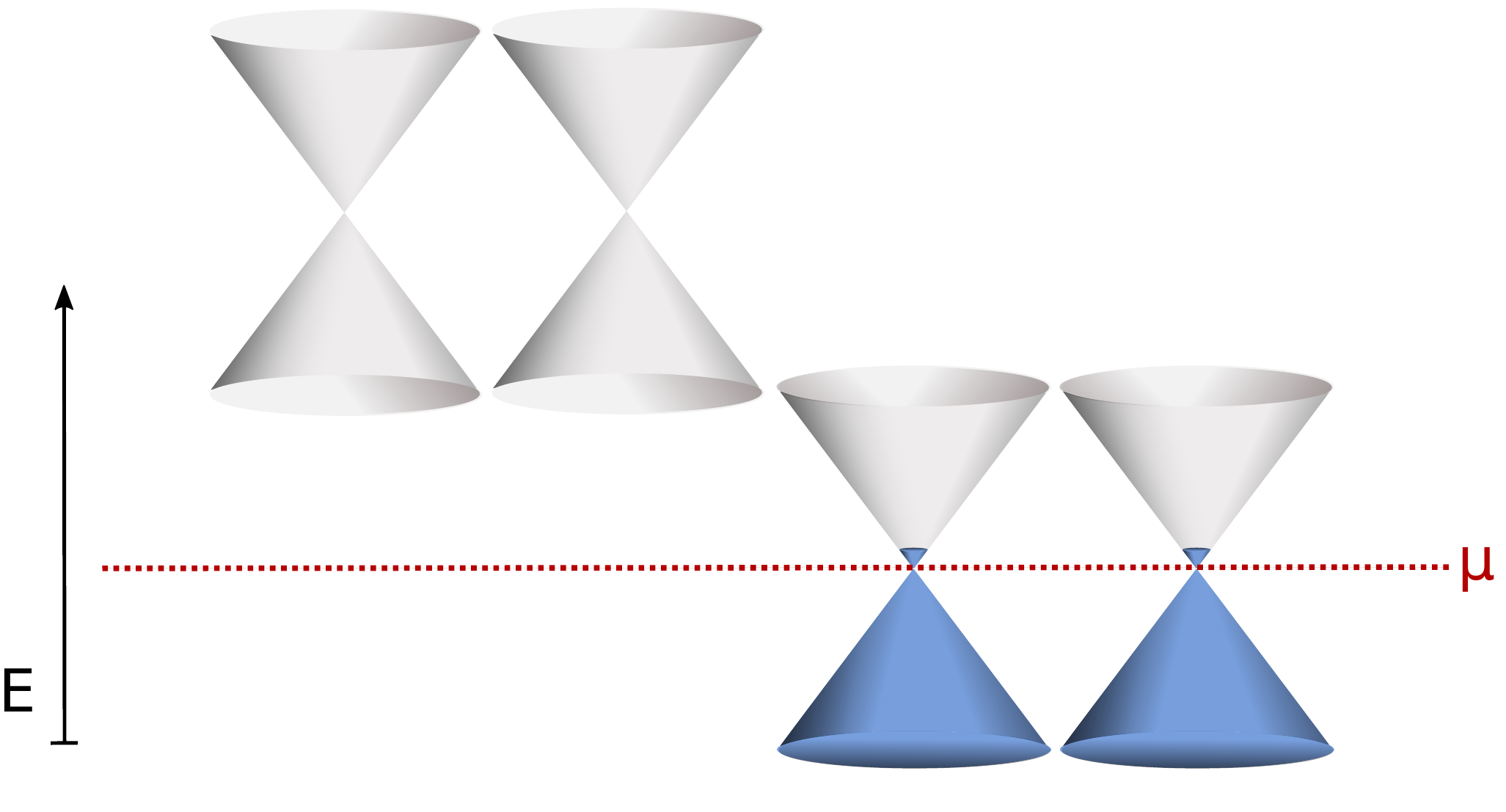}
         \caption{Degeneracy for $\nu=-2$.}
         \label{nueq2}
     \end{subfigure}
     \begin{subfigure}[b]{0.145\textwidth}
         \centering
         \includegraphics[width=\textwidth]{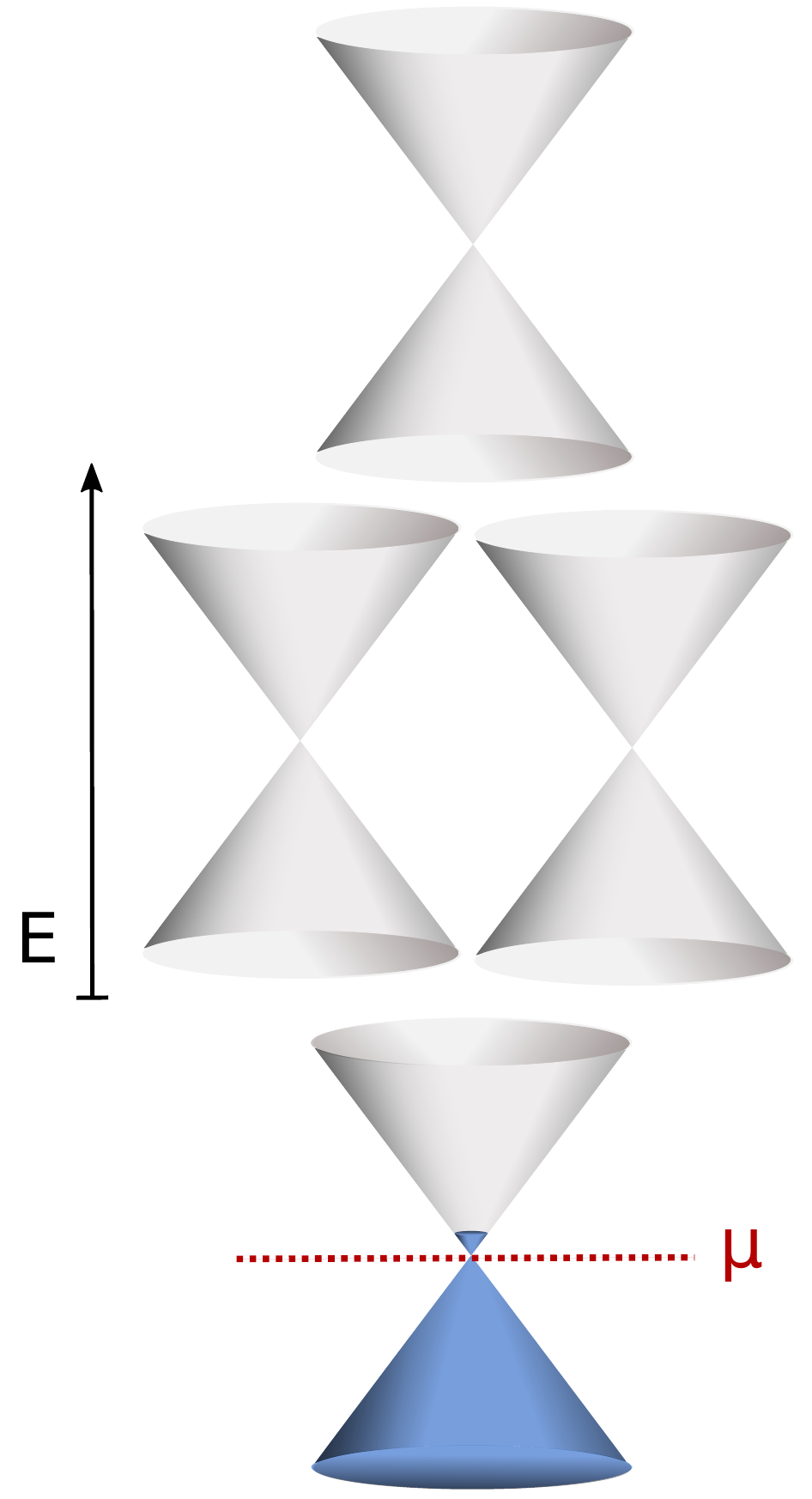}
         \caption{Degeneracy for $\nu=-3$ with eigenvalues $\{+1,0,0,-1\}$.}
         \label{nueq1}
     \end{subfigure}
     \begin{subfigure}[b]{0.145\textwidth}
         \centering
         \includegraphics[width=\textwidth]{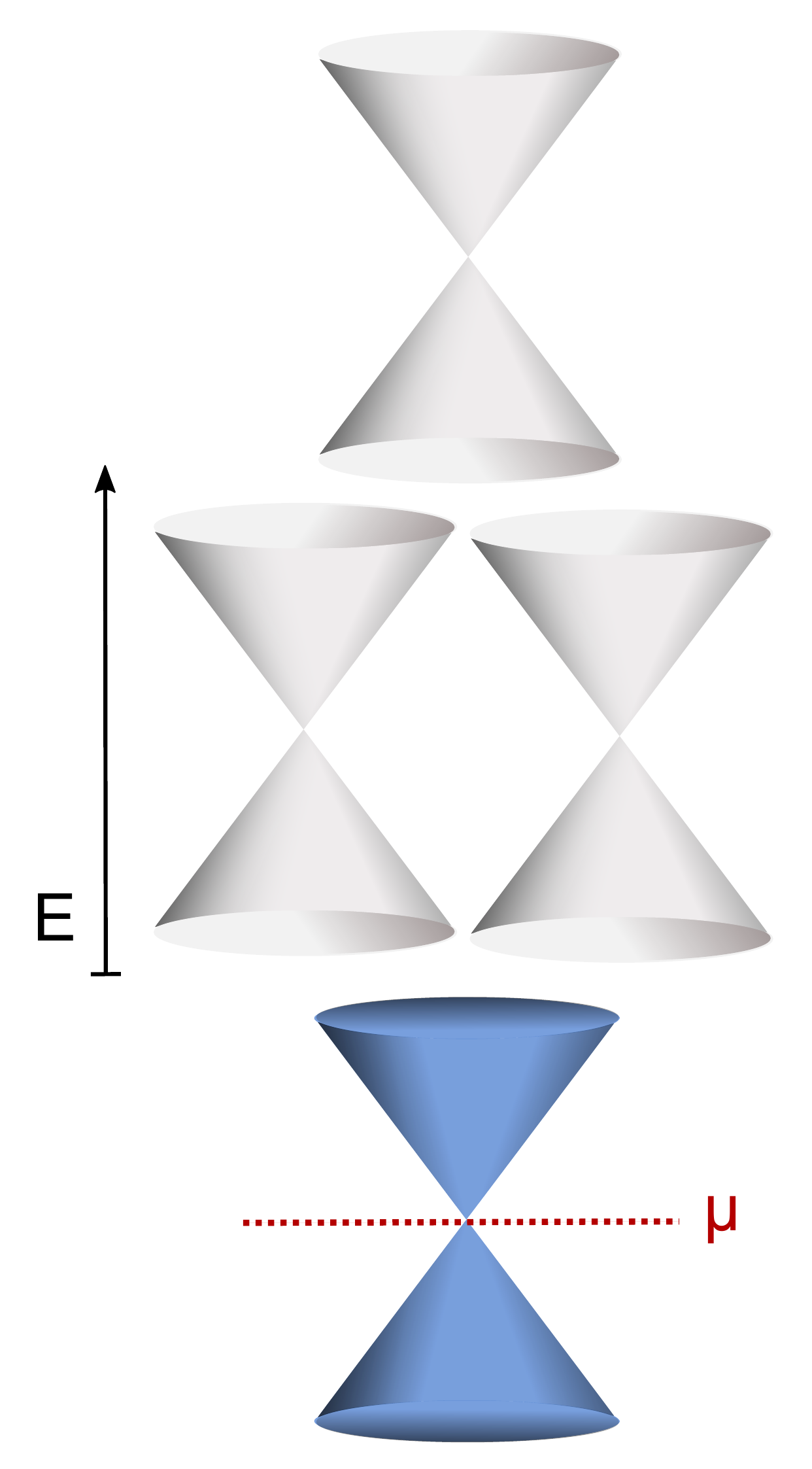}
         \caption{Degeneracy for $\nu=-2$ with eigenvalues $\{+1,0,0,-1\}$.}
         \label{nueq2bad}
     \end{subfigure}
        \caption{Dirac cones (each cone is 2-fold degenerate) and their filling (blue). (a) $\nu=0$ with no additional symmetry breaking; (b) Dirac revival at $\nu=-2$ due to a high-temperature order parameter $M$ with two 8-dimensional eigenspaces; (c) $\nu=-3$ with an $M$ with 3 eigenspaces labeled by eigenvalues $\{+1,0,0,-1\}$; (d) $\nu=-2$ also with an $M$ with 3 eigenspaces labeled $\{+1,0,0,-1\}$---there are no active Dirac cones and an $M$ with this structure will only work for $\nu=\pm 3$. For simplicity, we show only half of the 8 Dirac cones (associated with spin, valley, and mini-valley) and do not explicitly display that the Dirac cones are part of a bandstructure in the moir\'e Brillouin zone with finite bandwidth, as in Fig.~\ref{AdditionalFermiSurfaces}.}
        \label{GappingOutBands}
\end{figure}
\begin{figure}[tb]
     \centering
     \begin{subfigure}[b]{0.3\textwidth}
         \centering
         \includegraphics[width=\textwidth]{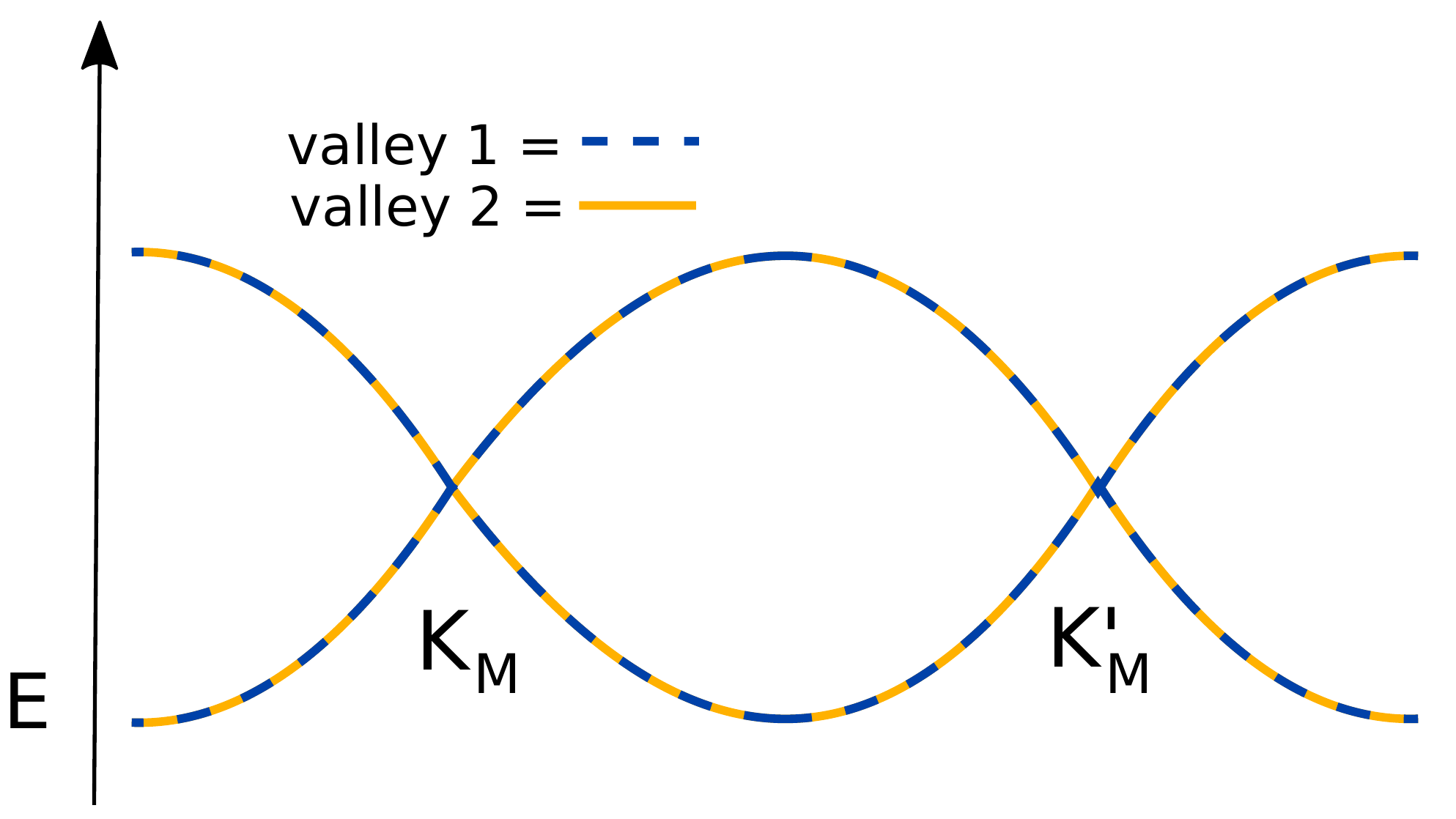}
         \caption{Bands without an $M$.}
         \label{Bandsuncrossed}
     \end{subfigure}
     \hfill
    \begin{subfigure}[b]{0.3\textwidth}
         \centering
         \includegraphics[width=\textwidth]{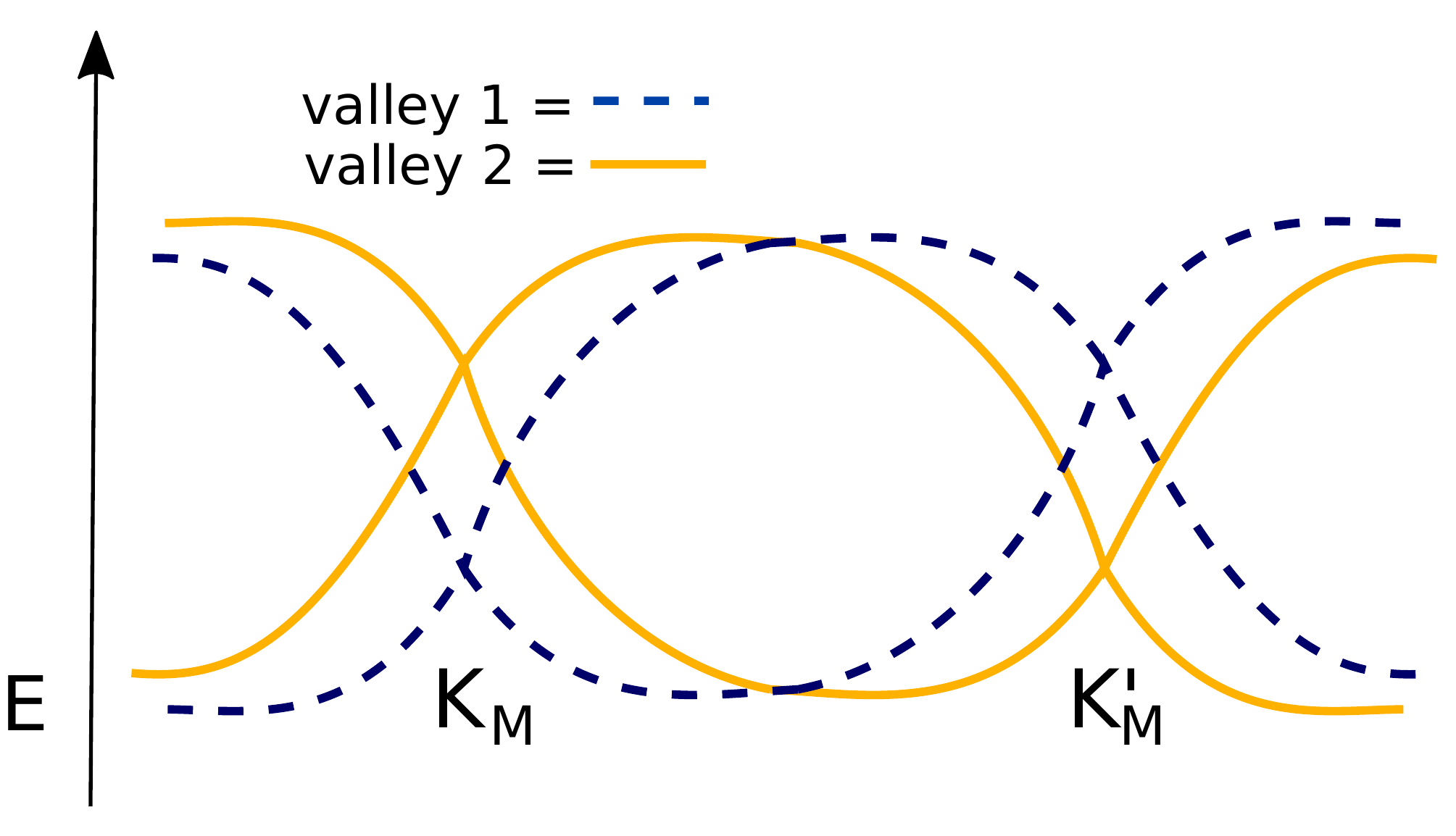}
         \caption{Bands with $M=\tau_z\mu_z$.}
         \label{bandcrossings}
     \end{subfigure}
        \caption{How additional Fermi surfaces emerge for the high-temperature order $M=\tau_z\mu_z$. Their absence requires the valleys $v=\pm$ to mix away from the $\text{K}_\text{M}$ ($p=+$) and $\text{K'}_\text{M}$ ($p=-$) points. The same is true for most $M\propto\mu_z$ in \tableref{HighTemperatureOrders}. (a) shows the bare band structure around the $\text{K}_\text{M}$ and $\text{K'}_\text{M}$ points without $M$ and how they connect along a one-dimensional momentum cut. Part (b) is the same with $M=\tau_z\mu_z$ added, clearly exhibiting additional Fermi surfaces. For simplicity, we have depicted the bandstructure here to be the same for either valley for all momenta, although they are only required to be mirror images of each other. Including this splitting away from the high-symmetry points does not alter our argument.}
        \label{AdditionalFermiSurfaces}
\end{figure}

To define the different options for this symmetry-broken high-temperature state, we will consider adding momentum-independent quadratic terms to \equref{FinalLowEnergyTheory2}, {\it i.e.\/}, the parent Hamiltonian $H_{\text{LE}}$ is replaced by
\begin{equation}
	\widetilde{H}_{\text{LE}} =  \sum_{\vec{q}}^{\Lambda} f^\dagger_{\vec{q}} \left[   q_x \gamma_x + q_y  \gamma_y + M \right] f^\pdagger_{\vec{q}}. \label{HighTempParentHam}
\end{equation}
Here $M$---a $16 \times 16$ matrix in valley, mini-valley, spin, and sublattice space---is the high-temperature order parameter. As the data of \refcite{Ilani19} indicates that Dirac cones re-emerge around integer fillings as a consequence of these high-energy orders, we want $M$ to commute with the Dirac matrices $\gamma_{x,y}$. Additionally, we require it to have only two different (and, hence, $8$-fold degenerate) eigenspaces to correctly reproduce the reduction of degeneracy of the Dirac cones by a factor of two at $\nu=\pm 2$, see \figref{GappingOutBands}(a,b). 
Because of this reduced degeneracy, the WZW terms in this Section will have $\mathcal{N}=1$ in Eqs.~(\ref{model3_coef}) and (\ref{SWZW}), and skyrmions in the partner orders will have charge $2e$ \cite{PhysRevLett.100.156804}.
Finally, to further reduce the number of possibilities, we will focus on order parameter configurations of $M$ that are minima of symmetry-restricted free-energy expansions and, as such, can be reached by a second order transitions from the high-temperature phase without $M$.

While the complete list of remaining $M$ is provided in \tableref{HighTemperatureOrders}, we next discuss the different classes of order parameters separately, organized by whether they respect time-reversal and/or spin-rotation symmetry.

\begin{table}[tb]
\begin{center}
\caption{All possible high-temperature symmetry breaking orders and how they transform. 
We take only one representative of states which are related by a U(1)$_v$ or SU(2)$_s$ rotation. We denote any state which breaks translation symmetry by MDW, any state which breaks U(1)$_v$ conservation by IVC and indicate whether it also breaks time-reversal symmetry. "FM" denotes ferromagnet, "Sp" denotes spin, "Mv" denotes mini-valley, and "V" stands for valley, one or more of which can be polarized ("P"). For example, we call the order $\tau_z\mu_z$ "MvVP". 
We also indicate which states will cause bands to cross and yield additional Fermi surfaces.}
\label{HighTemperatureOrders}
 \resizebox{.5\textwidth}{!}{%
 \begin{ruledtabular}
 \begin{tabular} {c|cccccc}
$M$ & SU(2)$_s$ & $\Theta$  & $T_{\vec{a}_r}$ & U(1)$_v$ & Extra FSs   & Type\\ \hline
$\tau_z\mu_z$ &  \cmark & \cmark & \cmark &\cmark& \cmark & MvVP \\ 
$\mu_x$ &  \cmark & \cmark & \xmark &\cmark& \xmark & MDW$_+$\\
$\tau_x\rho_y\mu_z$ &  \cmark & \cmark & \cmark &\xmark& \xmark & IVC$_+$\\
\hline
$\tau_z\sigma_z$ &  \xmark & \cmark & \cmark &\cmark& \xmark & SpVP\\
$\mu_z\sigma_z$ &  \xmark & \cmark & \cmark &\cmark& \cmark & SpMvP\\
$\tau_z\mu_x\sigma_z$ &  \xmark & \cmark & \xmark &\cmark& \xmark & MDW$_+$\\
$\rho_y\tau_x\sigma_z$ &  \xmark & \cmark & \cmark &\xmark& \xmark & IVC$_+$\\
$\rho_y\mu_x\tau_x\sigma_z$ &  \xmark & \cmark & \xmark &\xmark & \xmark& IVC-MDW$_+$\\
\hline
$\tau_z\mu_x$ &  \cmark & \xmark & \xmark &\xmark & \xmark & MDW$_-$\\
$\rho_y\tau_x$ &  \cmark & \xmark & \cmark &\xmark & \xmark& IVC$_-$\\
$\rho_y\tau_x\mu_x$ &  \cmark & \xmark & \xmark &\xmark & \xmark& IVC-MDW$_-$ \\
\hline
$\sigma_z$ &  \xmark & \xmark & \cmark &\cmark & \xmark& FM\\
$\mu_x\sigma_z$ &  \xmark & \xmark & \xmark &\cmark& \xmark & MDW$_-$\\
$\tau_z\mu_z\sigma_z$ &  \xmark & \xmark & \cmark &\cmark& \cmark & SpMvVP\\
$\rho_y\tau_x\mu_z\sigma_z$ &  \xmark & \xmark & \cmark &\xmark& \cmark & IVC$_-$
 \end{tabular} \end{ruledtabular} %
 }
\end{center}
\end{table}

\subsection{Preserving spin-rotation and time-reversal}
Let us start with states that preserve both spin-rotation invariance and time-reversal symmetry. To see that this is a particularly important class of $M$ for WZW terms, recall that all relevant pairing terms transform under $A_1$ or $A_2$ and, hence, are described by a single complex number ($N_s=2$ real numbers). We are, thus, left with three partner order parameters, and SU(2)$_s$ is the only symmetry with a three-dimensional IR. Therefore, we will be able to find cases without an anisotropy term between the different fluctuating partner order parameters only if SU(2)$_s$ is present. 
Taking into account the constraints mentioned above to correctly reproduce the Dirac revival, we are left with three classes of options
\begin{equation}
M = \tau_z \mu_z, \quad M = (\mu_x,\mu_y), \quad M = (\tau_x,\tau_y) \rho_y \mu_z,
\label{OptionsForOrderParameters}
\end{equation}
where we grouped together symmetry-related choices, with respect to translation and U(1)$_v$ symmetry.

Intuitively, the first one in \equref{OptionsForOrderParameters} simply corresponds to pushing down (up) in energy those states where valley and mini-valley are identical (opposite). The second one can be thought of as an ``inter-mini-valley-coherent state'' or a time-reversal symmetric moir\'e density wave (MDW$_+$), breaking moir\' e-translation symmetry. Note that the actual system only has a discrete translational symmetry, corresponding to a discrete rotational symmetry of the vector $(\mu_x,\mu_y)$, see \tableref{RepresentationsOfSymmetries}; therefore, it is associated with a discrete set of symmetry-inequivalent configurations---in this case, $M=\mu_x$ and $M=\mu_x+\sqrt{3}\mu_y$, as can be derived by minimizing the free energy (see \appref{FreeEnergyExpansion}). Despite being inequivalent from the point of view of the symmetries of the microscopic model, we can focus only on one of these two options, say $M=\mu_x$, as they are related by the continuous symmetry, $e^{i\varphi \mu_z}$, which is an emergent symmetry of our low-energy model (\ref{FinalLowEnergyTheory2}), including all the superconducting states we consider. 
Finally, the third term in \equref{OptionsForOrderParameters} is an ``inter-valley-coherent'' (IVC) state, that preserves translational symmetry. As a result of the continuous U(1)$_v$ symmetry, we can choose $M=\tau_x\rho_y\mu_z$ without loss of generality.

There is one additional restriction concerning these high-temperature orders, which is related to the connectivity of the bands in the moir\'e Brillouin zone away from the Dirac cones, that we have not taken into account yet. This is most clearly illustrated by way of an example:
as illustrated in \figref{AdditionalFermiSurfaces}, $M=\tau_z \mu_z$ requires mixing of the valleys away from the high-symmetry points, since otherwise additional Fermi surfaces will necessarily appear in some parts of the Brillouin zone. 
In this sense, this choice of $M$ and all other high-temperature order parameters that require additional mixing of the bands, which we indicate in \tableref{HighTemperatureOrders}, are less natural candidate orders to explain the behavior seen in experiment \cite{Yazdani19,Ilani19}. However, for completeness, we study all of them. When analyzing whether a state will give rise to extra Fermi surfaces, we allow for arbitrary mixing of the bands that is not prohibited by the symmetries of the system. For instance, while one might think that $\tau_x \rho_y\mu_z$ will lead to the same band structure as shown in \figref{AdditionalFermiSurfaces}, the bands can hybridize since U(1)$_v$ is broken (so is the emergent valley symmetry, with $e^{i\varphi \rho_y\tau_x}$, away from the Dirac cones) and unwanted Fermi surfaces can be avoided.

Next, we discuss the resulting possible WZW terms between superconducting orders and correlated insulators born out of the high-temperature parent Hamiltonian (\ref{HighTempParentHam}) for the different possible $M$. Note that there are two crucial consequences of having an additional high-temperature order parameter: first, it can remove some of the options in \tableref{SummaryOrderParamters} of partner order parameters that were possible without $M$ since these order parameters vanish upon projection to the low- (relevant for $\nu=-2$) or high-energy ($\nu=+2$) eigenspace of $M$. However, by virtue of reducing the number of active degrees of freedom and by breaking certain symmetries, $M$ can also provide additional options that were not possible without it.

Because the physics will be the easiest, let us begin by illustrating this with the first high-temperature order, $M=\tau_z \mu_z$, in \equref{OptionsForOrderParameters}. It is readily seen that it transforms under $A_2$ and, hence, reduces $D_6$ to $C_6$. We can write down an effective model that only contains the 4 ``active'' Dirac cones, see \figref{GappingOutBands}(b), by replacing $f_{\vec{q},\sigma,v,s,p} \rightarrow \delta_{p,v} \widetilde{f}_{\vec{q},\sigma,v,s}$. The low-energy theory is now given by the four Dirac cones described by $\sum_{\vec{q}}^{\Lambda} \widetilde{f}^\dagger_{\vec{q}}[q_x \widetilde{\gamma}_x + q_y  \widetilde{\gamma}_y] \widetilde{f}^\pdagger_{\vec{q}}$, with $8\times 8$ reduced Dirac matrices $\widetilde{\gamma}_x=\tau_z\sigma_0\rho_x$ and $\widetilde{\gamma}_y=\tau_0\sigma_0\rho_y$ (note: no $\mu_0$-matrix anymore); the superconducting states become $H_{\text{SC}} = \Delta\sum_{\vec{q}} \widetilde{f}^\dagger_{\vec{q}} i \sigma_y \tau_x\widetilde{f}^\dagger_{-\vec{q}} + \text{H.c.}$, for both $A_1$ and $A_2$. They become identical upon projection, as expected since $M=\tau_z \mu_z$ transforms under $A_2$. It is straightforward to project the partner order parameters in \tableref{SummaryOrderParamters} and one finds that only one set, $(\sigma_x,\sigma_y,\sigma_z)\tau_z\rho_z$, the three component QSH order parameter, survives projection; this is due to the fact that it is the only set of $m_j$ in \tableref{SummaryOrderParamters} that commutes with $M=\tau_z\mu_z$. We indicate this in the last column of \tableref{SummaryOrderParamters} and conclude that the QSH insulator is the only correlated insulator that can provide a WZW term for singlet pairing and both at $\nu=0$ as well as $\nu=\pm 2$ with high-temperature $M=\tau_z \mu_z$.

Rather than projecting the orders from the full space, a simpler and a more general approach is to repeat the procedure of \secref{ProcedureOfFindingTerms} to find WZW terms directly in the reduced ($8\times 8$) eigenspaces of $M$. First of all, this reproduces the above finding that the QSH order parameter remains a partner order parameter.
Second, it also shows that $M=\tau_z \mu_z$ allows for further partner order parameters that were not included already in \tableref{SummaryOrderParamters}: for instance, as can be seen in first and seventh line in \tableref{SymmBrokenOrdersSinglet}, the projection to one of the eigenspaces of $M=\tau_z \mu_z$ allows for a set of partner order parameters consisting of a (one-component) sublattice polarized (SP), $\rho_z$, and a (two-component) state, $\rho_x\mu_x(\tau_+,\tau_-)$, which can be interpreted as either and IVC or MDW in the full space. In the full space, this set of $m_j$ is incomplete due to translational symmetry, similar to the example in \equref{ExampleOfThreems} discussed above. In either of the two eigenspaces of $M=\tau_z \mu_z$, however, the action of translation can be represented on the three components $\rho_x\mu_x(\tau_+,\tau_-);\rho_z$ since $M\rho_x\mu_x(\tau_x,\tau_y)=\rho_x\mu_y(-\tau_y,\tau_x)$.

\begin{table*}[htb]
\begin{center}
\caption{Partner orders for singlet pairing which are not included in \tableref{SummaryOrderParamters} and are candidates for $\nu=\pm 2$. We only list partner orders which are disallowed without the additional symmetry breaking of an $M$. We indicate the $M$ for which the partner order $m_j$ is a candidate and note that the partner orders are only defined up to multiplication by $M$ in the projected space. For instance, the orders $m_j=\rho_z\tau_z\mu_y(\sigma_x,\sigma_y)$ with corresponding $M=\mu_z\sigma_z$ can also be expressed as $m_j=\rho_z\tau_z\sigma_x(\mu_x,\mu_y)$, $m_j=\rho_z\tau_z\sigma_y(\mu_x,\mu_y)$, or $m_j=\rho_z\tau_z\mu_x(\sigma_x,\sigma_y)$; all of these anti-commute and are orders that survive projection.
$M$ which will lead to additional Fermi surfaces as described in \tableref{HighTemperatureOrders} and are thus less likely are denoted with square brackets. 
The full set of orders for singlet pairing, including those in \tableref{SummaryOrderParamters} (up to projection) can be found in \tableref{SummaryOrderParamtersWithMs} and \tableref{SummaryTripletOrderParamtersWithMsTR}. We only include those orders for which at least two components are related by a valley, mini-valley, or spin rotation. In the last column that indicates the type of partner order, "Sp" denotes spin, "S" denotes sublattice, "V" denotes valley, one or more of which can be polarized ("P"). "SBO" denotes spin-bond ordering, and "AFM" denotes antiferromagnetism (see \secref{summary} for more info). We note that the labeling of the symmetries of each $m_j$ are only well defined up to multiplication by the corresponding $M$. Also note for the cases in this table only, we distinguish between $\mu_x$ and $\mu_y$ for orders which have at least two distinct $M$'s proportional to both $\mu_x$ and $\mu_y$.}
\label{SymmBrokenOrdersSinglet}
\begin{ruledtabular}
 \begin{tabular} {c|ccccccc}
$M$ & Partner Orders $m_j$ & Partner SC & SU(2)$_s$ & $\Theta$  & $T_{\vec{a}_r}$ & U(1)$_v$ & Type \\ \hline

$\mu_x;\tau_z\mu_y\sigma_z;[\tau_z\mu_z]$ & $\rho_x\mu_x(\tau_+,\tau_-);\rho_z$ & $A_1$&\cmark & $+$ & \xmark & \cmark & IVC-MDW$_+$; SP \\ 

$\tau_z\mu_x\sigma_z;\tau_z\sigma_{x/y};[\mu_z\sigma_z]$ & $\rho_x\mu_z\sigma_z(\tau_+,\tau_-);\rho_z$ & $A_1$&\xmark & $+$ & \cmark & \cmark & IVC$_+$; SP \\ 
$\rho_y\tau_x\sigma_z;\tau_z\sigma_z;[\mu_z\sigma_{x/y}]$ & $\tau_z\rho_z\sigma_z(\mu_+,\mu_-);\rho_z\tau_z\mu_z$ & $A_1$&\xmark & $+$ & \cmark & \cmark & MDW$_+$; VP \\ 

$\tau_z\mu_x\sigma_z;\rho_y\mu_x\tau_x\sigma_z;\mu_y;[\mu_z\sigma_z]$ & $\rho_z\tau_z\mu_y(\sigma_x,\sigma_y);\rho_z\tau_z\sigma_z$ & $A_1$&\xmark & $+$ & \xmark & \cmark & AFM$^\perp_+$; SpVSP \\ 

$\rho_y\mu_x\tau_x\sigma_z;\tau_z\sigma_z;\rho_y\tau_y\sigma_z;\tau_y\rho_y\mu_z$ & $\tau_y\mu_z\rho_x(\sigma_x,\sigma_y);\rho_z\tau_z\sigma_z$ & $A_1$&\xmark & $+$ & \cmark & \xmark & SBO-AFM$^\perp_+$; SpVSP \\ 

$\rho_y\tau_x\sigma_z;\tau_z\mu_x\sigma_z;[\tau_z\mu_z]$ & $\rho_z\mu_z(\sigma_x,\sigma_y);\rho_z\tau_z\sigma_z$ & $A_1$&\xmark & $+$ & \cmark & \cmark & SBO-AFM$^\perp_+$; SpVSP \\ 
\hline

$\tau_z\mu_y;\mu_x\sigma_z;[\tau_z\mu_z]$ & $\rho_x\mu_x(\tau_+,\tau_-);\rho_z$ & $A_2$&\cmark & $+$ & \xmark & \cmark & IVC-MDW$_+$; SP \\ 

$\mu_x\sigma_z;\tau_z\sigma_{x/y};[\mu_z\sigma_z]$ & $\rho_x\sigma_z(\tau_+,\tau_-);\rho_z$ & $A_2$&\xmark & $-;+$ & \cmark & \cmark & IVC$_-$;SP\\
$\tau_z\sigma_z;[\rho_y\tau_x\mu_z\sigma_z;\mu_z\sigma_{x/y}]$ & $\rho_z\sigma_z(\mu_+,\mu_-);\rho_z\tau_z\mu_z$ & $A_2$&\xmark & $-;+$ & \cmark & \cmark & MDW$_-$; VP\\

$\mu_x\sigma_z;\rho_y\mu_y\tau_x\sigma_z;\tau_z\mu_y;[\mu_z\sigma_z]$ & $\rho_z\mu_y(\sigma_x,\sigma_y);\rho_z\tau_z\sigma_z$ & $A_2$&\xmark & $-;+$ & \xmark & \cmark & AFM$^\perp_-$; SpVSP\\

$\rho_y\mu_x\tau_x\sigma_z;\tau_z\sigma_z;\tau_y\rho_y;[\rho_y\tau_x\mu_z\sigma_z]$ &$\tau_x\rho_x(\sigma_x,\sigma_y);\rho_z\tau_z\sigma_z$ & $A_2$&\xmark & $-;+$ & \cmark & \xmark & SBO-AFM$^\perp_-$; SpVSP \\ 

$\mu_x\sigma_z;[\tau_z\mu_z;\rho_y\tau_x\mu_z\sigma_z]$ & $\rho_z\mu_z(\sigma_x,\sigma_y);\rho_z\tau_z\sigma_z$ & $A_2$&\xmark & $+$ & \cmark & \cmark & SBO-AFM$^\perp_-$; SpVSP
 \end{tabular}\end{ruledtabular}
\end{center}
\end{table*}

In a similar way, the other two $M$ in \equref{OptionsForOrderParameters} can be analyzed. As for the $m_j$ in \tableref{SummaryOrderParamters}, which are already possible without any $M$, these two $M$ only work for $A_1$ pairing but are both associated with two different partner orders. As can be seen in \tableref{SymmBrokenOrdersSinglet}, $M=\mu_{x/y}$ ($M=\rho_y\tau_{x,y}\sigma_{x,y,z}$) makes one (two) additional set(s) of $m_j$ possible at $\nu=\pm 2$. When breaking spin-rotation symmetry in the next subsection below, we will see that high-temperature orders $M$ can stabilize many more partner orders with WZW terms.

Before turning to this, we mention there is some ambiguity in presenting the partner order in the presence of a given $M$. This follows from the observation that there are several partner orders in the full $16$-dimensional space that project to the same orders in the relevant $8$-dimensional eigenspaces of $M$. {\it E.g.\/}, both the regular QSH insulator with $m_j = \sigma_j \tau_z\rho_z$ as well as $m'_j = Mm_j = \sigma_j \mu_z\rho_z$ are equally valid for $M=\tau_z\mu_z$. Here and in the main text, we always only show one of these equivalent options. To this end, we will always show the unique form of the order parameter that will have $\mathcal{N}\neq 0$ in \equref{model3_coef} and, hence, can give rise to a WZW term in the full space (but, in some cases, will require additional broken symmetries). For completeness, we provide in \tableref{SummaryOrderParamtersWithMs} a complete list that also contains these alternative and redundant choices explicitly.

\subsection{Breaking spin-rotation invariance}\label{BreakingSpinRotationInvariance}
Let us next generalize our discussion of high-temperature orders $M$ to include the breaking of spin-rotation invariance, while keeping time-reversal symmetry.
In this case, we are left with the following combinations of Pauli matrices 
\begin{equation}
\begin{split}
    M = \tau_z\vec{\sigma}, \quad M = \mu_z\vec{\sigma}, \quad M = \tau_z(\mu_x,\mu_y)\vec{\sigma}, \\ \quad M = \rho_y(\tau_x,\tau_y)\vec{\sigma}, \quad M = \rho_y(\tau_x,\tau_y)(\mu_x,\mu_y)\vec{\sigma}.\label{MjSes}
\end{split}
\end{equation}
As before, we have already grouped them together as multi-component order parameters such that different components transform into each other under the symmetries of the system. While for the first two options in \equref{MjSes} all possible orientations of these vector order parameters are symmetry-equivalent, we have to analyze the possible stable phases for the remaining three choices; these are matrix- and third-rank-tensor-valued order parameters. 
This analysis can be performed systematically by writing down the most general free-energy expansion in terms of these components, see \appref{FreeEnergyExpansion}. We find that of the multitude of options, only some of the configurations for each of the last three order parameters in \equref{MjSes} will have the correct eigenspace degeneracy needed for four degenerate Dirac cones at $\nu=2$ (or $\nu=-2$).
For example, for the high-temperature order $\rho_y(\tau_x,\tau_y)\vec{\sigma}$, both $\rho_y\tau_x\sigma_z$ and $\rho_y(\tau_x\sigma_x+\tau_y\sigma_y)$ are stable minima of the most general free energy. While the first of these two options, does have only two eigenvalues, $\pm 1$, (each $8$-fold degenerate) and, hence, shifts the Dirac cones as shown in \figref{GappingOutBands}(b), the second one has eigenvalues $\pm 1$ ($4$-fold each) and $0$ ($8$-fold) and, hence, can only work for filling $\nu=\pm 3$, see \figref{GappingOutBands}(c). Here we take the simplest minima with the correct eigenspectrum for each $M$ and discuss any other possibilities in \appref{FreeEnergyExpansion}. We note, in passing, that no WZW terms are possible starting from a parent theory with an $M$ of this form that corresponds to filling $\nu=\pm 3$: in this case, the effective low-energy theory will be a theory with $4\times 4$ Dirac matrices. Since the maximal number of anti-commuting $4\times 4$ Hermitian matrices is $5$, this is not compatible with \equref{model3_coef}.

Returning to $\nu=\pm 2$, we conclude that \equref{MjSes} only leads to five different high-temperature orders to consider, which are summarized in line $4$ to $8$ in \tableref{HighTemperatureOrders}. We not only list their symmetries, but also whether they require additional mixing away from the $\text{K}_\text{M}$ and $\text{K'}_\text{M}$ point, to avoid unwanted Fermi surfaces.

We analyze these terms in the same way as above. As before, we find that some of the partner orders which are already possible at $\nu=0$ (without any $M$) remain, as indicated in the last column of \tableref{SummaryOrderParamters}. In addition, the presence of these high-temperature orders leads to additional options, summarized in \tableref{SymmBrokenOrdersSinglet} (the full list of redundant options is given in \tableref{SummaryOrderParamtersWithMs}); these latter cases are, thus, only possible around $\nu=\pm 2$. As anticipated above, for all of the WZW terms with $M$ breaking spin-rotation symmetry, the lack of three-dimensional IRs implies that not all three partner orders can transform under the same IR and anisotropy terms between the two distinct classes of partner orders are generically expected. 
In fact, for $M=\rho_y\mu_x\tau_x\sigma_z$, a WZW term is possible with all three particle-hole partner orders transforming under different IRs (see \tableref{SummaryOrderParamtersWithMs}). Since this requires more fine-tuning, we do not include this option in \tableref{SymmBrokenOrdersSinglet}. 

\subsection{Breaking time-reversal symmetry}\label{BreakingTimeReversalSymmetry}
Finally, we can also repeat the same analysis for high-temperature order parameters that are odd under time-reversal symmetry.
We find that all of these terms are incompatible with the $A_1$ pairing term, \textit{i.e.}, the projection of the $A_1$ pairing term onto the eigenspaces of any of these order parameters vanishes. For $A_2$, the following four classes of time-reversal odd $M$ are possible
\begin{equation}
    \tau_z(\mu_x,\mu_y), \,\,\, \rho_y(\tau_x,\tau_y),\,\,\, (\mu_x,\mu_y)\vec{\sigma},  \,\,\, \rho_y\mu_z\vec{\sigma}(\tau_x,\tau_y).
\label{SetOfTROStates}
\end{equation}
A discussion of all stable configurations of these multi-component orders can be found in \appref{FreeEnergyExpansion}. But we find, as before, that the additional options which involve linear combinations of the different components do not have the correct degeneracies of eigenspaces required for $\nu=\pm 2$. The four $M$ associated with \equref{SetOfTROStates} together with the additional possible $M$ with the right degeneracies to describe $\nu=\pm 2$, but that lead to vanishing pairing, can be found in the last $7$ lines of \tableref{HighTemperatureOrders}. The different partner orders can be read off from \tableref{SummaryOrderParamters} and \tableref{SymmBrokenOrdersSinglet} as before.

\section{Generalization to triplet pairing}\label{TripletPairing}
In this section, we extend the previous discussion to also include triplet pairing.

\subsection{Possible triplets states}
We can repeat the same procedure of determining possible WZW partners for triplet pairing. To this end, let us begin by discussing the different possible triplet states. These are characterized by an order-parameter $\Delta_{\vec{q}}$ in \equref{GeneralFormOfPairing} involving the spin Pauli matrices $\sigma_{x,y,z}$. As in singlet pairing, we restrict possible pairing terms to those which pair electrons with opposite momenta, and between opposite valleys and mini-valleys, \textit{i.e.}, only the off-diagonal matrix elements of $\Delta_{\vec{q}}$ in valley and mini-valley space, $p'=-p$ and $v'=-v$, are non-zero. Keeping only the momentum-independent terms around the $\text{K}_\text{M}$ and $\text{K'}_\text{M}$ points, $\Delta_{\vec{q}}\rightarrow \Delta^t$, we obtain the different triplet states listed in \tableref{TripletPairingTerms} according to the IRs of the spatial point group $D_6$. Similar to the singlet case above, we see that the property derived in \refcite{2019arXiv190603258S} of triplet states even under $C_2$ not giving rise to a gap in isolated (valley- and/or spin-degenerate) bands, carries over to the Dirac points: the $A_1$ and $A_2$ triplets do not induce a gap in our Dirac theory either. We also point out that triplet pairing cannot be ruled out \textit{a priori} due to the presence of disorder, such as variations of the local twist angles, as triplet pairing can be protected by an Anderson theorem, special to graphene moir\'e superlattices, as has recently been shown \cite{MicroscopicAndDisorder}.  

In \tableref{TripletPairingTerms}, we have focused on the regular SU(2)$_s$ spin symmetry and neglected the admixture of spin-singlet and triplet, possible due to the proximity to an enhanced spin symmetry \cite{2019arXiv190603258S}. 
Contrary to the case of singlet pairing, the IR of the complete symmetry group is thus three-dimensional for $A_{1,2}$ and $B_{1,2}$: as is well known, there are two distinct types of stable triplet vectors, which we will choose as
\begin{eqnarray}
\vec{d} = (1,0,0)^T, \qquad \vec{d} = (0,1,i)^T \label{TwoDistinctTriplets}
\end{eqnarray}
and refer to as ``unitary'' and ``non-unitary'' triplets, respectively. 

In the case of the IR $E_1$, there are two different forms of momentum-independent order parameters with the same symmetries and the associated basis functions are superpositions, $\chi^{E_1}_{\vec{k},1,j} = (a \rho_y+ b \mu_z\rho_x)\sigma_j$ and $\chi^{E_1}_{\vec{k},2,j} = (a \tau_z\rho_x- b \tau_z\mu_z\rho_y)\sigma_j$, $j=1,2,3$, $a,b\in \mathbb{R}$. Here, the superconducting order parameter has the form $\Delta_{\vec{k}} = \sum_{\mu=1,2,j=1,2,3} \eta_{\mu,j} \chi^{E_1}_{\vec{k},\mu,j}$. Since it transforms as the product of a two- and a three-dimensional IR, the set of symmetry-inequivalent order parameters becomes quite rich and has been discussed in detail in \refcite{2019arXiv190603258S} for twisted bilayer graphene. 

Here, we will not need further details about these triplet phases since only $\tau_z\sigma_j$ ($B_1$) and $\mu_z\sigma_j$ ($B_2$) satisfy the condition of $\Delta T \gamma_j=-\gamma_j^T\Delta T$ with $T=i\sigma_y\tau_x\mu_x$, for $\gamma_y=-\gamma_y^T=\rho_y$, and  $\gamma_x=\gamma_x^T=\tau_z\rho_x$ which is the criterion (\ref{FirstPartOfCond}) for anti-commuting with the kinetic terms in Nambu space.

The unitary triplet in \equref{TwoDistinctTriplets} corresponds to $N_s=2\times 3$ real components; as this is already more than the five components forming the WZW term in \equref{SWZW}, it cannot give rise to WZW terms as long as spin-rotation invariance is preserved. Similarly, the manifold SO(3) of the non-unitary triplet is not consistent with the WZW term in \equref{SWZW} either. This is different if spin-rotation invariance is broken by high-temperature orders $M$, as we will discuss next.

\begin{table}[tb]
\begin{center}
\caption{Summary of the triplet pairing states according to the IRs of the spatial point group $D_6$. The last column indicates whether the superconducting state can gap out the Dirac cones. The allowed triplet vectors for the one-dimensional IRs are given in \equref{TwoDistinctTriplets}, while we refer to \refcite{2019arXiv190603258S} for $E_1$ and the gap structure of the pairing states in the entire Brillouin zone at generic $\nu$.}
\label{TripletPairingTerms}
\begin{ruledtabular}
 \begin{tabular} {c|ccc} 
    Order parameter $\Delta^t$   & Transform as & IR of $D_6$ & Gap \\ \hline
$\mu_z\rho_z\vec{d}\cdot\vec{\sigma}$    & const., $z^2$   & $A_1$ & \xmark  \\ 
$\tau_z\rho_z\vec{d}\cdot\vec{\sigma}$    & $z$   & $A_2$ & \xmark  \\ 
$\tau_z\vec{d}\cdot\vec{\sigma}$    & $x(x^2-3y^2)$   & $B_1$ & \cmark  \\ 
$\mu_z\vec{d}\cdot\vec{\sigma}$    & $y(3x^2-y^2)$   & $B_2$  & \cmark \\ 
$(\rho_y,\tau_z\rho_x)\sigma_j$    & $(xz,yz)$   & $E_1$ & \xmark \\ 
$\mu_z(\rho_x,-\tau_z\rho_y)\sigma_j$    & $(xz,yz)$   & $E_1$ &\xmark
 \end{tabular}\end{ruledtabular}
\end{center}
\end{table}

\subsection{High-temperature orders and WZW terms}
We repeated the same analysis discussed in detail in \secref{SymmetryBreakingAtHighT} above for singlet pairing, but now for the two unitary and non-unitary triplets transforming under $B_1$ and $B_2$; we went through all $M$ that lead to a Dirac revival at $\nu=\pm 2$, investigated whether the respective pairing states survive projection to their eigenspaces, and searched for all partner order parameters in this reduced space which will give rise to joint WZW terms (\ref{SWZW}), with $\mathcal{N}=1$. The results are summarized in \tableref{SummaryTripletOrderParamters} and will be discussed next.

First, as explained above, only $M$ that break SU(2)$_s$ are possible. In principle, there are two different ways of breaking it: using the conventions for the triplet vectors in \equref{TwoDistinctTriplets}, $M$ could correspond to a polarization along $\sigma_z$. Then, as a consequence of the residual spin-rotation symmetry along the $\sigma_z$ axis, both the unitary and non-unitary triplet have three independent real components. For instance, the unitary triplet can be parametrized in this case as
\begin{align}\begin{split}
 \vec{d} &= \Delta e^{i\varphi} (\cos\theta \vec{e}_x + \sin\theta \vec{e}_y) \\
 & = (n_1 + i\, n_2) \vec{e}_x + (n_3 + i\, n_4) \vec{e}_y, 
\end{split}\end{align}
where we introduce the unit vectors $\vec{e}_j$ and, in the second line, a redundant parameterization with the four $n_a$ associated with mass terms $\mathcal{M}_a$ in the Nambu-Dirac theory (\ref{NambuDiracTheory}). Even if we ignore the additional constraint ($n_1/n_2=n_3/n_4$) accounting for the fact that only three of them are independent, this does not correspond to the scenario we are interested where skyrmions in the three-component partners orders carry electric charge and form the Cooper pairs. This is why we will not further discuss this spin polarization of $M$.

\begin{table*}[tb]
\begin{center}
\caption{Possible partner order parameters, $m_j$, $j=1,2,3$, see  \equref{PartnerOrderParamsDef}, for unitary ($B^a_{1,2}$) and non-unitary triplet pairing pairing ($B_{1,2}^b$) with triplet vectors defined in \equref{TwoDistinctTriplets}.
The final two columns correspond to the high-temperature orders required to break the spin-rotation symmetry, \textit{i.e.}, all of these options only work at $\nu=\pm 2$. The second to last column, $M^a$, refers to unitary and the last column, $M^b$, to non-unitary triplet pairing.$M^a$'s which will lead to additional Fermi surfaces as described in \tableref{HighTemperatureOrders} and are thus less likely are denoted with square brackets. The full set of orders for triplet pairing, including including those related by multiplication by $M$ are listed in \tableref{SummaryTripletOrderParamtersWithMs} and \tableref{SummaryTripletOrderParamtersWithMsTR}.
}
\label{SummaryTripletOrderParamters}
\begin{ruledtabular}
 \hspace*{-8em}\begin{tabular} {c|ccccccccc}
Pairing  & $m_j$ & $D_6$ & $\Theta$  & $T_{\vec{a}_r}$ & U(1)$_v$ & SU(2)$_s$ & Type & $M^a$ & $M^b$\\ \hline
$B^a_1/B_1^b$ &  $(\tau_+,\tau_-)\mu_z\rho_x$; $\rho_z$ & $B_2/A_2$; $B_2$ & $-$; $+$ & $1$ & $m=1$; $m=0$ & \cmark & IVC$_-$; SP & $\tau_z\mu_x\sigma_x;[\mu_z\sigma_x]$ & $\sigma_x$\\
$B^a_1/B_1^b$ &  $(\mu_+,\mu_-)\rho_z$; $\tau_z\rho_z\mu_z$ & $B_2$/$A_1$; $B_1$ & $+$ & $(\omega_r,\omega_r^*)$; $1$ & $m=0$ &  \cmark & MDW$_+$; VP  & $\tau_z\sigma_x;[\rho_y\tau_x\mu_z\sigma_x]$ & $\sigma_x$\\
$B^a_1/B_2^b$ &  $(\tau_+,\tau_-)\rho_x\sigma_x$; $\rho_z$ & $A_1/B_1$; $B_2$ & $-$; $+$ & $1$ & $m=1$; $m=0$ &  \xmark & spIVC$_-$; SP & $[\mu_z\sigma_x]$ & $\sigma_x$\\
$B^a_1/B_2^b$ &  $(\mu_+,\mu_-)\tau_z\rho_z\sigma_x$; $\tau_z\rho_z\mu_z$ & $A_2$/$B_1$; $B_1$ & $+$ & $(\omega_r,\omega_r^*)$; $1$ & $m=0$ &  \xmark & MDW$_+$; VP & $\tau_z\sigma_x$ & $\sigma_x$\\
$B^a_1$ &  $(\sigma_y,\sigma_z)\rho_z$; $\tau_z\rho_z\sigma_x$ & $B_2$/$B_2$; $A_2$ & $-$;$+$ &  $1$ & $m=0$ &  \xmark & AFM$^\perp_-$; SpVSP & $[\rho_y\tau_x\mu_z\sigma_x]$ & -\\
$B^a_1$ &  $(\sigma_y,\sigma_z)\rho_z\tau_z\mu_z$; $\tau_z\rho_z\sigma_x$ & $B_1$/$B_1$; $A_2$ & $-$;$+$ &  $1$ & $m=0$ &  \xmark & AFM$^\perp_-$; SpVSP & $\tau_z\mu_x\sigma_x$ & -\\
$B^a_1$& $\rho_z\mu_x(\sigma_y,\sigma_z);\rho_z\tau_z\sigma_x$ & $B_2/B_2;A_2$  & $-;+$ &$(\omega_r,\omega_r^*)$; $1$& $m=0$& \xmark  & \hspace*{-1em}MDW-AFM$^\perp_-$; SpVSP &$[\mu_z\sigma_x]$& - \\ 

$B^a_1$ & $\rho_x\mu_z\tau_x(\sigma_y,\sigma_z);\rho_z\tau_z\sigma_x$ & $B_2/B_2;A_2$  & $+$&$1$ & $m=1$ & \xmark & IVC-SBO$^\perp_+$; SpVSP & $\tau_z\sigma_x$ & - \\

\hline 

$B^a_2/B_2^b$ &  $(\tau_+,\tau_-)\rho_x$; $\rho_z$ & $A_1/B_1$; $B_2$ & $+$ & $1$ & $m=1$; $m=0$ &  \cmark & IVC$_+$; SP & $\mu_{x}\sigma_x;[\mu_z\sigma_x]$ & $\sigma_x$\\
$B^a_2/B_2^b$ &  $(\mu_+,\mu_-)\tau_z\rho_z$; $\rho_z\tau_z\mu_z$ & $A_2$/$B_1$; $B_1$ & $-$; $+$ & $(\omega_r,\omega_r^*)$; $1$ & $m=0$ &  \cmark & MDW$_-$; VP & $\rho_y\tau_x\sigma_x;\tau_z\sigma_x$ & $\sigma_x$ \\
$B^a_2/B_1^b$ &  $(\tau_+,\tau_-)\rho_x\sigma_x\mu_z$; $\rho_z$ & $B_2/A_2$; $B_2$ & $+$ & $1$ & $m=1$; $m=0$ &  \xmark & IVC$_+$; SP& $[\mu_z\sigma_x]$ & $\sigma_x$\\
$B^a_2/B_1^b$ &  $(\mu_+,\mu_-)\rho_z\sigma_x$; $\tau_z\rho_z\mu_z$ & $B_2$/$A_1$; $B_1$ & $-$; $+$ & $(\omega_r,\omega_r^*)$; $1$ & $m=0$ &  \xmark& MDW$_-$; VP & $\tau_z\sigma_x$ & $\sigma_x$ \\
$B^a_2$ &  $(\sigma_y,\sigma_z)\rho_z$; $\tau_z\rho_z\sigma_x$ & $B_2$/$B_2$; $A_2$ & $-$; $+$ &  $1$ & $m=0$ &  \xmark & AFM$^\perp_-$; SpSVP& $\rho_y\tau_x\sigma_x$ & - \\

$B^a_2$ &  $(\sigma_y,\sigma_z)\rho_z\tau_z\mu_z$; $\tau_z\rho_z\sigma_x$ & $B_1$/$B_1$; $A_2$ & $-$; $+$ &  $1$ & $m=0$ & \xmark & AFM$^\perp_-$; SpVSP& $\mu_x\sigma_x$ & - \\

 $B^a_2$& $\rho_x\tau_x(\sigma_y,\sigma_z);\rho_z\tau_z\sigma_x$ & $B_2/B_2;A_2$ & $-;+$ &$1$& $m=1$& \xmark & IVC-SBO$^\perp_-$; SpVSP & $\tau_z\sigma_x$ & -\\ 

 $B^a_2$ & $\rho_z\mu_x\tau_z(\sigma_y,\sigma_z);\rho_z\tau_z\sigma_x$ & $A_2/A_2;A_2$ & $+$ &$(\omega_r,\omega_r^*)$; $1$& $m=0$ & \xmark &  \hspace*{-1em}MDW-AFM$^\perp_+$; SpVSP  & $[\mu_z\sigma_x]$ & -
 \end{tabular}
 \end{ruledtabular}
\end{center}
\end{table*}

The second way of breaking spin-rotation symmetry by $M$ corresponds to having $M \propto \sigma_x$, \textit{i.e.}, along $\vec{d}$ for the unitary and perpendicular to it for the non-unitary triplet state in \equref{TwoDistinctTriplets}. While the non-unitary triplet transforms as $\vec{d} \rightarrow e^{i\varphi} \vec{d}$ under the residual spin rotation (by $\varphi$ along $\sigma_x$ here) and, thus, will remain distinct from any of the singlets when introducing $M$, the unitary triplet is explicitly invariant under the residual spin rotation. For this reason, one might be tempted to conclude that it becomes equivalent to one of the singlets in \tableref{DifferentIndependentPairingTerms}, mix with it, and will not have to be discussed separately. This is, however, not the case and again related to the special role of $C_2$ symmetry in two spatial dimensions \cite{2019arXiv190603258S,scheurer2017selection}: we have seen that only singlet (triplet) even (odd) under $C_2$ can give rise to a gap and fulfill the criteria for WZW terms. Consequently, as long as $C_2$ is a symmetry, also the unitary triplets transforming under $B_{1,2}$ in \tableref{SummaryTripletOrderParamters} are distinct from the singlets $A_{1,2}$ with WZW terms studied above.  

We also note that the partner orders discussed in the spinless model in \refcite{2020arXiv200400638K}---in our notation $\mu_z\rho_x\sigma_x(\tau_x,\tau_y);\rho_z$, see \appref{ConnectionToAshvin} for more details---are among our possibilities for triplet pairing. As can be read off from \tableref{SummaryTripletOrderParamters}, these partner orders are possible for unitary triplet pairing, with high-temperature order $M=\mu_z\sigma_x$ and for non-unitary pairing with $M=\sigma_x$. Out of these two different $M$, only the first one will lead to additional Fermi surfaces if no further mixing between the bands occurs far away from the $\text{K}_\text{M}$ and $\text{K'}_\text{M}$ points.
As can be seen in \tableref{SummaryTripletOrderParamters}, our analysis reveals that there are many more options for triplet pairing and associated partner orders in the presence of $M$.

We finally note that all of the partner order parameters for the triplets which are not spin polarized were already present in \tableref{SummaryOrderParamters} above and can, thus, also be possible partner states for both singlet and triplet phases, and both at $\nu=0$ (without $M$, singlet only) as well as $\nu=\pm 2$ (with the appropriate $M$). Clearly, the QSH state, $\tau_z\rho_z\sigma_j$ in \tableref{SummaryOrderParamters}, can only provide the partner order parameter for singlet superconductivity, as triplet will necessarily require broken spin-rotation symmetry. If we also take into account the partner orders for singlets in \tableref{SymmBrokenOrdersSinglet}, which only work for $\nu=\pm 2$, we see that all the partner orders in \tableref{SummaryTripletOrderParamters} that are possible for both unitary and non-unitary triplet pairing also work for singlet.


\section{Summary and Discussion}
\label{summary}

Experimental studies of the low-temperature phase diagram for TBG show superconducting domes separated by correlated insulators at integer filling fractions \cite{SenthilJournalClub,Efetov19,Young19,Li20,PabloNematicity,WSe2TBG}. We connected spin-singlet superconductivity to possible order parameters for correlated insulators, referred to as partner orders $m_j$, by WZW terms in Sec.~\ref{WZWNoOtherOrders}. More recent STM observations \cite{Yazdani19,Ilani19} have argued for further symmetry breaking in a high-temperature parent state with Dirac fermions at each integer filling fraction. Only at $\nu=0$ is no symmetry breaking required in this parent state for the Dirac fermions to appear, as illustrated in Fig.~\ref{GappingOutBands}, and so the results in Sec.~\ref{WZWNoOtherOrders} and the order parameters $m_j$ in Table~\ref{SummaryOrderParamters} can be applied at this filling.
We considered order parameters $M$ for the high-temperature symmetry breaking in the vicinity of $\nu=\pm 2$ in Sec.~\ref{SymmetryBreakingAtHighT} and Table~\ref{HighTemperatureOrders}; these additional orders have two consequences for WZW terms. First, for a given $M$, they can rule out certain combinations of partner orders and superconductivity, since some of these order parameters vanish upon projection to one of the eigenspaces of $M$. However, all of the candidate orders for $\nu=0$ still remain possible for $\nu=\pm 2$, if an appropriate $M$ is present, as indicated in the last column in \tableref{SummaryOrderParamters}. Second, the high-temperature order parameter will reduce the number of low-energy degrees of freedom and break a certain subset of the symmetries of the system, which will allow for additional combinations of $m_j$ and superconductivity with a WZW term; these options, which are thus only possible at $\nu=\pm 2$, are listed \tableref{SymmBrokenOrdersSinglet}.

In \secref{TripletPairing}, we have repeated the same analysis for triplet pairing. Here, spin-rotation invariance has to be explicitly broken in the high-temperature phase to obtain a WZW term. Therefore, the proposed connection between correlated insulators and a triplet superconductor will not be possible around $\nu=0$. Additional $M$ around $\nu=\pm 2$, however, can reduce the spin symmetry and lead to the various possible combinations of triplet pairing and $m_j$ summarized in \tableref{SummaryTripletOrderParamters}.

Our comprehensive discussion of allowed combinations of superconductivity and correlated insulators in the absence or presence of possible $M$ involves a variety of different order parameters.
Recalling that our starting point is the low-energy Dirac theory (\ref{FinalLowEnergyTheory2}) with $\gamma_{x,y}$ representing $16\times 16$ matrices in valley ($\tau_i$), mini-valley ($\mu_i$), spin ($\sigma_i$), and generalized sublattice space ($\rho_i$), we studied the following types of orders:

\begin{itemize}
    \item IVC$_+$: time-reversal even intervalley coherent state, which has density modulations on the graphene lattice scale.
    \item IVC$_-$: as in IVC$_+$, but time-reversal odd.
    \item SP ($\rho_z$): moir\'e sublattice polarized state which is partner to either an IVC$_+$ or IVC$_-$ as $m_j$.
    \item MDW$_+$: time-reversal even, density modulations on the moir\'e lattice scale.
    \item MDW$_-$: as in MDW$_+$, but time-reversal odd.
    \item VP ($\rho_z\tau_z\mu_z$): valley, mini-valley, and moir\'e sublattice polarized state which is partner to either an MDW$_\pm$ as $m_j$.
    \item AFM$^\perp_+$: time-reversal even, in-plane, two-sublattice antiferromagnet on the moir\'e lattice scale.
    \item AFM$^\perp_-$: as in AFM$_+$, but time-reversal odd.
    \item SBO$^\perp_+$: time-reversal even spin-bond ordering on the moir\'e lattice scale.
    \item SBO$^\perp_-$: as in SBO$_+$, but time-reversal odd.
    \item SpVSP ($\rho_z\tau_z\sigma_z$): valley, spin, and moir\'e sublattice polarized state which is partner to either an AFM$_\pm$ or SBO$_\pm$ as $m_j$.
    \item QSH: quantum spin Hall, leading to opposite Chern number bands for spin up 
    and down.
\end{itemize}

We finally make a few remarks on the structure and implications of our central results in Tables~\ref{SummaryOrderParamters}-\ref{SymmBrokenOrdersSinglet} and \ref{SummaryTripletOrderParamters}. Let us first note that only the superconducting states transforming under one-dimensional IRs of $D_6$ can give rise to WZW terms, irrespective of $M$ and filling. In fact, for singlet only $A_1$ or $A_2$ and for (both unitary or non-unitary) triplet only $B_1$ or $B_2$ are possible. 
If, indeed, the correlated insulators and superconductors are intimately related by a WZW term, the number of pairing states is thus fairly constrained, as the two-dimensional IRs give rise to the majority of different superconducting order parameters \cite{2019arXiv190603258S}. On top of this, the superconducting domes closest to charge neutrality will have to be singlet in that scenario. If the superconductor is due to electron-phonon coupling, we know from the general analysis of \cite{Scheurer_2016} and \cite{MicroscopicAndDisorder} that the superconducting order parameter must be spin singlet and transform trivially under all symmetries. Consequently, only the particle-hole order parameters in the lines with pairing $A_1$ in \tableref{SummaryOrderParamters} and \tableref{SymmBrokenOrdersSinglet} are possible. One would then view the electron-phonon coupling having ``tipped the balance'' towards a particular type of superconductivity. The WZW term, which is a Berry phase term independent of a specific Hamiltonian \cite{PhysRevB.82.245117}, will continue to apply and constrain the partner insulating orders.  

It is also worth pointing out that, while we have identified $15$ possible high-temperature orders, four of them have to be regarded as less natural choices: they require additional symmetry breaking away from the $\text{K}_\text{M}$ and $\text{K'}_\text{M}$ points to avoid spurious Fermi surfaces coexisting with the Dirac points (see \tableref{HighTemperatureOrders}). This also has implications for the partner orders as it, \textit{e.g.}, makes $M=\mu_z\sigma_x$ and, hence, the MDW-AFM$_{\pm}^\perp$ and spIVC$_{-}$ partner orders less plausible for unitary triplet pairing.

Furthermore, we emphasize that our relation between $M$ and the associated sets of superconducting and partner order parameters could give crucial insights. For instance, if future experiments establish that the parent state around $\nu=\pm 2$ is characterized by the MDW$_+$ order parameter $M=\mu_x$, the pairing state must be the $A_1$ singlet and the partner order parameters have to be either the IVC$_+$ and SP phases, $m_j=(\tau_x\rho_x,\tau_y\rho_x;\rho_z)$, or the QSH state with $m_j = \tau_z\rho_z \sigma_j$ or the IVC-MDW$_+$ and SP phases, $m_j=(\rho_x\mu_x\tau_x,\rho_x\mu_x\tau_y;\rho_z)$. Furthermore, if a correlated insulating state $m_j$ breaks time-reversal symmetry then the pairing cannot be singlet pairing and transform under $A_1$; in that case, electron-phonon coupling alone cannot be responsible for superconductivity as mentioned above. However, if the high-temperature order $M$ breaks only translational symmetry, but preserves all others in \tableref{RepresentationsOfSymmetries}, the pairing must be the $A_1$ singlet. 

We note that QSH is the only example of a set of partner order parameters where all three components are related by symmetry and, as such, requires the least amount of fine tuning of all $m_j$. As can be read off in \tableref{SummaryOrderParamters}, it is relevant to both $A_1$ and $A_2$ singlet pairing at $\nu=0$ and $\nu =\pm 2$ with five possible $M$ (four of which will not give rise to extra Fermi surfaces); for all other partner orders, two different IRs have to  be energetically close in energy for the connection of correlated insulator and superconductivity to be physically plausible. To provide another example, if future experiments establish that $M=\sigma_x$ around $\nu=\pm 2$ is realized (not realized), the superconducting state will have to be (cannot be) a non-unitary triplet. 

We point out that our key results---the sets of partner orders and high-temperature order parameters
$M$---are not altered when three-fold rotation symmetry, $C_3$, is broken due to the presence of strain
and electronic nematic order \cite{PasupathySTM,NadjPergeSTM,AndreiSTM,YazdaniSTM,PabloNematicity,STMReview}; the broken $C_3$ symmetry can also explain the observed Landau-level degeneracy near charge neutrality
\cite{2019PhRvB.100l5104Z,Vafek20}. To see why it does not affect our results, first note that removing the $C_3$ symmetry will allow the Dirac cones to move away from the $K$ and $K'$ points, but we can still write
down a low-energy theory as in \secref{ModelsSymmetries} by expanding around the shifted Dirac cones. The only modifications are anisotropic Dirac velocities and that the momentum transfers and, hence, the MDW states become incommensurate with the moir\'e lattice. However, because none of the relevant
superconducting states transform non-trivially under $C_3$ and we did not use this symmetry to exclude further partner orders or high-temperature orders,
Tables~\ref{SummaryOrderParamters}-\ref{SymmBrokenOrdersSinglet} and \ref{SummaryTripletOrderParamters} still apply when $C_3$, is broken (with the sole exception of the transformation behavior of the MDW
states under $T_{\vec{a}_r}$ in Tables~\ref{SummaryOrderParamters} and
\ref{SummaryTripletOrderParamters}).
 While our mechanism thus still applies when electronic nematic order (and strain) breaks $C_3$, nematic order itself cannot be a partner order parameter for any superconductor, as it is inconsistent with \equref{LastPartOfGenCrti}.

A recent Monte-Carlo study \cite{RafaelMengMonteCarlo} has found evidence of the VP state, with order parameter $\rho_z\tau_z\mu_z$, around $\nu=0$ (referred to as quantum valley Hall state in \refcite{RafaelMengMonteCarlo}). As can be seen in \tableref{SummaryOrderParamters}, this order together with MDW$_{\pm}$ can provide the three partner order parameters for both singlet pairing states around charge neutrality. However, we caution that the two mini-valley Dirac nodes have opposite chiralities in Refs.~\onlinecite{RafaelMengMonteCarlo}, while those in \equref{FinalLowEnergyTheory2} have the same chiralities. It is not clear whether the short-range non-local interactions in their models are sufficient to include the effects of the WZW terms of the same chirality Dirac nodes that we have investigated here.

The WZW connection between the superconductivity and the correlated insulator order also has interesting consequences for the structure of the core of a superconducting vortex which could be explored in scanning tunneling microscopic experiment. By analogy to vortices in the valence bond solid order of insulating antiferromagnets carrying unpaired spins \cite{LevinSenthil04,KaulMelko2008}, superconducting vortices would carry quanta of the partner order. 

Taken together, we have proposed a mechanism by which  superconductivity and the correlated insulators are intimately related in TBG (see also the work of Khalaf {\it et al.\/} \cite{2020arXiv200400638K} discussed in Appendix~\ref{ConnectionToAshvin}). While future experiments will have to establish whether this is realized in the system or not, we believe that our systematic discussion of the different microscopic realizations of this physics can help constrain the order parameters of superconductivity, the correlated insulators, and the high-temperature parent in TBG and, potentially, also related moir\'e superlattices.
Numerical studies of models with WZW terms \cite{Mong20} perturbed by symmetry-breaking and chemical potential terms will also be useful.

\subsection*{Acknowledgements}

We acknowledge useful discussions with S.~Chatterjee, E.~Berg, P.~Jarillo-Herrero, E.~Khalaf, Shang Liu, R.~Samajdar, T.~Senthil, A.~Yazdani, O.~Vafek, and A.~Vishwanath.
This research was supported by the National Science Foundation under Grant No.~DMR-2002850.

\appendix

\section{Free-energy expansions for $M$  }\label{FreeEnergyExpansion}
Some of the high-temperature order parameters $M$, given in Eqs.~(\ref{OptionsForOrderParameters}), (\ref{MjSes}), and (\ref{SetOfTROStates}) of the main text, have several components and are vectors, matrices, or third-rank tensors. Assuming that these phases are reached by a single, second order phase transition, each of them can only assume certain discrete configurations. We here derive these configurations by writing down the most general free-energy expansions and minimizing them.

As this is a standard procedure and the analysis is very similar for the different cases, we illustrate it with a few instructive examples and collect the final results in \tableref{Stable Phases}.

We begin with $M= \sum_{i=x,y} v_i \mu_i$, $v_i\in\mathbb{R}$, which is just a vector-valued order parameter. For convenience, we rewrite our real vector $v$ as a complex scalar:
\begin{align*}
    M&=\frac{1}{2}(\mu_x+i\mu_y)(v_1-iv_2)+\frac{1}{2}(\mu_x-i\mu_y)(v_1+iv_2)\\ &\equiv \mu^+v^*+\mu^-v
\end{align*}
so that under our discrete translation symmetry, the phase $\phi$ of $v=e^{i\phi}|v|$ transforms as $\phi\rightarrow\phi+\frac{2\pi}{3}$. To quartic order, the most general form the free energy may take that obeys translation symmetry is
\begin{equation}
    \mathcal{F}\propto a|v|^2+b_1\text{Re}[v^3]+b_2\text{Im}[v^3]+c|v|^4
\end{equation}
In minimizing this free energy, we note that only the third order terms proportional to $b_1$ and $b_2$ here will fix the phase $\phi$. We constrain these two coefficients by considering how $v$ transforms under our remaining point group symmetries:
\begin{equation}
    C_2:v\rightarrow v^*
\end{equation}
\begin{equation}
    c_3:v\rightarrow v
\end{equation}
\begin{equation}
    C_{2x}:v\rightarrow v^*
\end{equation}

Allowing us to set $b_2=0$. Then the phase of $v$ is fixed by maximizing (minimizing) $\cos3\phi$ which has maxima (minima) at $\phi=0,\frac{\pi}{3}+\frac{2\pi n}{3}$. Then we have two distinct solutions for this option which cannot be related by translation:
\begin{equation}
    m=\mu_x \qquad m=\frac{1}{2}\mu_x+\frac{\sqrt{3}}{2}\mu_y 
\end{equation}

As second example, we study $M=\rho_y(\tau_x,\tau_y)\sigma_i$, and write $M = \sum_{i=x,y}\sum_{j=1}^3v_{ij} \rho_y \tau_i \sigma_j$, \textit{i.e.}, parametrize it by the matrix-valued real order parameter $v$. Deriving the action of all symmetries in \tableref{RepresentationsOfSymmetries}, it is easy to show that the most general expression for the free energy $\mathcal{F}$ up to quartic order reads as
\begin{equation}
    \mathcal{F} = a\,\text{tr}\left[ v^T v\right] + b_1 \left( \text{tr}\left[ v^T v\right] \right)^2 + b_2 \text{tr}\left[ v^T v v^T v\right],
\end{equation}
with unknown real-valued coefficients $a$, $b_1$, and $b_2$. Minimizing (conveniently done via singular-value decomposition) and taking the symmetry-inequivalent minima yields the options in the first row of \tableref{Stable Phases}. The analysis $M=\tau_z(\mu_x,\mu_y)\vec{\sigma}$ is similar. However, note that one has to go to sixth order to find all symmetry in-equivalent states in order to determine which ground states are not equivalent by translation.

Finally, we briefly summarize how we obtained the phases for the more complicated, tensor-valued case $\rho_y(\tau_x,\tau_y)(\mu_x,\mu_y)\vec{\sigma}$. We again parametrize as
\begin{equation}
    M=v_{lmn}\rho_y\mu_l\tau_m \sigma_n, \qquad v_{lmn} \in \mathbb{R},
\end{equation}
with summation over repeated indices assumed and rewrite
\begin{equation}
\begin{split}
        &M=\rho_y(v_{xmn}\mu_x\tau_m\sigma_n+v_{ymn}\mu_y\tau_m\sigma_n)\\&=\left[\frac{1}{2}(v_{xmn}+iv_{ymn})(\mu_x-i\mu_y)\right.\\&\left.+\frac{1}{2}(v_{xmn}-iv_{ymn})(\mu_x+i\mu_y)\right]\tau_m\sigma_n\rho_y\\&\equiv \left[v_{mn}^*\mu^+ + v_{mn}\mu^-\right]\tau_m\sigma_n\rho_y
\end{split}
\end{equation}
and write the most general rotationally invariant free energy we can construct from a complex rank 2 tensor that obeys the symmetries of our system.
\begin{equation}
\begin{split}
    &\mathcal{F}=a\text{tr}\left[v^\dagger v \right]+b_1(\text{tr}\left[v^\dagger v \right])^2+b_2\text{tr}\left[v^\dagger vv^\dagger v\right]\\&+b_3\text{tr}\left[v^\dagger v^*v^T v\right]+b_4\text{tr}\left[v^\dagger vv^T v^*\right]+b_5|\text{tr}\left[v^\dagger v^*\right]|^2
\end{split}
\end{equation}
or equivalently with re-definition of couplings $b_i$:
\begin{equation}
\begin{split}
    \mathcal{F}=a\text{tr}\left[v^\dagger v \right]+b_1(\text{tr}\left[v^\dagger v \right])^2+b'_2(v^*_{mn}v_{mp}\epsilon_{npq})^2\\+b'_3(v^*_{nm}v_{pm}\epsilon_{npq})^2+b'_4(\epsilon_{mkl}\epsilon_{npq}v^*_{mn}v_{kp})^2\\+b'_5(\epsilon_{abl}\epsilon_{rsq}\epsilon_{mkl}\epsilon_{npq}v^*_{mn}v^*_{kp}v_{ar}v_{bs})
\end{split}
\end{equation}
We note under the point group symmetries:
\begin{equation}
   \text{$C_2:$ } v_{mn}\rightarrow -\sigma^z_{mm'}v_{m'n}^*
\end{equation}
\begin{equation}
   \text{$C_3:$ } v_{mn}\rightarrow v_{mn}
\end{equation}
\begin{equation}
   \text{$C_{2x}:$ } v\rightarrow -v^*,
\end{equation}
all of which leave the free energy invariant. We can consider what $v_{mn}$ will minimize the energy depending on the values of the coefficients and obtain the following ground states:
\begin{equation}
\begin{split}
&v_1=\begin{pmatrix} 1&i & 0\\ 0& 0 &0  \end{pmatrix}e^{i\phi} \qquad v_2=\begin{pmatrix} 1&0 & 0\\ i& 0 &0  \end{pmatrix}e^{i\phi} \qquad \\&v_3=\begin{pmatrix} 1&i & 0\\ i& -1 &0  \end{pmatrix}e^{i\phi} \qquad v_4=\begin{pmatrix} 1&0 & 0\\ 0& 0 &0  \end{pmatrix}e^{i\phi}
\\&v_5=\begin{pmatrix} 1&0 & 0\\ 0& 1 &0  \end{pmatrix}e^{i\phi} \qquad v_6=\begin{pmatrix} 1&0 & 0\\ 0& i &0  \end{pmatrix}e^{i\phi}
\\&v_7=\begin{pmatrix} a&0 & i\\ 0& b &0  \end{pmatrix}e^{i\phi} \qquad v_8=\begin{pmatrix} a&ib & 1\\ ia& -b &-i  \end{pmatrix}e^{i\phi}
\end{split}
\end{equation}
The final two ground states, depend on the couplings $b_i$ and and have $a,b\neq 0$ (however we note for most possible couplings $b_i$, the ground states are one of the first six options). We verify the above states are ground states by scanning the space of couplings and verifying that minimizing the free energy over many points in the phase space does not yield any new minima. Therefore, while the above states are true ground states for some region of the parameter space of the $b_i$, it is possible that there are additional ground states in some small region of parameter space that is missed by this minimization procedure.  The phase $\phi$ is fixed by adding a cubic or sixth (or higher) order term to the free energy. The most general third order term we can add to fix $\phi$ is:
\begin{equation}
    \Delta\mathcal{F}_3=\epsilon_{abc}\epsilon_{ijk}v_{ai}v_{bj}v_{ck}
\end{equation}
However, we note this term is 0 always since our tensor is a 2$\times$3 matrix. We then consider the most general sixth order term we can add that will depend on the phase $\phi$,
\begin{equation}
    \Delta\mathcal{F}_6=b_5\text{tr}[v^Tv]^3+b_6\text{tr}[v^Tvv^Tv]\text{tr}[v^Tv]+b_7\text{tr}[v^Tvv^Tvv^Tv]
\end{equation}
We find the states $v_1$, $v_2$, and $v_3$ can be made independent of $\phi$ via a valley or spin rotation. For $v_4$, $v_5$, and $v_7$ the 6th order contribution requires minimizing (or maximizing) $\cos(6\phi)$, $v_6$ and $v_8$ require a 12th order term which minimizes (maximizes) $\cos(12\phi)$, yielding the 17 orders listed for this option in \tableref{Stable Phases}. We note that the options which are relevant for our analysis at $\nu=2$ are those which have an eigenspectrum $\{+1,+1,-1,-1\}$ as shown in \figref{GappingOutBands}. Of the options in \tableref{Stable Phases}, the only ones with this spectrum are  $\rho_y\sigma_x\tau_x(\sqrt{3}\mu_x+\mu_y)$ and $\rho_y(\sigma_x\tau_x\mu_x+\sigma_y\tau_y\mu_y)$. Therefore in the rest of the text and in particular in \tableref{SymmBrokenOrdersSinglet}, $M=\rho_y\tau_x\mu_x\sigma_x$ and corresponding orders $m_j$ will also have additional distinct options for $m_j$  which may be obtained by applying the rotations $e^{-i\mu_z\frac{\pi}{12}}$ and $e^{-i\tau_z\mu_z\sigma_z\frac{\pi}{4}}$ to the $m_j$ for this $M$ only.

\begin{table*}[tb]
\begin{center}
\caption{Ground states of free energy expansion for high-temperature orders. We omit states which are related to another state in the table by a spin or U(1)$_v$ rotation or by $T_{\vec{a}_r}$.  }
\label{Stable Phases}

\hspace*{-2 em}\begin{tabular} {c|c}
\hline\hline
$M$ & Possible Ground States   \\ \hline
$\rho_y(\tau_x,\tau_y)\vec{\sigma}$ & $\rho_y\tau_x\sigma_z$\\
 & $\rho_y(\tau_x\sigma_x+\tau_y\sigma_y)$\\
\hline
$\tau_z(\mu_x,\mu_y)\vec{\sigma}$ & $\tau_z\mu_x\sigma_z$\\

 & $\tau_z(\mu_x\sigma_x+\mu_y\sigma_y)$\\
 & $\tau_z\sigma_x(\sqrt{3}\mu_x+\mu_y)$\\
\hline

$\rho_y(\tau_x,\tau_y)(\mu_x,\mu_y)\vec{\sigma}$ & $\rho_y\tau_x(\mu_x\sigma_x+\mu_y\sigma_y)$  \\
 & $\rho_y\sigma_x(\mu_x\tau_x+\mu_y\tau_y)$  \\
 & $\rho_y(\tau_x\mu_x\sigma_x+\tau_x\mu_y\sigma_y-\tau_y\mu_x\sigma_y+\tau_y\mu_y\sigma_x)$  \\
& $\rho_y\tau_x\sigma_x\mu_x$  \\
& $\rho_y\tau_x\sigma_x(\sqrt{3}\mu_x+\mu_y)$\\
& $\rho_y\mu_x(\tau_x\sigma_x+\tau_y\sigma_y)$\\

& $\rho_y(\tau_x\sigma_x+\tau_y\sigma_y)(\sqrt{3}\mu_x+\mu_y)$  \\

& $\rho_y(\mu_x\tau_x\sigma_x+\mu_y\tau_y\sigma_y)$\\\
& $\rho_y(\mu_x(\sqrt{3}\tau_x\sigma_x-\tau_y\sigma_y)+\mu_y(\tau_x\sigma_x+\sqrt{3}\tau_y\sigma_y))$\\
& $\rho_y\mu_x((1+\sqrt{3})\sigma_x\tau_x-(\sqrt{3}-1)\tau_y\sigma_y)+\rho_y\mu_y((\sqrt{3}-1)\sigma_x\tau_x+(\sqrt{3}+1)\tau_y\sigma_y)$\\
& $\rho_y\mu_x((1+\sqrt{3})\sigma_x\tau_x+(\sqrt{3}-1)\tau_y\sigma_y)+\rho_y\mu_y(-(\sqrt{3}-1)\sigma_x\tau_x+(\sqrt{3}+1)\tau_y\sigma_y)$\\
& $\rho_y(\mu_x(a\tau_x\sigma_x+b\tau_y\sigma_y)+\mu_y(\tau_x\sigma_z))$\\
& $\rho_y(\mu_x(\sqrt{3}(a\tau_x\sigma_x+b\tau_y\sigma_y)-\tau_x\sigma_z)+\mu_y(((a\tau_x\sigma_x+b\tau_y\sigma_y)+\sqrt{3}\tau_x\sigma_z))$\\
& $\rho_y(\mu_x(a\tau_x\sigma_x-b\tau_y\sigma_y+\tau_x\sigma_z)+\mu_y(a\tau_y\sigma_x+b\tau_x\sigma_y-\tau_y\sigma_z))$\\
& $\rho_y((a\tau_x\sigma_x-b\tau_y\sigma_y+\tau_x\sigma_z)(\sqrt{3}\mu_x+\mu_y)+(a\tau_y\sigma_x+b\tau_x\sigma_y-\tau_y\sigma_z)(-\mu_x+\sqrt{3}\mu_y))$\\

& $\rho_y((a\tau_x\sigma_x-b\tau_y\sigma_y+\tau_x\sigma_z)(\mu_x(\sqrt{3}+1)+\mu_y(\sqrt{3}-1))+(a\tau_y\sigma_x+b\tau_x\sigma_y-\tau_y\sigma_z)(-\mu_x(\sqrt{3}-1)+\mu_y(\sqrt{3}+1)))$\\
& $\rho_y((a\tau_x\sigma_x-b\tau_y\sigma_y+\tau_x\sigma_z)(\mu_x(\sqrt{3}+1)-\mu_y(\sqrt{3}-1))+(a\tau_y\sigma_x+b\tau_x\sigma_y-\tau_y\sigma_z)(\mu_x(\sqrt{3}-1)+\mu_y(\sqrt{3}+1)))$\\

 \hline
$\tau_z(\mu_x,\mu_y)$ & $\tau_z\mu_x$\\
 \hline
$\rho_y(\mu_x,\mu_y)(\tau_x,\tau_y)$ &  $\rho_y\tau_x\mu_x$\\
& $\rho_y(\mu_x\tau_x+\mu_y\tau_y)$\\
 & $\rho_y\tau_x(\sqrt{3}\mu_x+\mu_y)$\\
\hline
$(\mu_x,\mu_y)\vec{\sigma}$ & $\sigma_x\mu_x$\\
& $\mu_x\sigma_x+\mu_y\sigma_y$\\ 
 & $\sigma_x(\sqrt{3}\mu_x+\mu_y)$\\
\hline
$\rho_y\mu_z(\tau_x,\tau_y)\vec{\sigma}$ & $\rho_y\mu_z\tau_x\sigma_z$\\
&$\rho_y\mu_z(\tau_x\sigma_z+\tau_y\sigma_y)$\\
\hline
\hline
 \end{tabular}
\end{center}
\end{table*}

\section{Alternative condition for WZW term}\label{DerivationWZWTerm}

We here provide a derivation of the set of conditions in \equref{CriterionAlt} of the main text for a WZW term,
\begin{enumerate}[label=c.\arabic*]
    \item For $\gamma_i$ we have $\gamma_i \Delta T = -\Delta T \gamma_i^T \neq 0$, $i=1,2$,\label{cond1} 
    \item For $m_j$ we have $m_j \Delta T = \Delta T m_j^T \neq 0$, $j=1,2,3$,\label{cond2}
    \item $\text{tr}[\gamma_{i_1}\gamma_{i_2}m_{j_1}m_{j_2}m_{j_3}]\propto \epsilon_{j_1,j_2,i_1,i_2,i_3}$,\label{cond3}
\end{enumerate}
by showing that they are equivalent to \equref{model3_coef}.
To recall our notation, $\gamma_i$ denote the Dirac matrices,  $\Delta i\sigma_y \tau_x\mu_x=\Delta T$ is our pairing order parameter, $\Delta$, multiplied by the unitary part of the time-reversal operator $\textit{T}=i\sigma_y\tau_x\mu_x$, and $m_j$, $j=1,2,3$, are the partner order candidates. 

We can construct all possible partners in the non-redundant Nambu bases specified by \equref{FirstNambuSpinor} and \equref{AlternativeNambuSpinors} by first considering all possible orders in a redundant Nambu space given by the spinor:
\begin{equation}
    \Psi_{\vec{q}}=\begin{pmatrix}f_{\vec{q}}\\ f^\dagger_{-\vec{q}}\end{pmatrix}
\end{equation}
provided that all partner orders we can construct in a redundant Nambu basis will survive projection to some non-redundant Nambu basis via projection $P_C=\frac{1}{2}(1+\eta_zC)$ where $\eta_z$ acts here on the redundant Nambu space, $C$ is real and symmetric, and $\eta_zC$ commutes with all the terms in our Hamiltonian in the redundant Nambu basis. For example, for singlet pairing $A_1$ and the first option in \tableref{SummaryOrderParamters} $(\tau_+,\tau_-)\rho_x;\rho_z$, we may choose $C=\mu_z$ or $C=\sigma_z$ for the definition of our projection operator. These two choices of $C$ correspond to the spin-Nambu and mini-valley-Nambu spinor, given by \equref{FirstNambuSpinor} and by the first choice in \equref{AlternativeNambuSpinors}, respectively. The second Nambu spinor in \equref{AlternativeNambuSpinors}, corresponding to $C=\tau_z$, would not work in this case, as the IVC state under consideration does not commute with $\tau_z$. 

In the redundant basis, our kinetic terms take the form:
\begin{equation}
    \mathcal{H}_{\text{kin}}=\sum_{\vec{q}}^\Lambda \Psi_{\vec{q}}^\dagger[ \rho_x\tau_zq_x+\rho_y\eta_zq_y ]\Psi_{\vec{q}} 
\end{equation}
Our pairing term is given by:
\begin{equation}
\begin{split}
        \mathcal{H}_{\text{pair}}=&\sum_{\vec{q}}^\Lambda \Psi_{\vec{q}}^\dagger[ \text{Re}[\delta](-\sigma_y\tau_x\mu_x\Delta\eta_y)\\&+\text{Im}[\delta](-\sigma_y\tau_x\mu_x\Delta\eta_x) ]\Psi_{\vec{q}} 
\end{split}
\end{equation}
where $\delta=e^{i \phi}$ captures the phase of the superconducting order parameter. 

We assume conditions (\ref{cond1}), (\ref{cond2}), (\ref{cond3}) and show that they imply \equref{model3_coef}. The transformation which takes partners $m_j$, kinetic terms $\gamma_i$ and $\Delta T$ to a (non-redundant) Nambu basis is as follows:
\begin{equation}
    m_{j}\rightarrow \mathcal{M}_{j}=\left[ \frac{1}{2}(m_j+m_j^T)\eta_z+\frac{1}{2}(m_j-m_j^T) \right]P_C
\end{equation}
\begin{equation}
    \gamma_i\rightarrow \Gamma_i=\left[ \frac{1}{2}(\gamma_i+\gamma_i^T)+\frac{1}{2}(\gamma_i-\gamma_i^T)\eta_z \right]P_C
\end{equation}
\begin{equation}
    \Delta T\rightarrow \mathcal{M}_{4,5}=i\Delta T\eta_{x/y}P_C
\end{equation}
where $\eta_z$ acts on redundant Nambu space and $\eta_zC$ in the projector $P_C=\frac{1}{2}(1+\eta_zC)$ is chosen to commute with all $\gamma_i$ and $m_j$ and to anti-commute with $\Delta T$. Furthermore, $C$ must be real and symmetric as already stated above. Such a choice $C$ always exists for each of our candidate pairings for some $C=\sigma_{0/x}\mu_{0/z}\tau_{0/z}$ (assuming for our triplet states that they are polarized along the $\sigma_x$-direction). This is guaranteed by (\ref{cond3}) which requires $m_1m_2m_3\propto\rho_z\tau_z$, indicating that an even number of $m_j$ must anti-commute with $\sigma_x$, $\mu_z$, and $\tau_z$ respectively and that we then may always find some product of $\mu_z\tau_z\sigma_x$ that commutes with the $m_j$.

Given the anti-commutation relations of the $m_j$ and $\gamma_i$ in \eqref{cond3}, we see the $\mathcal{M}_j$ $j=1,2,3$ and $\Gamma_i$ will anti-commute and that (\ref{cond1}) and (\ref{cond2}) similarly imply that $\mathcal{M}_{4}$ and $\mathcal{M}_{5}$ will anti-commute with $\mathcal{M}_{1,2,3}$ and $\Gamma_{1,2}$. We note that (\ref{cond1}), (\ref{cond2}), and (\ref{cond3}) imply that any partner $m_j$ will commute with all the $\gamma_i$ and superconducting term in an extended Nambu basis, which we may see by applying the above transformations to the $\gamma_i$, $m_j$ and pairing term, without the projection operator $P_c$. Finally we show that the trace in \equref{model3_coef} is nonzero. Note that in the non-redundant Nambu basis our two components of the superconducting order satisfy $\mathcal{M}_4\mathcal{M}_5\propto \frac{1}{2}(1+\eta_zC)\eta_z$ and we have:
\begin{equation}
\begin{split}
        &\gamma_1\gamma_2m_1m_2m_3\text{Re}[\Delta T]\text{Im}[\Delta T]\rightarrow\Gamma_1\Gamma_2\mathcal{M}_1\mathcal{M}_2\mathcal{M}_3\mathcal{M}_4\mathcal{M}_5\\&=\frac{1}{2}(1+\eta_zC)\left[ \frac{1}{2}(\gamma_1\gamma_2m_1m_2m_3+\gamma_1^T\gamma_2^Tm_1^Tm_2^Tm_3^T)\eta_z\right.\\&\left.+\frac{1}{2}(\gamma_1\gamma_2m_1m_2m_3-\gamma_1^T\gamma_2^Tm_1^Tm_2^Tm_3^T)\right]\eta_z 
\end{split}
\end{equation}

\begin{equation}
\begin{split}
        \text{tr}[ \Gamma_1\Gamma_2\mathcal{M}_1\mathcal{M}_2\mathcal{M}_3\mathcal{M}_4\mathcal{M}_5]\propto\text{tr}\left[\frac{1}{2}(1+\eta_zC) \gamma_1\gamma_2m_1m_2m_3\right]
\end{split}
\end{equation}
Where the anti-symmetric part of $\Gamma_1\Gamma_2\mathcal{M}_1\mathcal{M}_2\mathcal{M}_3$ does not contribute to the trace and we have used the condition $\text{tr}[\gamma_1\gamma_2m_1m_2m_3]=\text{tr}[\gamma_1^T\gamma_2^Tm_1^Tm_2^Tm_3^T]$ which holds given the anti-commutation relations of the $\gamma_i$ and $m_j$.  Then we see given $\text{tr}[\gamma_1\gamma_2m_1m_2m_3]\propto I_d$ that:
\begin{equation}
\begin{split}
       &\text{tr}[ \Gamma_1\Gamma_2\mathcal{M}_1\mathcal{M}_2\mathcal{M}_3\mathcal{M}_4\mathcal{M}_5]\propto\text{tr}\left[\frac{1}{2}(1+\eta_zC) \gamma_1\gamma_2m_1m_2m_3\right]\\&=\text{tr}\left[\frac{1}{2}\gamma_1\gamma_2m_1m_2m_3\right]\propto I_d
\end{split}
\end{equation}
We have, thus, shown that (\ref{cond1}), (\ref{cond2}), and (\ref{cond3}) imply \equref{model3_coef}. 

To verify the converse of this, we assume the condition \equref{model3_coef}
\begin{equation}
\begin{split}
      &\text{tr}\left[\Gamma_{i_1}\Gamma_{i_2}\mathcal{M}_{i_1}\mathcal{M}_{i_2}\mathcal{M}_{i_3}\mathcal{M}_{i_4}\mathcal{M}_{i_5}\right] \propto \epsilon_{j_1,j_2,i_1,i_2,i_3,i_4,i_5} 
\end{split}
\end{equation}
and show that (\ref{cond1}), (\ref{cond2}), and (\ref{cond3}) follow. We note that the anti-commutation  relations of $\mathcal{M}_i$ and $\Gamma_i$ imply that \equref{cond1} and \equref{cond2} are satisfied and also that $m_j$ and $\gamma_i$ must anti-commute. Note that this is where the requirement that $C^T=C$ and $C$ be real become relevant. It can be verified that $\mathcal{M}_1$ and $\mathcal{M}_2$ anti-commuting yields the conditions $(1+C)\{m_1,m_2\}=0$ and $(1-C)\{m_1^T,m_2^T\}=0$, which requires $C$ be symmetric to insure $\{m_1,m_2\}=0$. Similar arguments hold for the $\Gamma_i$ and $\mathcal{M}_4$ and $\mathcal{M}_5$ with the requirement that $C^*=C$. We then establish the trace in \equref{cond3} is nonzero by:
\begin{equation}
\begin{split}
      &\text{tr}\left[\Gamma_1\Gamma_2\mathcal{M}_1\mathcal{M}_2\mathcal{M}_3\mathcal{M}_4\mathcal{M}_5 \right] \propto I_d\\&\implies \text{tr}\left[ \frac{1}{4}(1+\eta_zC)(\gamma_1\gamma_2m_1m_2m_3\right.\\&\qquad\qquad \left.+\gamma_1^T\gamma_2^Tm_1^Tm_2^Tm_3^T)\right]\propto I_d  
\end{split}
\end{equation}
The anti-commutation  relations of $m_j$ and $\gamma_i$ ensure $\text{tr}[\gamma_1\gamma_2m_1m_2m_3+\gamma_1^T\gamma_2^Tm_1^Tm_2^Tm_3^T]=2\text{tr}[\gamma_1\gamma_2m_1m_2m_3]$. Then we have:
\begin{equation}
     \text{tr}\left[ \frac{1}{2}(1+\eta_zC)\gamma_1\gamma_2m_1m_2m_3\right]\propto\text{tr}\left[ \gamma_1\gamma_2m_1m_2m_3\right]\propto I_d
\end{equation}
Then with the anti-commutation relations of $m_j$ and $\gamma_i$ we have that \equref{model3_coef} implies:
\begin{equation}
     \text{tr}[\gamma_{i_1}\gamma_{i_2}m_{j_1}m_{j_2}m_{j_3}]\propto \epsilon_{i_1,i_2,j_1,j_2,j_3}
\end{equation}
and \eqref{cond1}, \eqref{cond2} and (\ref{cond3}) are true as desired.

\section{Derivation of WZW term for Majorana Fermions}\label{MajoranaWZW}
While WZW terms have been previously derived for complex fermions in several works, see, e.g., \refcite{XuThomson} for a recent study, we here provide a derivation for Majorana fermions. The latter are important for our analysis as they automatically emerge when using redundant Nambu spinors. Our beginning action is:
\begin{equation}
\begin{split}
        &\mathcal{S}=\int dt\int d^2\textbf{q}\, f_\textbf{q}^\dagger \left[\partial_t+ \gamma_i\textbf{q}^i +m\sum_{a=3}^5n_{a-2}m_{a-2}\right] f_\textbf{q} \\&+ m f_\textbf{q}^\dagger \text{Re}[\delta]\Delta T  f_{-\textbf{q}}^*+ im f_\textbf{q}^\dagger \text{Im}[\delta]\Delta T  f_{-\textbf{q}}^*+\text{H.c.},
\end{split}
\end{equation}
where $\delta\in \mathbb{C}$ captures the complex phase of the superconducting order parameter and $n_4=\text{Re}[\delta]$ and $n_5=\text{Im}[\delta]$.
The matrices $m_j$ square to unity, $\vec{n}$ is a 5 component unit vector and $m$ is the magnitude of the orders $m_j$ and superconducting orders which transform under SO(5). 

We then may define a redundant Majorana spinor as:
\begin{equation}
    \eta=\begin{pmatrix}\eta_{1\textbf{q}}\\\eta_{2\textbf{q}}\end{pmatrix}, \quad 
    \eta_{1\textbf{q}}=\frac{1}{\sqrt{2}}(f^*_\textbf{q}+f_{-\textbf{q}}), \quad \eta_{2\textbf{q}}=\frac{i}{\sqrt{2}}(f^*_\textbf{q}-f_{-\textbf{q}}).
\end{equation}
 We introduce the Pauli matrices $\eta_i$ which act on redundant Majorana space. We define $\gamma_0=\rho_z\tau_z$ and the following additional operators:
\begin{equation}
\begin{split}
       &\Tilde{\gamma_0}=\gamma_0\eta_y \qquad \Tilde{\gamma}_i= -\frac{i\gamma_0}{2}((\gamma_i+\gamma_i^T)\eta_y-(\gamma_i-\gamma_i^T)) \\& \hspace{-1em}\Tilde{m}_{1,2,3}= -\frac{\gamma_0}{2}((m_j+m_j^T)-(m_j-m_j^T)\eta_y) \quad \Tilde{m}_{4,5} =-\gamma_0\eta_{x/z}\Delta T
\end{split}
\end{equation}

So that our action may be re-written as:
\begin{equation}
\begin{split}
        &\mathcal{S}=\int d^3q\frac{1}{2} \overline{\eta_q}\left[i\Tilde{\gamma_\mu}q^\mu +m\sum_{a=3}^7n_{a-2}\Tilde{m}_{a-2}\right] \eta_{-q}\\&\equiv \frac{1}{2}\int d^3q\eta_q^TM \eta_{-q} 
\end{split}
\end{equation}
where $M=\Tilde{\gamma_0}\left[ i\Tilde{\gamma_\mu}q^\mu +m\sum_{a=3}^7n_{a-2}\Tilde{m}_{a-2}\right]$ and $\overline{\eta}_q=\eta^T_q\Tilde{\gamma}_0$
and the time coordinate is included in the measure $\int d^3 q$. When integrating out fermion fields, we have for Majorana fields:
\begin{equation}
    \text{Pf}(M)=\int d\eta e^{-\frac{1}{2}\eta^T M \eta}
\end{equation}
It may be verified that $M$ is a skew-symmetric matrix. We may then re-exponentiate to get an effective action with the help of the identity:
\begin{equation}
    \text{Pf}(M)=i^{n^2}e^{\frac{1}{2}\text{tr}[\text{ln}((\eta_y\otimes I_n)^T M)]}
\end{equation}
where $i^{n^2}$ is a prefactor that will not change our result. Then we have:
\begin{equation}
    \mathcal{S}_{\text{eff}}=-\frac{1}{2}\text{tr}[\text{ln}((\eta_y\otimes I_n)^T M)]
\end{equation}
We then expand about a small $\delta \vec{n}\cdot \vec{\Tilde{m}}$ and find:
\begin{equation}
\begin{split}
        \mathcal{S}_{\text{eff}}&=-\frac{1}{2}\text{tr}[\text{ln}((\eta_y\otimes I_n)^T M)]\\&=-\frac{1}{2}\text{tr}[\text{ln}(-\eta_y M)]+\frac{1}{2}\text{tr}[\text{ln}(1+\frac{\eta_y\Tilde{\gamma}_0m\delta \vec{n}\cdot \vec{\Tilde{m}}}{\eta_y M})]\\&\approx\frac{1}{2}\text{Tr}\left[ \frac{m\delta \vec{n}\cdot \vec{\Tilde{m}} M^\dagger\Tilde{\gamma}_0}{M^\dagger M} \right]
\end{split}
\end{equation}
where we ignore the first term in the expansion as it will not contribute to the topological term. Expanding $M^\dagger M$, the term which is of the correct order in $ \vec{n}\cdot \vec{\Tilde{m}}$ and momenta to contribute to the WZW term we are interested in is:
\begin{equation}
\begin{split}
        &\delta\mathcal{S}_{\text{eff}}= \frac{1}{2}\text{tr}\left[m^2\delta (-\partial^2+m^2)^{-1} \right.\\&\left.\times((-\partial^2+m^2)^{-1}m\Tilde{\gamma}_\mu\partial_\mu ( \vec{n}\cdot \vec{\Tilde{m}}))^3\vec{n}\cdot \vec{\Tilde{m}}\right]
\end{split}
\end{equation}
Here $\delta\mathcal{S}_{\text{eff}}$ is the variation of the action with respect to $ \delta\vec{n}\cdot \vec{\Tilde{m}}$. Inserting complete bases of momenta eigenstates and neglecting momenta indices of $\vec{n}\cdot \vec{\Tilde{m}}$, in Fourier space we find:
\begin{equation}
\begin{split}
    &\delta\mathcal{S}_{\text{eff}}= -i\int \frac{d^3 q}{(2\pi)^2}
    \int \frac{d^3 k_1}{(2\pi)^2} \int \frac{d^3k_2}{(2\pi)^2}\int \frac{d^3k_3}{(2\pi)^2}\\&\times(k_1)_\mu (k_2)_\nu (k_3)_\rho\frac{1}{q^2+m^2}
    \frac{1}{(q+k_1)^2+m^2}\\&\times\frac{1}{(q+k_1+k_2)^2+m^2}\frac{1}{(q+k_1+k_2+k_3)^2+m^2}
    \\&\times\frac{m^5}{2}\text{tr}\left[\Tilde{\gamma}_\mu\Tilde{\gamma}_\nu\Tilde{\gamma}_\rho\delta (\vec{n}\cdot \vec{\Tilde{m}}) 
    ( \vec{n}\cdot \vec{\Tilde{m}})^3\vec{n}\cdot \vec{\Tilde{m}}\right]
\end{split}
\end{equation}
We then take in the denominators $k_1,k_2,k_3\rightarrow0$ so that we may integrate out the variable $q$:
\begin{equation}
    \begin{split}
    \delta\mathcal{S}_{\text{eff}}&\approx -i\int \frac{d^3 q}{(2\pi)^2}
    \int \frac{d^3 k_1}{(2\pi)^2} \int \frac{d^3k_2}{(2\pi)^2}\int \frac{d^3k_3}{(2\pi)^2} (k_1)_\mu (k_2)_\nu (k_3)_\rho
    \\&\times\frac{1}{(q^2+m^2)^4}\frac{m^5}{2}\text{tr}\left[\Tilde{\gamma}_\mu\Tilde{\gamma}_\nu\Tilde{\gamma}_\rho\delta (\vec{n}\cdot \vec{\Tilde{m}}) (
    ( \vec{n}\cdot \vec{\Tilde{m}}))^3\vec{n}\cdot \vec{\Tilde{m}}\right]
    \\&=\frac{-i}{64\pi}\int \frac{d^3 k_1}{(2\pi)^2} \int \frac{d^3k_2}{(2\pi)^2}\int \frac{d^3k_3}{(2\pi)^2} (k_1)_\mu (k_2)_\nu (k_3)_\rho
    \\&\times\frac{1}{2}\text{tr}\left[\Tilde{\gamma}_\mu\Tilde{\gamma}_\nu\Tilde{\gamma}_\rho\delta (\vec{n}\cdot \vec{\Tilde{m}}) (
    \vec{n}\cdot \vec{\Tilde{m}})^3\vec{n}\cdot \vec{\Tilde{m}}\right]
\end{split}
\end{equation}
Then Fourier transforming back we find:
\begin{equation}
\begin{split}
       \delta \mathcal{S}_{\text{eff}}&=\frac{1}{128\pi}\int d^3x\text{tr}\left[\Tilde{\gamma}_\mu\Tilde{\gamma}_\nu\Tilde{\gamma}_\rho\delta (\vec{n}\cdot \vec{\Tilde{m}}) 
    ( \partial_\mu \vec{n}\cdot \vec{\Tilde{m}})\right.\\&\left.\times ( \partial_\nu \vec{n}\cdot \vec{\Tilde{m}}) ( \partial_\rho \vec{n}\cdot \vec{\Tilde{m}})\vec{n}\cdot \vec{\Tilde{m}}\right]
    \\&=\frac{1}{128\pi}\int d^3x\int_0^1 du\text{tr}\left[\Tilde{\gamma}_\mu\Tilde{\gamma}_\nu\Tilde{\gamma}_\rho\delta (\vec{n}\cdot \vec{\Tilde{m}}) 
    ( \partial_\mu \vec{n}\cdot \vec{\Tilde{m}})\right.\\&\left.\times ( \partial_\nu \vec{n}\cdot \vec{\Tilde{m}}) ( \partial_\rho \vec{n}\cdot \vec{\Tilde{m}})\partial_u(\vec{n}\cdot \vec{\Tilde{m}})\right]
\end{split}
\end{equation}
where in the final step, we introduce auxiliary coordinate $u$, and extend our unit vector $\Vec{n}$ to depend on the additional coordinate $u$ such that $\vec{n}(\tau,x,y,1)=\vec{n}(\tau,x,y)$ and $\vec{n}(\tau,x,y,0)=(1,0,0,0,0)$. Finally, we divide by $4$ so that the coordinate $u$ will also be anti-symmetrized and take the trace over our $\Tilde{m}_j$ we obtain:
\begin{equation}
\begin{split}
       \hspace{-2em} \delta \mathcal{S}_{\text{eff}}&=\frac{16\mathcal{N}i\epsilon_{\mu\nu\rho \sigma }}{512\pi}\int d^3x\int_0^1 du\epsilon_{\alpha\beta\gamma\delta\epsilon}\text{tr}\left[n_\alpha 
     \partial_\mu n_\beta  \partial_\nu n_\gamma \partial_\rho n_\delta \partial_\sigma n_\epsilon\right]
     \\&=\frac{4!\mathcal{N}i}{32\pi}\int d^3x\int_0^1 du\epsilon_{\alpha\beta\gamma\delta\epsilon}\text{tr}\left[n_\alpha 
     \partial_\tau n_\beta  \partial_x n_\gamma \partial_\rho n_\delta \partial_y n_\epsilon\right]
     \\&=2\pi\frac{3\mathcal{N}i}{8\pi^2}\int d^3x\int_0^1 du\epsilon_{\alpha\beta\gamma\delta\epsilon}\text{tr}\left[n_\alpha 
     \partial_u n_\beta  \partial_\tau n_\gamma \partial_x n_\delta \partial_y n_\epsilon\right],
\end{split}
\end{equation}
which is the topological term we wished to derive at level $\mathcal{N}$ where $\mathcal{N}$ is 1 for $\nu=2$ and 2 for $\nu=0$.

\section{Connection to Khalaf {\it et al.\/}}\label{ConnectionToAshvin}

The work of Khalaf {\it et al.\/} \cite{2020arXiv200400638K} appeared while our work was in progress. They discuss a specific scenario for the WZW term in a spinless model of twisted bilayer graphene. Using our labelling of Pauli matrices---$\tau$ for valley, $\mu$ for mini-valley, and $\rho$ for sublattice---they considered
\begin{equation}
\widetilde{m}_1 = \tau_x \rho_y, \quad \widetilde{m}_2 = \tau_y \rho_y, \quad \widetilde{m}_3 = \rho_z. \label{AshvinsOP}
\end{equation}
However, before being able to connect to our sets of possible WZW terms, we have to make sure that we use the same conventions (as indicated by the tildes in the equation above). Comparing our Dirac matrices defining the non-interacting Hamiltonian in \equref{FinalLowEnergyTheory2}, $\gamma_x = \rho_x \tau_z$, $\gamma_y = \rho_y$, with theirs, $\widetilde{\gamma}_x = \mu_z\rho_x$, $\widetilde{\gamma}_y = \mu_z\rho_y\tau_z$, we find that the field operators are related by $\widetilde{f}_{\vec{q}} = U_2U_1f_{\vec{q}}$, where
\begin{equation}
U_1 = \frac{\mu_0+\mu_z}{2} + \frac{\mu_0-\mu_z}{2}\rho_z, \,\,\, U_2 = \frac{\tau_0+\tau_z}{2} + \frac{\tau_0-\tau_z}{2}\rho_z.
\end{equation}
With this, we can rewrite \equref{AshvinsOP} in our conventions, yielding
\begin{equation}
m_1 = -\mu_z \tau_y \rho_x, \quad m_2 = \mu_z \tau_x \rho_x, \quad m_3 = \rho_z.
\end{equation}
The first two components transform into each other under U(1)$_v$ and are the IVC part and the third order parameter is the valley Hall state of \refcite{2019arXiv191102045B}. Note that $m_1$, $m_2$ break time-reversal symmetry, while $m_3$ preserves it but breaks $C_2$, in agreement with the statements in \refcite{2019arXiv191102045B}.

Since \refcite{2020arXiv200400638K} considers a model without spin, there is no unique mapping to our spinful description of the system. In fact, there are two natural microscopic realizations: first, reinserting spin as $(m_1,m_2) = \mu_z\rho_x\vec{\sigma}(\tau_+,\tau_-)$, $m_3=\rho_z$ we recover the partner orders in the 11th line of \tableref{SummaryTripletOrderParamters} and the second line in \tableref{SymmBrokenOrdersSinglet} (in both tables labelled as type IVC$_+$; SP). Our analysis, thus, shows that it can form WZW terms with both singlet and triplet pairing (although in the unitary triplet case, additional Fermi surfaces will appear), and which $M$ have to be realized in the respective cases. 

Second, we see from the fourth line of \tableref{SummaryOrderParamters} (labelled as type IVC$_-$; SP) that also the spinless realization, $(m_1,m_2) = \mu_z\rho_x(\tau_+,\tau_-)$, $m_3=\rho_z$, can form a WZW term with the $A_2$ singlet for $M=\tau_z\mu_y$, \textit{i.e.}, only if translation invariance is broken in the high-temperature phase (note $M=\mu_z\sigma_z$ will lead to unwanted Fermi surfaces).


\begin{widetext}

\section{Full set of partner order parameters}\label{FullSetOfOrderParameters}
We list the full set of possible orders for singlet pairings $A_1$ and $A_2$ in \tableref{TableOfAllSinglets} and triplet pairings $B_1$ and $B_2$ in \tableref{tripletallorders} and \tableref{unitaryallorders} in the full space with no additional symmetry breaking. These include options we eliminated in \tableref{SummaryOrderParamters} and \tableref{SummaryTripletOrderParamters} for breaking translational, valley rotation, or spin rotation symmetry but are still mathematically viable options in that they satisfy \equref{model3_coef} in the full space. We also list which of the above orders survive projection to a subspace defined by a high-temperature order $M$ with projection $\frac{1}{2}(1+M)$ for singlet pairing in \tableref{surviveMsinglet} and \tableref{surviveMsingletTR} and for triplet pairing in \tableref{surviveMtriplet} and \tableref{surviveMtripletTR}. 

\begin{table*}[h]
\begin{center}\caption{Forms of possible orders for singlet pairing. Note that the options which do not appear by the main text either require additional symmetry breaking beyond the options for $\nu=2$ or have three independently fluctuating options.}\label{TableOfAllSinglets}
\begin{tabular}{|c c c c c|} 
 \hline
 Partner Orders & Partner 1 & Partner 2 & Partner 3 & SC Partner\\ [0.5ex] 
 \hline\hline
 1 & $\mu_{x/y}\rho_z$ & $\mu_{y/x}\tau_{z}\rho_z\sigma_j$ & $\mu_z\rho_z\sigma_j$  & $A_1$   \\
 2 & $\rho_x\tau_{x/y}$ & $\tau_{y/x}\mu_z\rho_x\sigma_j$ & $\mu_z\rho_z\sigma_j$& $A_1$   \\
 3 & $\rho_x\tau_{x/y}$ & $\rho_x\tau_{y/x}\mu_{x/y}$ &  $\rho_z\mu_{x/y}$& $A_1$ \\
 4 & $\tau_{x/y}\mu_{x,y}\rho_x$ & $\tau_z\mu_{y,x}\rho_z\sigma_j$ & $\tau_{x/y}\mu_z\rho_x\sigma_j$  & $A_1$ \\

 \hline
 5 & $\tau_{x/y}\mu_{x/y}\rho_x$ & $\tau_{y/x}\rho_x\sigma_j$ & $\mu_{x/y}\rho_z\sigma_j$  & $A_2$  \\
 
  6 & $\tau_{x/y}\mu_z\rho_x$ & $\tau_{y/x}\rho_x\sigma_j$ & $\mu_z\rho_z\sigma_j$  & $A_2$  \\
  \hline\hline
 7 &\multicolumn{2}{c}{ $\rho_x(\tau_x,\tau_y)$} & $\rho_z$ &  $A_1$ \\
8 & \multicolumn{2}{c}{$\rho_x\mu_z\sigma_j(\tau_x,\tau_y)$} & $\rho_z$ & $A_1$\\
9 & \multicolumn{2}{c}{$\rho_x\mu_{x/y}(\tau_x,\tau_y)$} & $\rho_z$ &  $A_1$\\
10 & \multicolumn{2}{c}{$\rho_z(\mu_x,\mu_y)$} & $\rho_z\tau_z\mu_z$ &$A_1$ \\
11 & \multicolumn{2}{c}{$\tau_z\rho_z\sigma_j(\mu_x,\mu_y)$ }& $\rho_z\tau_z\mu_z$ & $A_1$\\
12 &\multicolumn{2}{c}{ $\rho_x\tau_{x,y}(\mu_x,\mu_y)$} & $\rho_z\tau_z\mu_z$ & $A_1$\\
13 &\multicolumn{2}{c}{ $\tau_z\rho_z\mu_{x,y}(\sigma_i,\sigma_j)$} & $\rho_z\tau_z\sigma_k$ &  $A_1$\\
14 &\multicolumn{2}{c}{ $\rho_z\mu_z(\sigma_i,\sigma_j)$ }& $\rho_z\tau_z\sigma_k$ &  $A_1$\\
15 &\multicolumn{2}{c}{ $\rho_x\mu_z\tau_{x,y}(\sigma_i,\sigma_j)$} & $\rho_z\tau_z\sigma_k$ &  $A_1$\\

\hline
16 & \multicolumn{2}{c}{$\rho_x\sigma_j(\tau_x,\tau_y)$} & $\rho_z$ &  $A_2$ \\
17 &\multicolumn{2}{c}{ $\rho_x\mu_a(\tau_x,\tau_y)$} & $\rho_z$ &  $A_2$\\
18 & \multicolumn{2}{c}{$\rho_z\sigma_j(\mu_a,\mu_b)$} & $\rho_z\tau_z\mu_c$ & $A_2$\\
19 & \multicolumn{2}{c}{$\rho_x\tau_{x,y}(\mu_i,\mu_j)$} & $\rho_z\tau_z\mu_k$ &  $A_2$\\
20 &\multicolumn{2}{c}{ $\tau_z\rho_z(\mu_x,\mu_y)$}& $\rho_z\tau_z\mu_z$ &  $A_2$  \\
21 &\multicolumn{2}{c}{ $\rho_z\mu_a(\sigma_i,\sigma_j)$ }& $\rho_z\tau_z\sigma_k$ &  $A_2$\\
22 & \multicolumn{2}{c}{$\rho_x\tau_{x,y}(\sigma_i,\sigma_j)$} & $\rho_z\tau_z\sigma_k$ &  $A_2$\\

\hline
\end{tabular}
\end{center}
\end{table*}

\begin{table*}[h]
\begin{center}
\caption{Forms of possible orders for triplet pairing, choosing a triplet state polarized along $\sigma_x$. Note that the options which do not appear by the main text either require additional symmetry breaking beyond the options for $\nu=2$ or have three independently fluctuating options. }\label{tripletallorders}
\begin{tabular}{|c c c c c |} 
 \hline
 Partner Orders & Partner 1 & Partner 2 & Partner 3 &  SC Partner \\ [0.5ex] 
 \hline\hline
1 & $\tau_{x/y}\mu_z\rho_x$ & $\tau_{x/y}\mu_{x/y}\sigma_x\rho_x$ & $\tau_z\mu_{y/x}\rho_z\sigma_x$ &  $B_1$   \\
2 & $\tau_{x/y}\mu_z\rho_x$ & $\tau_{y/x}\sigma_x\rho_x$ & $\mu_{z}\rho_z\sigma_x$ &   $B_1$   \\
3 & $\tau_{x/y}\mu_z\rho_x$ & $\tau_{y/x}\mu_z\rho_x\sigma_{y/z}$ & $\rho_z\sigma_{y/z}$ &  $B_1$   \\
4 & $\rho_z\mu_{x/y}$ & $\tau_{x/y}\sigma_x\rho_x$ & $\tau_{y/x}\mu_{x/y}\sigma_x\rho_x$ &  $B_1$   \\
5 & $\tau_{x/y}\mu_{x/y}\sigma_x\rho_x$ & $\tau_{y/x}\mu_z\sigma_{y/z}\rho_x$ &$\mu_{y/x}\sigma_{z/y}\rho_z$ & $B_1$   \\
6 & $\tau_{x/y}\sigma_x\rho_x$ & $\tau_{x/y}\rho_x\sigma_{y/z}\mu_z$ & $\tau_z\mu_z\sigma_{z/y}\rho_z$ &   $B_1$   \\
7 & $\rho_z\mu_{x/y}$ & $\rho_z\mu_{y/x}\sigma_{y/z}$ & $\tau_z\rho_z\mu_z\sigma_{y/z}$ &   $B_1$   \\
8 & $\rho_x\mu_{x/y}$ & $\rho_z\mu_{z}\sigma_{x}$ & $\tau_z\mu_{y/x}\sigma_{x}\rho_z$ &   $B_1$   \\
9 & $\rho_z\sigma_{y/z}$ & $\tau_z\mu_z\sigma_{z/y}\rho_z$ & $\mu_z\rho_z\sigma_x$ &   $B_1$   \\
10 & $\tau_z\mu_{x/y}\sigma_x\rho_z$ & $\sigma_{y/z}\rho_z$ & $\mu_{x/y}\sigma_{z/y}\rho_z$ &   $B_1$   \\
\hline
11 & $\tau_{x/y}\rho_x$ & $\tau_{y/x}\sigma_x\rho_x\mu_a$ & $\sigma_x\rho_z\mu_a$ &   $B_2$   \\
12 & $\tau_{x/y}\rho_x$ & $\tau_{y/x}\sigma_{y/z}\rho_x$ & $\sigma_{y/z}\rho_z$ &   $B_2$   \\
13 & $\tau_{x/y}\rho_x\sigma_x\mu_a$ & $\tau_{x/y}\sigma_{y/z}\rho_x$ & $\sigma_{z/y}\rho_z\tau_z\mu_a$ &   $B_2$   \\
14 & $\rho_x\sigma_x\mu_a$ & $\sigma_{y/z}\rho_z$ & $\tau_z\sigma_{z/y}\rho_z\mu_a$ &  $B_2$   \\
\hline
 \hline
 14 &\multicolumn{2}{c}{$\rho_x\mu_z(\tau_x,\tau_y)$} & $\rho_z$ & $B_1$ \\
 15 & \multicolumn{2}{c}{$\rho_x\sigma_x(\tau_x,\tau_y)$} & $\rho_z$ &  $B_1$\\
 16 & \multicolumn{2}{c}{$\rho_x\mu_z\sigma_{y/z}(\tau_x,\tau_y)$} & $\rho_z$ &   $B_1$\\
 17 & \multicolumn{2}{c}{$\rho_x\sigma_x\mu_{x/y}(\tau_x,\tau_y)$} & $\rho_z$ &  $B_1$\\
 18 & \multicolumn{2}{c}{$\rho_x\sigma_x\tau_{x/y}(\mu_x,\mu_y)$} & $\rho_z\tau_z\mu_z$ &  $B_1$\\
 19 &\multicolumn{2}{c}{$\rho_z\tau_z\sigma_x(\mu_x,\mu_y)$} & $\rho_z\tau_z\mu_z$ &$B_1$\\
 20 & \multicolumn{2}{c}{$\rho_z\sigma_{y/z}(\mu_x,\mu_y)$}& $\rho_z\tau_z\mu_z$ &  $B_1$  \\
 21 & \multicolumn{2}{c}{$\rho_z(\mu_x,\mu_y)$} & $\rho_z\tau_z\mu_z$& $B_1$\\
 22 & \multicolumn{2}{c}{$\tau_{x/y}\mu_z\rho_x(\sigma_y,\sigma_z)$} & $\rho_z\tau_z\sigma_x$ &  $B_1$\\
 23 & \multicolumn{2}{c}{$\rho_z(\sigma_y,\sigma_z)$} & $\rho_z\tau_z\sigma_x$ &  $B_1$ \\  
 24 & \multicolumn{2}{c}{$\rho_z\mu_{x/y}(\sigma_y,\sigma_z)$} & $\rho_z\tau_z\sigma_x$ &  $B_1$ \\
 25 & \multicolumn{2}{c}{$\rho_z\tau_z\mu_{z}(\sigma_y,\sigma_z)$} & $\rho_z\tau_z\sigma_x$ & $B_1$ \\
 
 \hline
  26 &\multicolumn{2}{c}{ $\rho_x(\tau_x,\tau_y)$} & $\rho_z$ &  $B_2$ \\
  27 & \multicolumn{2}{c}{$\rho_x\sigma_x\mu_a(\tau_x,\tau_y)$} & $\rho_z$ &   $B_2$ \\
  28 &\multicolumn{2}{c}{ $\rho_x\sigma_{y/z}(\tau_x,\tau_y)$} & $\rho_z$ &   $B_2$ \\
  29 & \multicolumn{2}{c}{$\rho_x\tau_{x/y}\sigma_x(\mu_a,\mu_b)$} & $\rho_z\tau_z\mu_c$ &  $B_2$ \\
  30 &\multicolumn{2}{c}{ $\rho_z\tau_z(\mu_x,\mu_y)$} & $\rho_z\tau_z\mu_z$ & $B_2$ \\
  31 & \multicolumn{2}{c}{$\rho_z\sigma_x(\mu_a,\mu_b)$} & $\rho_z\tau_z\mu_c$ &  $B_2$ \\
  32 & \multicolumn{2}{c}{$\rho_z\tau_z\sigma_{y/z}(\mu_a,\mu_b)$} & $\rho_z\sigma_x\mu_c$ &   $B_2$ \\
  33 & \multicolumn{2}{c}{$\rho_x\tau_{x/y}(\sigma_y,\sigma_z)$} & $\rho_z\tau_z\sigma_x$ &   $B_2$ \\
  34 & \multicolumn{2}{c}{$\rho_z(\sigma_y,\sigma_z)$} & $\rho_z\tau_z\sigma_x$ &   $B_2$ \\
  35 & \multicolumn{2}{c}{$\rho_z\tau_z\mu_i(\sigma_y,\sigma_z)$}& $\rho_z\tau_z\sigma_x$ &   $B_2$ \\
    
\hline
\end{tabular}
\end{center}
\end{table*}

\begin{table*}[h]
\begin{center}
\caption{Forms of possible orders for non-unitary triplet pairing which do not transform into one another with triplet state proportional to $\sigma_y+i\sigma_z$. Note additional symmetry breaking is required for many of these options.}\label{unitaryallorders}
\begin{tabular}{|c c c c c|} 
 \hline
 Partner Orders & Partner 1  & Partner 2 & Partner 3 &  SC Partner \\ [0.5ex] 
 \hline\hline
 1 & $\tau_{x/y}\mu_z\rho_x$ & $\tau_{y/x}\mu_z\rho_x\sigma_{x}$ & $\rho_z\sigma_{x}$ &   $B^U_1$   \\
 2 & $\rho_z\mu_{x/y}$ & $\rho_z\mu_{y/x}\sigma_{x}$ & $\rho_z\tau_z\mu_z\sigma_{x}$ &   $B^U_1$   \\
 \hline
 3 & $\tau_{x/y}\rho_x$ & $\tau_{y/x}\sigma_{x}\rho_x$ & $\sigma_{x}\rho_z$ &   $B^U_2$   \\
 \hline\hline
 4 &\multicolumn{2}{c}{$\rho_x\mu_z\sigma_x(\tau_x,\tau_y)$}  & $\rho_z$ &  $B^U_1$\\
 5 & \multicolumn{2}{c}{$\rho_z(\mu_x,\mu_y)$} & $\rho_z\tau_z\mu_z$ &  $B^U_1$\\
 6 & \multicolumn{2}{c}{$\rho_z\sigma_{x}(\mu_x,\mu_y)$}& $\rho_z\tau_z\mu_z$ &   $B^U_1$  \\
 7 & \multicolumn{2}{c}{$\rho_x\mu_z(\tau_x,\tau_y)$} & $\rho_z$ & $B^U_1$ \\
 \hline
 8 & \multicolumn{2}{c}{$\rho_x(\tau_x,\tau_y)$} & $\rho_z$ &  $B^U_2$ \\
 9 & \multicolumn{2}{c}{$\rho_z\tau_z\sigma_x(\mu_a,\mu_b)$} & $\rho_z\tau_z\mu_c$ &   $B^U_2$ \\
 10 & \multicolumn{2}{c}{$\rho_x\sigma_x(\tau_x,\tau_y)$} & $\rho_z$ & $B^U_2$\\
 11 & \multicolumn{2}{c}{$\tau_z\rho_z(\mu_x,\mu_y)$} & $\rho_z\tau_z\mu_z$ & $B^U_2$\\
  
\hline
\end{tabular}
\end{center}
\end{table*}

\begin{sidewaystable*}[h]
\begin{center}
\caption{The different time-reversal symmetric high-temperature order parameters, their symmetries, and the associated possible partner order parameters, $\mathcal{O}_j = \sum_{\vec{q}} f^\dagger_{\vec{q}} m_j f_{\vec{q}}$, $j=1,2,3$, for $A_1$ and $A_2$ pairing (if existing). Note that we write everything in the full basis, that contain both Dirac cones, and the two options in given in $\{\cdot\}$ can in general mix due to the broken symmetry related to the respective $M$. When there is more than one (symmetry-distinct) option, we separate them by ``or''. We note that $M^s_{3b}$ is related by a mini-valley rotation to $M^s_{3a}$ and the partners for this symmetry breaking term will be related by the same mini-valley rotation to $M^s_{3a}$'s partners, which is why we do not include it explicitly in this table.}
{ \label{SummaryOrderParamtersWithMs}
 \hspace*{-7em}\begin{tabular} {c|cc} \hline \hline
$M$  & $m_j$ for $A_1$ &  $m_j$ for $A_2$ \\ \hline
$\tau_z\mu_z$ & $\rho_z \biggl\{\begin{matrix} \tau_z \\ \mu_z \end{matrix}\biggr\} \sigma_j$\text{ or }$\rho_x(\mu_x,\mu_y)\biggl\{\begin{matrix} \tau_x \\ \tau_y \end{matrix}\biggr\};\rho_z\biggl\{\begin{matrix} \tau_0\mu_0 \\ \tau_z\mu_z \end{matrix}\biggr\}$  & $\rho_z \biggl\{\begin{matrix} \tau_z \\ \mu_z \end{matrix}\biggr\} \sigma_j$\text{ or }$\rho_x(\mu_x,\mu_y)\biggl\{\begin{matrix} \tau_x \\ \tau_y \end{matrix}\biggr\};\rho_z\biggl\{\begin{matrix} \tau_0\mu_0 \\ \tau_z\mu_z \end{matrix}\biggr\}$ \\
$\mu_x$  & $\rho_x \biggl\{\begin{matrix} \mu_0 \\ \mu_x \end{matrix}\biggr\} (\tau_x,\tau_y)$; $\rho_z \biggl\{\begin{matrix} \mu_0 \\ \mu_x \end{matrix}\biggr\}$ \text{or} $\rho_z\tau_z \biggl\{\begin{matrix} \mu_0 \\ \mu_x \end{matrix}\biggr\} \sigma_j$  & --- \\
$\tau_x\rho_y\mu_z$ &   $ \biggl\{\begin{matrix} \rho_z \\ \tau_x \rho_x \end{matrix}\biggr\} (\mu_x,\mu_y)$; $ \biggl\{\begin{matrix} \tau_z\rho_z\mu_z \\ \tau_y \rho_x \end{matrix}\biggr\} $ \text{or} $\biggl\{\begin{matrix} \tau_z\rho_z \\ \tau_y\rho_x\mu_z \end{matrix}\biggr\} \sigma_j$  & --- \\ \hline
$\tau_z\sigma_z$ &  $\rho_z(\mu_x,\mu_y)\biggl\{\begin{matrix} \sigma_0\tau_0 \\ \sigma_z\tau_z \end{matrix}\biggr\};\rho_z\mu_z\biggl\{\begin{matrix} \tau_z\sigma_0 \\ \tau_0\sigma_z \end{matrix}\biggr\}$\text{ or }$\rho_x\mu_z(\tau_x,\tau_y)\biggl\{\begin{matrix} \sigma_x \\ \sigma_y \end{matrix}\biggr\};\rho_z\biggl\{\begin{matrix} \tau_0\sigma_0 \\ \tau_z\sigma_z \end{matrix}\biggr\}$ & $\rho_z(\mu_x,\mu_y;\mu_z)\biggl\{\begin{matrix} \sigma_0\tau_z \\ \sigma_z\tau_0 \end{matrix}\biggr\}$\text{ or }$\rho_x(\tau_x,\tau_y)\biggl\{\begin{matrix} \sigma_x \\ \sigma_y \end{matrix}\biggr\};\rho_z\biggl\{\begin{matrix} \tau_0\sigma_0 \\ \tau_z\sigma_z \end{matrix}\biggr\}$ \\
$\mu_z\sigma_z$ &    $\rho_x (\tau_x,\tau_y) \biggl\{\begin{matrix} \mu_0\sigma_0 \\ \mu_z\sigma_z \end{matrix}\biggr\} $; $\rho_z\biggl\{\begin{matrix} \mu_0\sigma_0 \\ \mu_z\sigma_z \end{matrix}\biggr\} $\text{ or }$\rho_z\tau_z(\mu_x,\mu_y)\biggl\{\begin{matrix} \sigma_x \\ \sigma_y \end{matrix}\biggr\};\rho_z\tau_z\biggl\{\begin{matrix} \mu_z\sigma_0 \\ \mu_0\sigma_z \end{matrix}\biggr\}$  & \hspace{-1em} $\rho_x (\tau_x,\tau_y) \biggl\{\begin{matrix} \mu_0\sigma_z \\ \mu_z\sigma_0 \end{matrix}\biggr\} $; $\rho_z\biggl\{\begin{matrix} \mu_0\sigma_0 \\ \mu_z\sigma_z \end{matrix}\biggr\} $\text{ or }$\rho_z(\mu_x,\mu_y)\biggl\{\begin{matrix} \sigma_x \\ \sigma_y \end{matrix}\biggr\};\rho_z\tau_z\biggl\{\begin{matrix} \mu_z\sigma_0 \\ \mu_0\sigma_z \end{matrix}\biggr\}$ \\
$\tau_z\mu_x\sigma_z$ &  $\rho_x (\tau_x,\tau_y) \biggl\{\begin{matrix} \mu_y\sigma_0 \\ \mu_z\sigma_z \end{matrix}\biggr\} $; $\rho_z\biggl\{\begin{matrix} \mu_0\sigma_0\tau_0 \\ \mu_x\sigma_z\tau_z \end{matrix}\biggr\} $\text{ or }$\rho_z\mu_z(\sigma_x,\sigma_y) \biggl\{\begin{matrix} \tau_z\mu_x \\ \tau_0\mu_0 \end{matrix}\biggr\} ;\rho_z\biggl\{\begin{matrix} \tau_0\mu_x\sigma_0 \\ \tau_z\mu_0\sigma_z \end{matrix}\biggr\} $  & --- \\
$\rho_y\tau_x\sigma_z$  & $(\mu_x,\mu_y) \biggl\{\begin{matrix} \tau_y\rho_x \\ \tau_z\rho_z\sigma_z \end{matrix}\biggr\} $; $\mu_z\biggl\{\begin{matrix} \rho_x\sigma_z\tau_y \\
\rho_z\sigma_0\tau_z\end{matrix}\biggr\} $ or $\mu_z(\sigma_x\sigma_y)\biggl\{\begin{matrix} \rho_z \\ \rho_x\tau_x\end{matrix}\biggr\}$; $\biggl\{\begin{matrix} \tau_z\rho_z\sigma_z \\ \rho_x\tau_y\end{matrix}\biggr\}$  & --- \\
$\rho_y\mu_x\tau_x\sigma_z$  & $\biggl\{ \begin{matrix} \tau_z\rho_z\mu_x\sigma_z \\ \tau_y\rho_x \end{matrix}\biggr\}$; $\biggl\{ \begin{matrix} \rho_z\mu_y \\ \tau_x\rho_x\mu_z\sigma_z \end{matrix}\biggr\}$; $\biggl\{ \begin{matrix} \rho_z\mu_z\sigma_z \\ \tau_x\rho_x\mu_y \end{matrix}\biggr\}$  & \hspace{-1em} $\biggl\{ \begin{matrix} \mu_y\rho_z\sigma_z \\ \mu_z\rho_x\tau_x \end{matrix}\biggr\};\biggl\{ \begin{matrix} \rho_z\sigma_z\mu_z \\ \rho_x\tau_x\mu_y \end{matrix}\biggr\}$; $\biggl\{ \begin{matrix} \tau_z\rho_z\mu_x \\ \tau_y\rho_x\sigma_z \end{matrix}\biggr\}$ \\
$\rho_y\mu_x\tau_x\sigma_z$ &  $\tau_z\mu_y\rho_x(\sigma_x,\sigma_y)\biggl\{\begin{matrix} \rho_y\\ \tau_x\mu_x\sigma_z\end{matrix}\biggr\};\tau_z\rho_z\biggl\{\begin{matrix} \sigma_z \\ \tau_x\rho_y\mu_x\end{matrix}\biggr\}$  & \hspace{-1em} $\rho_z(\sigma_x,\sigma_y)\biggl\{\begin{matrix} \mu_x \\ \tau_x\rho_y\sigma_z\end{matrix}\biggr\};\rho_z\tau_z\biggl\{\begin{matrix} \sigma_z\\ \tau_x\rho_y\mu_x\end{matrix}\biggr\}$ \\
\hline \hline
 \end{tabular}}\label{surviveMsinglet}
\end{center}
\end{sidewaystable*}
\begin{sidewaystable*}[h]
\begin{center}
\caption{The different high-temperature order parameters which are odd under time reversal, their symmetries, and the associated possible partner order parameters, $\mathcal{O}_j = \sum_{\vec{q}} f^\dagger_{\vec{q}} m_j f_{\vec{q}}$, $j=1,2,3$, for $A_2$ pairings only.}
{ \label{SummaryOrderParamtersWithMsTRO}
 \begin{tabular} {c|cc} \hline \hline
$M$  & $m_j$ for $A_2$  \\ \hline
$\tau_z\mu_x$  & $\tau_z\rho_z\sigma_j \biggl\{\begin{matrix} \tau_0\mu_0 \\ \tau_z\mu_x \end{matrix}\biggr\} $ \text{ or } $(\tau_x,\tau_y)\rho_x\mu_z\biggl\{\begin{matrix} \tau_0\mu_0 \\ \tau_z\mu_x \end{matrix}\biggr\};\rho_z\biggl\{\begin{matrix} \tau_0\mu_0 \\ \tau_z\mu_x \end{matrix}\biggr\}$\\
$\tau_x\rho_y$  & $\tau_z\rho_z\sigma_j \biggl\{\begin{matrix} \tau_0\rho_0 \\ \tau_x\rho_y \end{matrix}\biggr\} $ \text{ or } $\tau_z\rho_z\mu_j\biggl\{\begin{matrix} \tau_0\rho_0 \\ \tau_x\rho_y \end{matrix}\biggr\}$\\
\hline
$\mu_x\sigma_z$ & $\mu_x\rho_x(\tau_x,\tau_y) \biggl\{\begin{matrix} \sigma_0\mu_0 \\ \sigma_z\mu_x \end{matrix}\biggr\};\rho_z\biggl\{\begin{matrix} \sigma_0\mu_0 \\ \sigma_z\mu_x \end{matrix}\biggr\} $ \text{ or } $(\sigma_x,\sigma_y)\rho_z\mu_z\biggl\{\begin{matrix} \sigma_0\mu_0 \\ \sigma_z\mu_x \end{matrix}\biggr\};\tau_z\rho_z\biggl\{\begin{matrix} \sigma_z \\ \mu_x \end{matrix}\biggr\}$\\

$\rho_y\tau_x\mu_z\sigma_z$ & $\rho_z\mu_z(\sigma_x,\sigma_y) \biggl\{\begin{matrix} \rho_0\tau_0\mu_0\sigma_0 \\ \rho_y\tau_x\mu_z\sigma_z \end{matrix}\biggr\};\rho_z\tau_z\sigma_z\biggl\{\begin{matrix} \rho_0\tau_0\mu_0\sigma_0 \\ \rho_y\tau_x\mu_z\sigma_z \end{matrix}\biggr\} $ \text{ or } $(\mu_x,\mu_y)\rho_z\sigma_z\biggl\{\begin{matrix} \rho_0\tau_0\mu_0\sigma_0 \\ \rho_y\tau_x\mu_z\sigma_z \end{matrix}\biggr\};\rho_z\tau_z\mu_z\biggl\{\begin{matrix} \rho_0\tau_0\mu_0\sigma_0 \\ \rho_y\tau_x\mu_z\sigma_z\end{matrix}\biggr\}$\\
\hline \hline
 \end{tabular}}\label{surviveMsingletTR}
\end{center}
\end{sidewaystable*}

\begin{sidewaystable*}[h]
\begin{center}
\caption{The different high-temperature order parameters which do not break time reversal symmetry, their symmetries, and the associated possible partner order parameters, $\mathcal{O}_j = \sum_{\vec{q}} f^\dagger_{\vec{q}} m_j f_{\vec{q}}$, $j=1,2,3$, for $B_1$ and $B_2$ pairing (if existing). We only include one of each set of high-temperature orders which are related by valley or mini-valley rotation. Here, our triplet state is aligned in the x direction in spin space ($\Delta\propto\sigma_x$). For orders which depend on spin, we include for each order a possibility aligned with the x direction and one in-plane order. }
{ \label{SummaryTripletOrderParamtersWithMs}
 \hspace*{-7em}\begin{tabular} {c|cc} \hline \hline
$M$  &  $m_j$ for $B_1$ &  $m_j$ for $B_2$ \\ \hline
$\tau_z\mu_z$ &  \hspace{-4em}$\rho_z(\sigma_y,\sigma_z)\biggl\{\begin{matrix} \tau_0\mu_0 \\ \tau_z\mu_z \end{matrix}\biggr\};\rho_z\sigma_x\biggl\{\begin{matrix} \tau_0\mu_z \\ \tau_z\mu_0 \end{matrix}\biggr\}$\text{ or }$\rho_x\sigma_x(\mu_x,\mu_y)\biggl\{\begin{matrix} \tau_x \\ \tau_y \end{matrix}\biggr\};\rho_z\biggl\{\begin{matrix} \tau_0\mu_0 \\ \tau_z\mu_z \end{matrix}\biggr\}$  & \hspace{-5em}$\rho_z(\sigma_y,\sigma_z)\biggl\{\begin{matrix} \tau_0\mu_0 \\ \tau_z\mu_z \end{matrix}\biggr\};\rho_z\sigma_x\biggl\{\begin{matrix} \tau_0\mu_z \\ \tau_z\mu_0 \end{matrix}\biggr\}$\text{ or }$\rho_x\sigma_x(\mu_x,\mu_y)\biggl\{\begin{matrix} \tau_x \\ \tau_y \end{matrix}\biggr\};\rho_z\biggl\{\begin{matrix} \tau_0\mu_0 \\ \tau_z\mu_z \end{matrix}\biggr\}$ \\
$\mu_x$ & $\rho_x\sigma_x(\tau_x,\tau_y) \biggl\{\begin{matrix} \mu_0 \\ \mu_x \end{matrix}\biggr\}$; $\rho_z \biggl\{\begin{matrix} \mu_0 \\ \mu_x \end{matrix}\biggr\}$ \text{or} $\rho_z(\sigma_y,\sigma_z) \biggl\{\begin{matrix} \mu_0 \\ \mu_x \end{matrix}\biggr\};\rho_z\tau_z\sigma_x\biggl\{\begin{matrix} \mu_0 \\ \mu_x \end{matrix}\biggr\}$  & --- \\
$\tau_x\rho_y\mu_z$ &   ---  & \hspace{-12em}$\sigma_x(\mu_x,\mu_y)\biggl\{\begin{matrix} \tau_x\rho_x\mu_0 \\ \tau_0\rho_z\mu_z \end{matrix}\biggr\};\biggl\{\begin{matrix} \tau_z\rho_z\mu_z \\ \tau_y\rho_x \end{matrix}\biggr\}$\text{ or }$\tau_y(\sigma_y,\sigma_z)\biggl\{\begin{matrix} \tau_0\rho_x\mu_0 \\ \tau_x\rho_z\mu_z \end{matrix}\biggr\};\sigma_x\biggl\{\begin{matrix} \tau_y\rho_x\mu_z \\ \tau_z\rho_z\mu_0 \end{matrix}\biggr\}$ \\ \hline

$\tau_z\mu_x\sigma_z$ & --- & \hspace{-12em}$\rho_x(\tau_x,\tau_y)\biggl\{\begin{matrix} \sigma_x\mu_x \\ \sigma_y\mu_0 \end{matrix}\biggr\};\rho_z\biggl\{\begin{matrix} \tau_0\mu_0\sigma_0 \\ \tau_z\mu_x\sigma_z \end{matrix}\biggr\}$\text{ or }$\rho_z(\mu_y,\mu_z)\biggl\{\begin{matrix} \tau_0\sigma_x \\ \tau_z\sigma_y \end{matrix}\biggr\};\rho_z\biggl\{\begin{matrix} \tau_0\mu_0\sigma_z \\ \tau_z\mu_x\sigma_0 \end{matrix}\biggr\}$ \\

$\rho_y\tau_x\sigma_z$ &  $\biggl\{\begin{matrix} \rho_x\tau_y\mu_z \\ \rho_z\tau_z\sigma_z\mu_z \end{matrix}\biggr\};\biggl\{\begin{matrix} \rho_x\sigma_x\tau_x \\ \rho_z\sigma_y \end{matrix}\biggr\};\biggl\{\begin{matrix} \rho_x\tau_x\sigma_y\mu_z \\ \rho_z\sigma_x\mu_z \end{matrix}\biggr\} $\text{ or }$(\mu_x,\mu_y)\biggl\{\begin{matrix} \rho_x\tau_x\sigma_x \\ \rho_z\tau_0\sigma_y \end{matrix}\biggr\};\mu_z\biggl\{\begin{matrix} \rho_x\tau_y\sigma_z \\ \rho_z\tau_z\sigma_0 \end{matrix}\biggr\}$  & --- \\

$\rho_y\mu_x\tau_x\sigma_z$ &  $ \biggl\{\begin{matrix} \rho_x\tau_x\mu_z\\ \rho_z\mu_y\sigma_z \end{matrix}\biggr\};\biggl\{\begin{matrix} \mu_x\rho_x\sigma_x\tau_x \\ \rho_z\sigma_y \end{matrix}\biggr\};\biggl\{\begin{matrix} \rho_x\sigma_y\tau_y\mu_z \\ \rho_z\sigma_x\mu_y\tau_z \end{matrix}\biggr\}$   &  $ \biggl\{\begin{matrix} \rho_x\tau_y\\ \rho_z\tau_z\mu_x\sigma_z \end{matrix}\biggr\};\biggl\{\begin{matrix} \mu_x\rho_x\sigma_x\tau_x \\ \rho_z\sigma_y \end{matrix}\biggr\};\biggl\{\begin{matrix} \rho_x\tau_x\sigma_y\\ \rho_z\sigma_x\mu_x\end{matrix}\biggr\}$ \\

$\rho_y\mu_x\tau_x\sigma_z$ &  $\biggl\{\begin{matrix}\rho_x\sigma_x\tau_x \\ \rho_z\sigma_y\mu_x \end{matrix}\biggr\};\biggl\{\begin{matrix}\rho_x\sigma_x\mu_y\tau_y \\ \rho_z\sigma_y\mu_z\tau_z \end{matrix}\biggr\};\biggl\{\begin{matrix} \rho_x\tau_x\mu_z\sigma_z \\ \rho_z\mu_y \end{matrix}\biggr\}$  & $(\mu_y,\mu_z) \biggl\{\begin{matrix} \rho_x\tau_y\sigma_x\\ \rho_z\tau_z\sigma_y \end{matrix}\biggr\};\biggl\{\begin{matrix} \rho_x\tau_y\sigma_z \\ \rho_z\tau_z\mu_x \end{matrix}\biggr\}$ \\
\hline
$\tau_z\sigma_x$ & \hspace{-4em}$\rho_x\mu_z(\tau_x,\tau_y)\biggl\{\begin{matrix} \sigma_y\\ \sigma_z \end{matrix}\biggr\};\rho_z\biggl\{\begin{matrix} \tau_0\sigma_0 \\ \tau_z\sigma_x \end{matrix}\biggr\}$ or $\rho_z(\mu_x,\mu_y)\biggl\{\begin{matrix} \tau_0\sigma_0 \\ \tau_z\sigma_x \end{matrix}\biggr\}$; $\rho_z\mu_z \biggl\{\begin{matrix}\tau_z\sigma_0 \\ \tau_0\sigma_x \end{matrix}\biggr\}$  & \hspace{-2em}$\rho_x(\tau_x,\tau_y)\biggl\{\begin{matrix} \sigma_y\\ \sigma_z \end{matrix}\biggr\};\rho_z\biggl\{\begin{matrix} \tau_0\sigma_0 \\ \tau_z\sigma_x \end{matrix}\biggr\}$ or $\rho_z(\mu_x,\mu_y)\biggl\{\begin{matrix} \tau_z\sigma_0 \\ \tau_0\sigma_x \end{matrix}\biggr\}$; $\rho_z\tau_z\biggl\{\begin{matrix} \tau_0\mu_z \\ \tau_z\mu_y \end{matrix}\biggr\}$ \\

$\mu_z\sigma_x$ &   \hspace{-4em}$\rho_x\mu_z(\tau_x,\tau_y)\biggl\{\begin{matrix} \mu_0\sigma_0\\ \mu_z\sigma_x \end{matrix}\biggr\};\rho_z\biggl\{\begin{matrix} \mu_0\sigma_0 \\ \mu_z\sigma_x \end{matrix}\biggr\}$ or $\rho_z(\mu_x,\mu_y)\biggl\{\begin{matrix} \sigma_y \\ \sigma_z \end{matrix}\biggr\}$; $\rho_z\tau_z \biggl\{\begin{matrix}\mu_z\sigma_0 \\ \mu_0\sigma_x \end{matrix}\biggr\}$  & \hspace{-2em}$\rho_x(\tau_x,\tau_y)\biggl\{\begin{matrix} \mu_0\sigma_0\\ \mu_z\sigma_x \end{matrix}\biggr\};\rho_z\biggl\{\begin{matrix} \mu_0\sigma_0 \\ \mu_z\sigma_x \end{matrix}\biggr\}$ or $\rho_z\tau_z(\mu_x,\mu_y)\biggl\{\begin{matrix} \sigma_y \\ \sigma_z \end{matrix}\biggr\}$; $\rho_z\tau_z\biggl\{\begin{matrix} \mu_z\sigma_0 \\ \mu_0\sigma_x \end{matrix}\biggr\}$ \\

$\tau_z\mu_x\sigma_x$ &  $\rho_x(\tau_x,\tau_y)\biggl\{\begin{matrix} \mu_z\sigma_0 \\ \mu_y\sigma_x \end{matrix}\biggr\};\rho_z\biggl\{\begin{matrix} \tau_0\mu_0\sigma_0 \\ \tau_z\mu_x\sigma_x \end{matrix}\biggr\}$ or $\rho_z(\sigma_y,\sigma_z)\biggl\{\begin{matrix} \tau_0\mu_y \\ \tau_z\mu_z \end{matrix}\biggr\}$; $\rho_z\biggl\{\begin{matrix} \tau_0\mu_x\sigma_0 \\ \tau_z\mu_0\sigma_x \end{matrix}\biggr\}$& ---  \\

$\rho_y\tau_x\sigma_x$ &  --- & $(\sigma_y,\sigma_z)\biggl\{ \begin{matrix} \rho_x\tau_x \\ \rho_z\tau_0 \end{matrix}\biggr\}$; $\biggl\{ \begin{matrix} \rho_x\tau_y\sigma_0 \\ \rho_z\tau_z\sigma_x \end{matrix}\biggr\}$ \text{ or } $(\mu_x,\mu_y;\mu_z)\biggl\{ \begin{matrix} \rho_x\tau_y\sigma_x \\ \rho_z\tau_z\sigma_0 \end{matrix}\biggr\}$   \\

\hline \hline
 \end{tabular}}\label{surviveMtriplet}
\end{center}
\end{sidewaystable*}

\begin{sidewaystable*}[h]
\begin{center}
\caption{The different high-temperature order parameters which break time reversal symmetry, their symmetries, and the associated possible partner order parameters, $\mathcal{O}_j = \sum_{\vec{q}} f^\dagger_{\vec{q}} m_j f_{\vec{q}}$, $j=1,2,3$, for $B_1$ and $B_2$ pairing (if existing). Again, we only include one of each set of high-temperature orders which are related by valley or mini-valley rotation. Here, our triplet state is aligned in the x direction in spin space ($\Delta\propto\sigma_x$). For orders which depend on spin, we include for each order a possibility aligned with the x direction and one in-plane order.}
{\label{SummaryTripletOrderParamtersWithMsTR}
 \hspace*{-6em}\begin{tabular} {c|cccccc} \hline \hline
$M$   & $m_j$ for $B_1$ &  $m_j$ for $B_2$ \\ \hline

$\tau_z\mu_x$  & ---  &  \hspace{-6em}$\rho_x\sigma_x(\mu_y,\mu_z)\biggl\{\begin{matrix} \tau_x \\ \tau_y \end{matrix}\biggr\};\rho_z\biggl\{\begin{matrix} \tau_0\mu_0 \\ \tau_z\mu_x \end{matrix}\biggr\}$\text{ or }$\rho_z(\sigma_y,\sigma_z)\biggl\{\begin{matrix} \tau_0\mu_0 \\ \tau_z\mu_x \end{matrix}\biggr\};\sigma_x\rho_z\biggl\{\begin{matrix} \tau_0\mu_x \\ \tau_z\mu_y \end{matrix}\biggr\}$ \\

$\rho_y\tau_x$  &  $\sigma_x(\mu_x,\mu_y)\biggl\{\begin{matrix} \rho_x\tau_y \\ \rho_z\tau_z \end{matrix}\biggr\};\mu_z\biggl\{\begin{matrix} \rho_z\tau_z \\ \rho_x\tau_y \end{matrix}\biggr\}$\text{ or }$\mu_z(\sigma_y,\sigma_z)\biggl\{\begin{matrix} \rho_x\tau_y \\ \rho_z\tau_z \end{matrix}\biggr\};\sigma_x\biggl\{\begin{matrix} \rho_z\tau_z \\ \rho_x\tau_y \end{matrix}\biggr\}$ & ---\\

$\rho_y\tau_x\mu_x$   &  $\biggl\{\begin{matrix} \rho_x\tau_x\mu_z \\ \rho_z\mu_y \end{matrix}\biggr\};\biggl\{\begin{matrix} \rho_x\sigma_x\tau_x\mu_y \\ \rho_z\sigma_x\mu_z \end{matrix}\biggr\};\biggl\{\begin{matrix} \rho_x\sigma_x\tau_y \\ \rho_z\tau_z\mu_x\sigma_x \end{matrix}\biggr\}$\text{ or }$(\sigma_y,\sigma_z)\biggl\{\begin{matrix} \rho_x\tau_x\mu_z \\ \rho_z\tau_0\mu_y \end{matrix}\biggr\};\sigma_x\biggl\{\begin{matrix} \rho_x\tau_y\mu_x \\ \rho_z\tau_z\mu_0 \end{matrix}\biggr\}$ & $\sigma_x(\mu_y,\mu_z)\biggl\{\begin{matrix} \rho_x\tau_x \\ \rho_z\tau_0 \end{matrix}\biggr\};\biggl\{\begin{matrix} \rho_z\tau_z\mu_x \\ \rho_x\tau_y\mu_0 \end{matrix}\biggr\};$\text{ or }$(\sigma_y,\sigma_z)\biggl\{\begin{matrix} \rho_x\tau_y\mu_0 \\ \rho_z\tau_z\mu_x \end{matrix}\biggr\};\sigma_x\biggl\{\begin{matrix} \rho_z\tau_z\mu_0 \\ \rho_x\tau_y\mu_x \end{matrix}\biggr\}$\\

\hline
$\sigma_z$    &  $\rho_x\mu_z(\tau_x,\tau_y)\biggl\{\begin{matrix} \sigma_0 \\ \sigma_z \end{matrix}\biggr\};\rho_z\biggl\{\begin{matrix} \sigma_0 \\ \sigma_z \end{matrix}\biggr\}$\text{ or }$\rho_z(\mu_x,\mu_y)\biggl\{\begin{matrix} \sigma_0 \\ \sigma_z \end{matrix}\biggr\};\tau_z\rho_z\mu_z\biggl\{\begin{matrix} \sigma_0 \\ \sigma_z \end{matrix}\biggr\}$ & $\rho_x(\tau_x,\tau_y)\biggl\{\begin{matrix} \sigma_0 \\ \sigma_z \end{matrix}\biggr\};\rho_z\biggl\{\begin{matrix} \sigma_0\\ \sigma_z \end{matrix}\biggr\};$\text{ or }$\rho_z\tau_z(\mu_x,\mu_y;\mu_z)\biggl\{\begin{matrix} \sigma_0 \\ \sigma_z \end{matrix}\biggr\}$\\

$\mu_x\sigma_z$    &  $\rho_x\sigma_x(\tau_x,\tau_y)\biggl\{\begin{matrix} \sigma_0\mu_y \\ \sigma_z\mu_z \end{matrix}\biggr\};\rho_z\biggl\{\begin{matrix} \sigma_0\mu_0 \\ \sigma_z\mu_x \end{matrix}\biggr\}$\text{ or }$\biggl\{\begin{matrix} \mu_x\rho_z \\ \rho_z\sigma_z \end{matrix}\biggr\};\biggl\{\begin{matrix} \sigma_x\mu_z\rho_z \\ \sigma_y\mu_y\rho_z \end{matrix}\biggr\};\biggl\{\begin{matrix} \sigma_x\mu_y\rho_z\tau_z \\ \sigma_y\mu_z\rho_z\tau_z \end{matrix}\biggr\}$ & ---\\

$\tau_z\mu_z\sigma_z$   &  \hspace{-1em}$\rho_x(\tau_x,\tau_y)\biggl\{\begin{matrix} \mu_0\sigma_x \\ \mu_z\sigma_y \end{matrix}\biggr\};\rho_z\biggl\{\begin{matrix} \tau_0\mu_0\sigma_0 \\ \tau_z\mu_z\sigma_z \end{matrix}\biggr\}$\text{ or }$\rho_z(\mu_x,\mu_y)\biggl\{\begin{matrix} \tau_z\mu_0\sigma_x \\ \tau_0\mu_z\sigma_y \end{matrix}\biggr\};\rho_z\biggl\{\begin{matrix} \tau_z\mu_z\sigma_0 \\ \tau_0\mu_0\sigma_z \end{matrix}\biggr\}$ & \hspace{-1em}$\rho_x(\tau_x,\tau_y)\biggl\{\begin{matrix} \tau_0\mu_z\sigma_x \\ \tau_z\mu_0\sigma_y \end{matrix}\biggr\};\rho_z\biggl\{\begin{matrix} \tau_0\mu_0\sigma_0 \\ \tau_z\mu_z\sigma_z \end{matrix}\biggr\}$\text{ or }$\rho_z(\mu_x,\mu_y)\biggl\{\begin{matrix} \tau_0\sigma_x \\ \tau_z\sigma_y \end{matrix}\biggr\};\rho_z\biggl\{\begin{matrix} \tau_z\mu_z\sigma_0 \\ \tau_0\mu_0\sigma_z \end{matrix}\biggr\}$\\

$\rho_y\tau_x\mu_z\sigma_z$ & ---  &  \hspace{-4em}$\biggl\{\begin{matrix} \rho_x\tau_y \\ \rho_z\tau_z\mu_z\sigma_z \end{matrix}\biggr\};\biggl\{\begin{matrix} \rho_x\sigma_x\tau_x\mu_z \\ \rho_z\sigma_y \end{matrix}\biggr\};\biggl\{\begin{matrix} \rho_x\tau_x\sigma_y \\ \rho_z\mu_z\sigma_x \end{matrix}\biggr\}$\text{ or }$(\mu_x,\mu_y)\biggl\{\begin{matrix}\rho_x\tau_y\sigma_x \\ \rho_z\tau_z\sigma_y \end{matrix}\biggr\};\biggl\{\begin{matrix}\rho_x\tau_y\mu_0\sigma_z \\ \rho_z\tau_z\mu_z\sigma_0 \end{matrix}\biggr\}$ \\
\hline

$\mu_x\sigma_x$ & ---  &  $\rho_x(\tau_x,\tau_y)\biggl\{\begin{matrix} \mu_0\sigma_0 \\ \mu_x\sigma_x \end{matrix}\biggr\};\rho_z\biggl\{\begin{matrix} \mu_0\sigma_0 \\ \mu_x\sigma_x \end{matrix}\biggr\}$\text{ or }$\rho_z\tau_z(\mu_y;\mu_z)\biggl\{\begin{matrix}\sigma_y \\ \sigma_z \end{matrix}\biggr\};\rho_z\tau_z\biggl\{\begin{matrix}\mu_x\sigma_0 \\ \mu_0\sigma_x \end{matrix}\biggr\}$ \\

$\rho_y\tau_x\mu_z\sigma_x$   &  $\biggl\{\begin{matrix} \rho_x\tau_y\mu_0\sigma_x \\ \rho_z\tau_z\mu_z\sigma_0 \end{matrix}\biggr\};(\mu_x,\mu_y)\biggl\{\begin{matrix} \rho_x\tau_x\sigma_x \\ \rho_z\tau_0\sigma_0 \end{matrix}\biggr\}$\text{ or }$(\sigma_y,\sigma_z)\biggl\{\begin{matrix} \rho_x\tau_x\mu_z \\ \rho_z\tau_0\mu_0 \end{matrix}\biggr\};\biggl\{\begin{matrix} \rho_x\tau_y\mu_z\sigma_0 \\ \rho_z\tau_z\mu_0\sigma_x \end{matrix}\biggr\}$& ---  \\

\hline \hline
 \end{tabular}}\label{surviveMtripletTR}
\end{center}
\end{sidewaystable*}

\end{widetext}

\clearpage
%

\end{document}